\documentstyle[11pt,epsfig]{article}
\begin{document}
\renewcommand{\thefootnote}{\fnsymbol{footnote}}
\sloppy
\newcommand{\rp}{\right)}
\newcommand{\lp}{\left(}
\newcommand \be  {\begin{equation}}
\newcommand \bea {\begin{eqnarray}}
\newcommand \ee  {\end{equation}}
\newcommand \eea {\end{eqnarray}}

\title{Log-periodic power law bubbles in Latin-American and Asian markets
and correlated anti-bubbles in Western stock markets: An empirical study}
\thispagestyle{empty}

\author{Anders Johansen$^1$ and Didier Sornette$^{1,2,3}$\\
$^1$ Institute of Geophysics and
Planetary Physics\\ University of California, Los Angeles, California 90095\\
e-mail: anders@moho.ess.ucla.edu. \\
$^2$ Department of Earth and Space Science\\
University of California, Los Angeles, California 90095\\
$^3$ Laboratoire de Physique de la Mati\`{e}re Condens\'{e}e\\ CNRS UMR6622 and
Universit\'{e} de Nice-Sophia Antipolis\\ B.P. 71, Parc
Valrose, 06108 Nice Cedex 2, France \\
e-mail: sornette@unice.fr}

\maketitle

\begin{abstract}

Twenty-two significant bubbles followed by large crashes or by severe
corrections in
the Argentinian, Brazilian, Chilean, Mexican, Peruvian, Venezuelan, Hong-Kong,
Indonesian, Korean, Malaysian, Philippine and Thai stock markets indices
are identified and analysed for log-periodic signatures decorating an average
power law acceleration. We find that log-periodic power laws adequately
describe speculative bubbles on these emerging markets with very few exceptions
and thus extend considerably the applicability of the proposed rational
expectation model of bubbles and crashes which has previously been developed
for the major financial markets in the world. This model is essentially
controlled by a crash hazard rate becoming critical due to a collective
imitative/herding behavior of traders. Furthermore, three of the bubbles are
followed by a log-periodic ``anti-bubble'' previously documented for the decay
of the Japanese Nikkei starting in Jan. 1990 and the price of Gold starting in
Sept. 1980 thus rendering a qualitative symmetry of bubble and anti-bubble
around the date of the peak of the market. A set of secondary western
stock market indices (London, Sydney, Auckland, Paris, Madrid, Milan, Zurich) 
as well as the Hong-Kong stock market are also shown to exhibit well-correlated
log-periodic power law anti-bubbles over a period 6-15 months triggered by 
a rash of crises on emerging markets in the early 1994. As the US market 
declined by no more than $10\%$ during the beginning of that period and quickly
recovered, this suggests that these smaller stock western markets can ``phase 
lock'' (in a weak sense) not only because of the over-arching influence of Wall
Street but also independently of the current trends on Wall Street due to other
influences.

\end{abstract}

\newpage

\pagenumbering{arabic}

\section{Introduction}

A series of works have documented a robust and universal signature preceding
large crashes occuring in major financial stock markets, namely accelerated
price increase decorated by large scale log-periodic oscillations culminating
close to a critical point
\cite{SJB96,FF96,SJ97,predic,Thesis,SJ98,97crash,Gluzman,Van1,Van2,JS98.2,manisfesto,JLS,JSL,Drozdz}.
Specifically, in the simplest form, the index $I(t)$ can be represented by the
following time dependence
\be \label{lpeq}
I\lp t\rp  = A + B\lp t_c -t\rp^z + C\lp t-t_c\rp^z\cos\lp \omega
\log \lp t_c - t\rp - \phi\rp ~,
\ee
where $A$ is the terminal price at the critical time $t_c$, the exponent
$0 < z < 1$ describes an acceleration and $\omega$ and $\phi$ are
respectively the angular frequency of the log-periodic oscillations and
their phase (or time unit). The log-periodic oscillations, {\it i.e.}, periodic
in the variable $\log \lp t_c - t\rp$, are the hallmark of a discrete scale
invariance \cite{physreports} since the argument of the cosine is reproduced
at each time $t_n$ converging to $t_c$ according to a geometrical time series
$t_c-t_n \propto \lambda^{-n}$ where $\lambda = e^{2\pi / \omega}$.

The previously reported cases well-described by eq.(\ref{lpeq}) comprise
the Oct. 1929 US crash, the Oct. 1987 world market crash, the Oct. 1997
Hong-Kong crash, the Aug. 1998 global market events, the 1985 Forex event on
the US dollar, the correction on the US dollar against the Canadian dollar and
the Japanese Yen starting in Aug.~ 1998, as well as the bubble on the Russian
market and its ensuing collapse in June 1997 \cite{JSL}. Symmetrically, ``anti-bubbles''
with decelerating market devaluations following all-time highs have also been
found to carry strong log-periodic structures \cite{antibulle}, represented by
eq.(\ref{lpeq}) with $t_c -t$ changed into $t-t_c$,
\be \label{lpdeceq}
\log\lp I\lp t\rp \rp = A + B\lp t -t_c\rp^z + C\lp t-t_c\rp^z\cos\lp \omega
\log \lp t - t_c\rp - \phi\rp ~.
\ee
The use of the logarithm of the index instead of the index as in
(\ref{lpeq}) has been discussed in \cite{SJ97,JS98.2,JSL} and is related to 
the duration of the bubble/anti-bubble as well as to the dynamics controlling 
the amplitude of the collapse. A quite remarkable strong example of such an 
anti-bubble is given by the Japanese Nikkei stock index from 1990 to present 
as well as by the Gold future prices after 1980, both after their all-time 
highs. For the Nikkei, a theoretical formulation \cite{antibulle} allowed
us to issue a quantitative prediction in early Jan.1999 (when the Nikkei was 
at its low), that the index will exhibit a recovery over 1999 and 2000
\cite{antibulle,predicNikkei}.

The hypothesis to rationalize these empirical facts first proposed in 
\cite{SJB96} is that stock
market crashes are caused by the slow buildup of long-range correlations
between traders leading to the expansion of a speculative bubble that may
become unstable at a critical time and lead to a crash or to a drastic change
of market regime. Bubbles are considered to be natural occurrences of the
dynamics of stock markets, as argued persuasively by Keynes \cite{Keynes} and
illustrated intuitively in classroom experiments \cite{BallHolt}. It is
possible to be more quantitative and construct a rational expectation model of
bubbles and crashes based on Blanchard's model \cite{Blanchard}, which has two
main components\,: (1) we assume that a crash may be caused by {\em local}
self-reinforcing imitation processes between noise traders which can be
quantified within the frame-work of critical phenomena developed in the 
Physical
Sciences and (2) we allow for a remuneration of the risk of a crash by a higher
rate of growth of the bubble, which reflects that the crash is not a certain
deterministic outcome of the bubble and, as a consequence, it remains rational
for traders to remain invested provided they are suitably compensated. The
bubble price is then completely controlled by the crash hazard rate and we
have proposed that its acceleration and its log-periodic structures are the
hallmark of a discrete scale invariance, appearing as a result of
self-organising interactions between traders \cite{SJB96,JLS,JSL}.

These empirical facts have until now been restricted to the major financial
markets of the world (WMFMs), {\it i.e.}, the stock markets on Wall Street,
Tokyo and Hong Kong as well as the foreign exchange market (FOREX) and the
Gold market in the seventies and early eighties. Recently, it was established
\cite{JSL} that the Russian stock market in $\approx [1996.2-1997.6]$ exhibited
an extended bubble followed by a (relatively slow but) large crash, which had
strong characteristics of log-periodicity decorating a power law acceleration
of the index, similar to those found in the WMFMs \cite{JS98.2}. This raises
the question whether such behavior may be found in emerging markets in general
or if the Russian case is unique due to its rather special characteristics
\cite{Intriligator}.

The purpose of the present analysis is twofold. The main objective is to
answer this question concerning whether log-periodic power laws can be as
successfully applied to speculative bubbles on emerging markets as it has
been done on the WMFMs. This is done by analyzing a range of emerging stock
markets using the same tools as in the previous analysis of the WMFMs, as
well as comparable time scales. The second objective is to illustrate on a
qualitative level using log-periodic signatures that the smaller Western
stock markets are strongly influenced by the leading trends on Wall Street.
Furthermore, we will show quantitatively that these smaller stock markets can
``phase lock'' (in a weak sense) not only because of the over-all influence
of Wall Street but also independently of the current trends on Wall Street.

The methodology we adopt is the one used in our previous works on the WMFMs,
which consists in a combination of parametric fits with formulas like
(\ref{lpeq}) and of non-parametric log-frequency analysis
\cite{JS98.2,manisfesto,JLS,JSL}. We have established the reliability of
this approach by extensive numerical tests on synthetic data. The use of the
same method will allow us to test the hypothesis that emerging markets exhibit
bubbles and crashes with similar log-periodic signatures as in the WMFMs.
We stress from the beginning that the results obtained on the emerging markets
analysed here does not carry the same robustness as obtained for the WMFMs with
respect to identification of the bubble and the values obtained for the
exponent $z$ and the frequency $\omega$ of the log-periodic oscillations.
We expect in part the technical difficulties in maintaining a high-quality
stock markets index on these smaller emerging markets to be responsible for
this. More important is presumably the fact that these emergent markets are
strongly influenced by events not directly related to the economy and stock
market of that particular country due to their smaller size. This also explains
the fact that the life-time of the bubbles identified on these emergent markets
are in general of somewhat shorter time-span compared to the bubbles and
anti-bubbles previously identified on the WMFMs. Fundamentally, this is
related to the question over which time-scales can a given market be regarded
as a closed system with a good approximation and connects to the question of
finite-size effects, a question that has been much studied in relation to
critical phenomena in physical systems \cite{finitesize}.

\section{Emerging Markets}

\subsection{Speculative bubbles}

Emerging markets are often the focus of interest and also often exhibit large
financial crises \cite{Rand}. The story of financial bubbles and crashes has
repeated itself over the centuries and in many different locations since the
famous tulip bubble of 1636 in Amsterdam, almost without any alteration in
its main global characteristics \cite{Galbraith,Montroll}.
\begin{enumerate}
\item The bubble starts smoothly with some
increasing production and sales (or demand for some commodity), in an otherwise
relatively optimistic market.
\item The interest for investments with good potential gains then leads to
increasing
investments possibly with leverage coming from novel sources, often from
international
investors. This leads to price appreciation.
\item This in turn attracts less sophisticated investors and, in addition,
leveraging
is further developed with small down payment (small margins or binders), which
lead to a demand for stock rising faster than the rate at which real money
is put in the
market.
\item At this stage, the behavior of the market becomes weakly coupled or
practically
uncoupled from real wealth (industrial and service) production.
\item As the price skyrocket, the number of new investors entering the
speculative market
decreases and the market enters a phase of larger nervousness, until a
point when
the instability is revealed and the market collapses.
\end{enumerate}
This scenario applies essentially to all market crashes, including old ones
such as Oct. 1929 on Wall Street, for which the US market was considered to
be at that time an interesting ``emerging'' market with good investment
potentialities for national as well as international investors. The robustness
of this scenario is presumably deeply rooted in investors psychology and
involves a combination of imitative/herding behavior and greediness (for the
development of the speculative bubble) and over-reaction to bad news in
period of instabilities.

\subsection{Classification of markets}

The commonalities recalled above does not imply that different markets
exhibit the same price trajectories. There can be strong difference due to
local constraints, such as cash flow restrictions, government control and
so on. In our analysis of several emerging markets, we find three main classes.

\subsubsection{Latin-American markets}

The Latin-American stock markets, that we analyze in details below, seems to
display features reminiscent of the largest financial markets, however, with
much larger fluctuations in the values obtained for the exponent $z$ and
log-angular frequency $\omega$. In the next sections, we will see to what
extent this similarity can be quantified. Specifically, the accelerating
log-periodic power law (\ref{lpeq}) will be fitted to the various stock market
data preceding large crashes as well as large decreases. We do not posit that
a crash has to occur suddenly, only that it marks the end of an accelerating
bullish period and the beginning of a bearish regime cumulating in a
significant drop.

\subsubsection{Asian tigers}

The stock markets of Asian Tigers, specifically Korea, Malaysia and Thailand
as well as that of the Philippines and Indonesia also display approximate
accelerating power law bubbles and subsequent crashes. However, the
acceleration accompanying the observed bubbles in these markets often seem
incompatible with the requirement of either $0<z<1$ or, more important, that
of a {\it real} power law with $t<t_c$ in eq. (\ref{lpeq}). This since the
optimisation algorithm kept on insisting on a $t_c$ smaller that the last
data point, thus causing a floating point error. For the smaller Asian stock
markets studied here this problem could be cured working on the {\it
logarithm} stock market index instead with the exception of the Korean stock
market and the 1997 crash in Indonesia. Again we mention that depending on the 
price dynamics ending the bubble, the index or the logarithm of the index 
turns out to be the relevant observable quantifying the acceleration of the 
bubble \cite{JS98.2,JSL}. Naturally, the nature of the log-periodic 
oscillations does not dependent of which observable is used.

As an additional example of the difference between the WMFMs and the stock
markets of Indonesia, Korea, Malaysia, Philippines and Thailand, we find that
similarly to what is seen for the Latin-American markets, the values obtained
for the exponent $z$ and log-angular frequency $\omega$ from the fitting with
eq. (\ref{lpeq}) fluctuates considerably compared to the WMFMs, as
reported in table \ref{asitab1} and \cite{JS98.2}. A recent analysis shows that
the Asian stock returns exhibit characteristics of bubbles, which are however
incompatible in details with the prediction of the model of rational
speculative bubbles \cite{Chanetal}. This suggests that a different formulation
than simply using eq. (\ref{lpeq}) is needed in order to capture the
trends displayed by these South-East and East Asian stock markets prior to
large corrections and crashes. In this respect, we note that the Korean stock
market could not be shown to display bubbles following eq. (\ref{lpeq}) nor
the Indonesian crash of July 1997.

\subsubsection{East-European stock markets}

The East-European stock markets seems to be following a completely different
logic than their larger Western counterparts and their indices does not
resemble those of the other markets. In particular, we find that they do not
follow neither power law accelerations nor log-periodic patterns though 
large crashes certainly occurs.

\subsection{Latin-American markets}

\subsubsection{Identification of bubbles}

\begin{figure}
\begin{center}
\epsfig{file=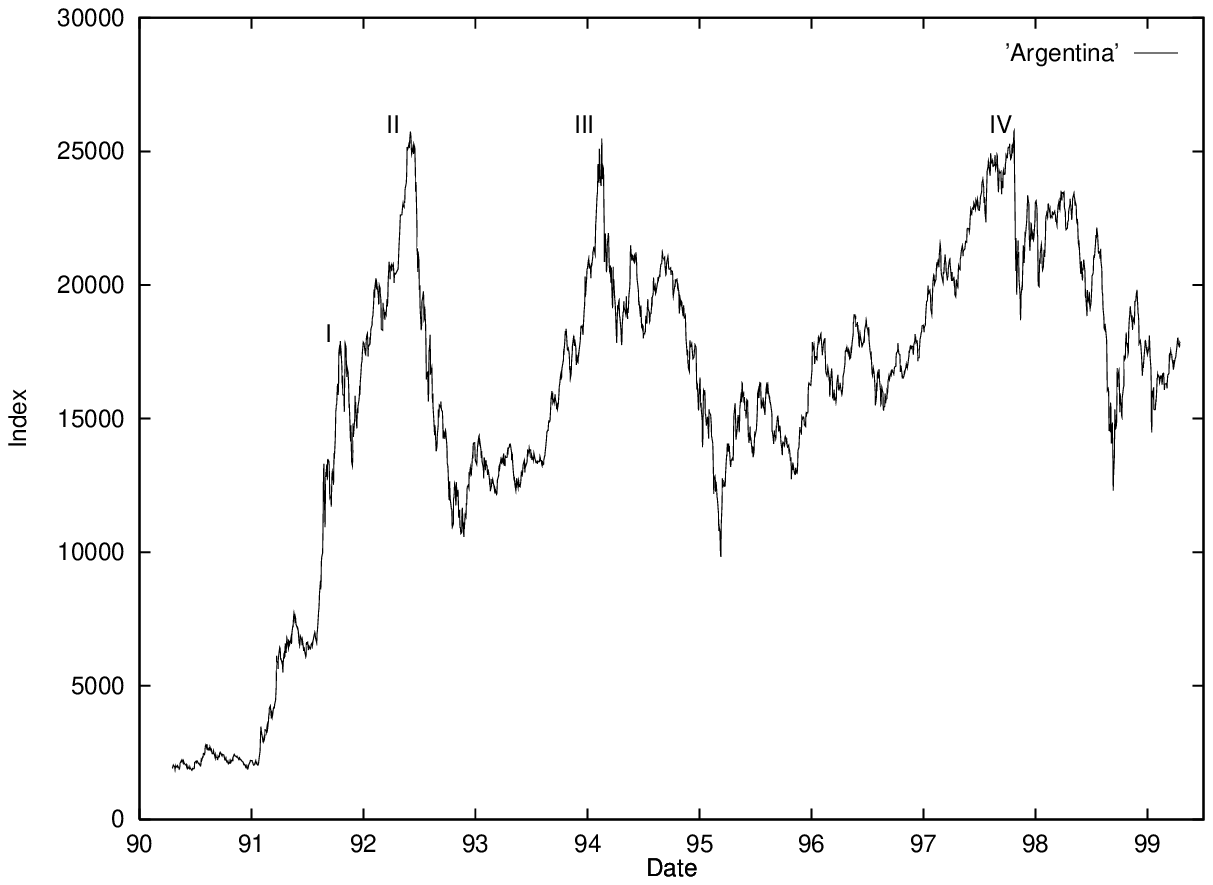,height=9cm,width=16cm}
\caption{\protect\label{arg} The Argentinian stock market index as a function
of date. 4 bubbles with a subsequent very large draw down can be identified.
The approximate dates are in chronological order mid-91 (I),  early 93 (II),
early 94 (III) and late 97 (IV).}
\vspace{1cm}
\epsfig{file=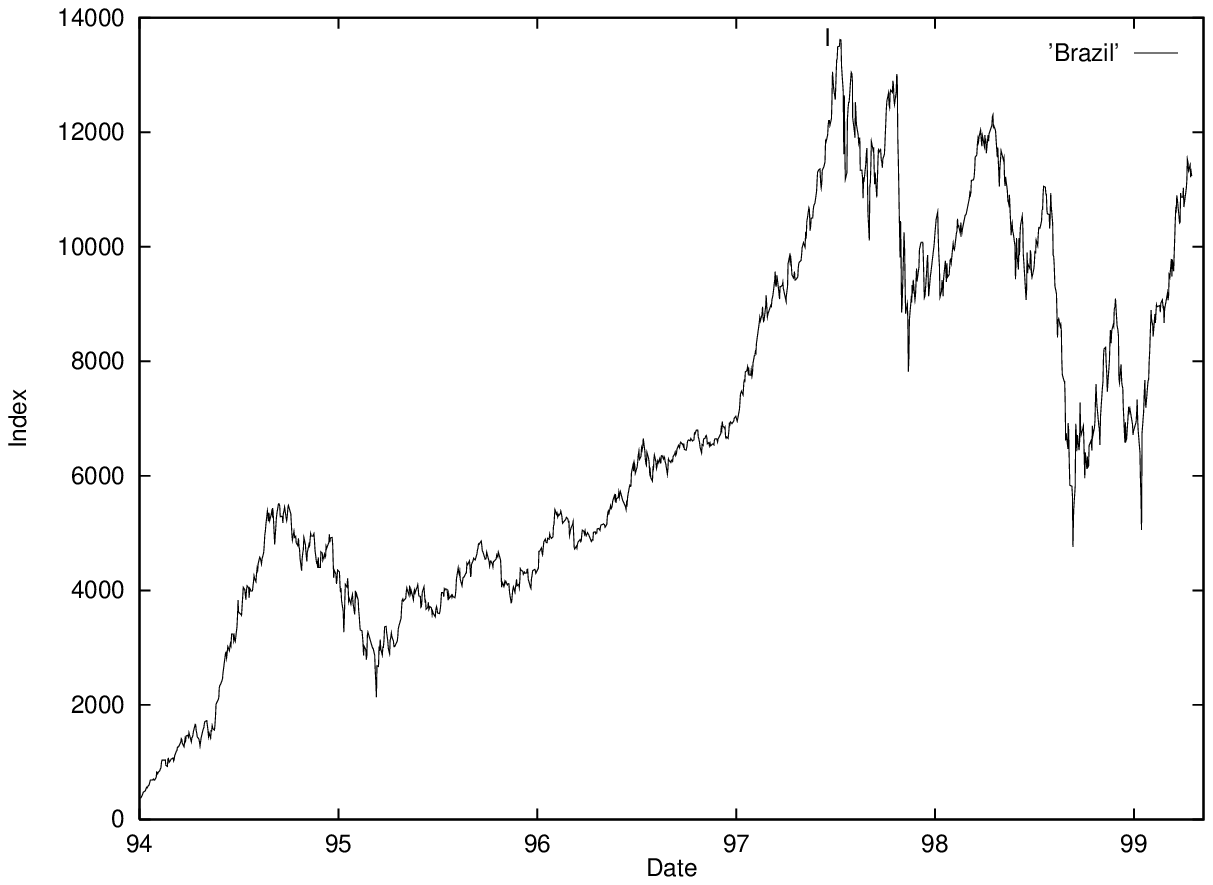,height=9cm,width=16cm}
\caption{\protect\label{bra}The Brazilian stock market index as a function of
date. 1 bubble with a subsequent very large draw down can be identified. The
approximate date is mid-97 (I).}
\end{center}
\end{figure}

\begin{figure}
\begin{center}
\epsfig{file=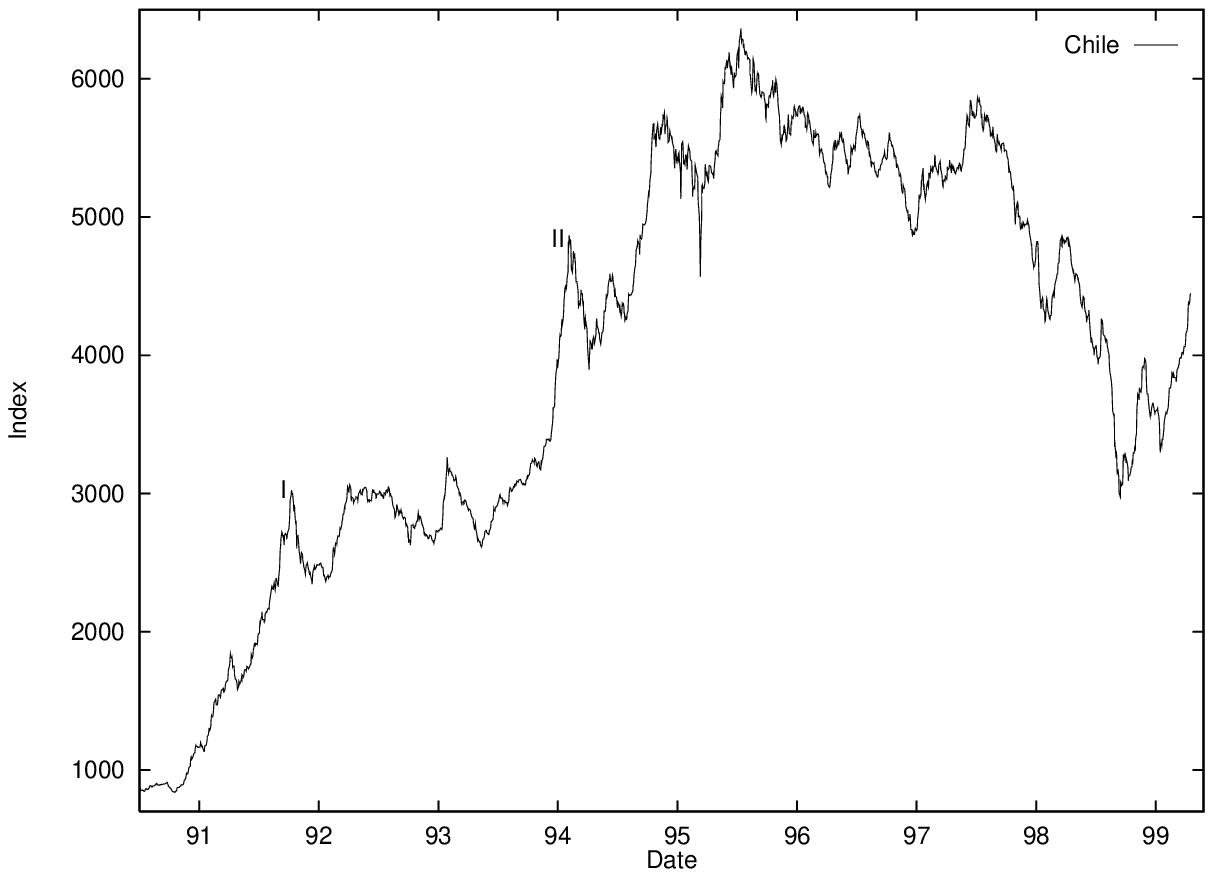,height=9cm,width=16cm}
\caption{\protect\label{chil} The Chilean stock market index as a function
of date. 2 bubbles with a subsequent very large draw down can be identified.
The approximate dates are in chronological order mid-91 (I) and early 94
(II).}
\vspace{1cm}
\epsfig{file=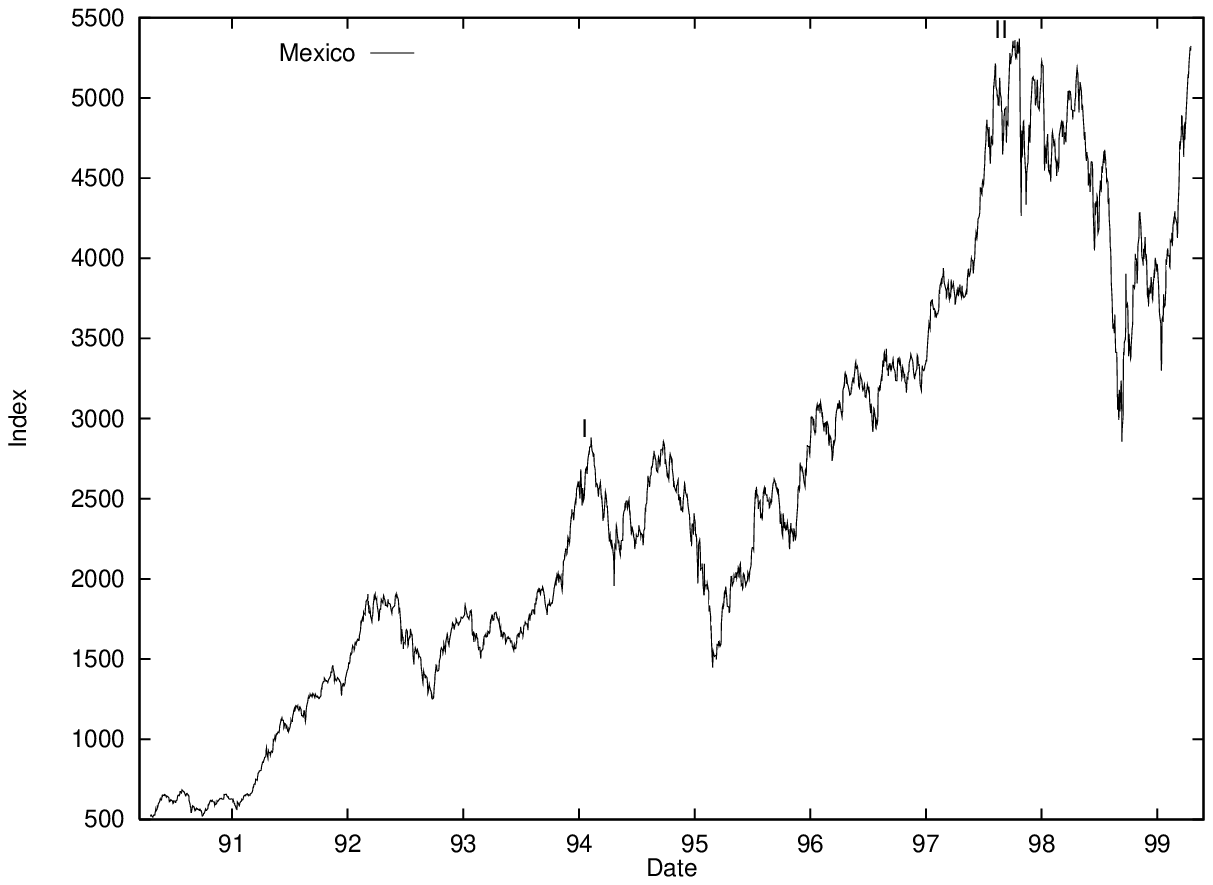,height=9cm,width=16cm}
\caption{\protect\label{mexi} The Mexican stock market index as a function
of date. 2 bubbles with a subsequent very large draw down can be identified.
The approximate dates are in chronological order early 94 (I) and mid-97
(II).}
\end{center}
\end{figure}

\begin{figure}
\begin{center}
\epsfig{file=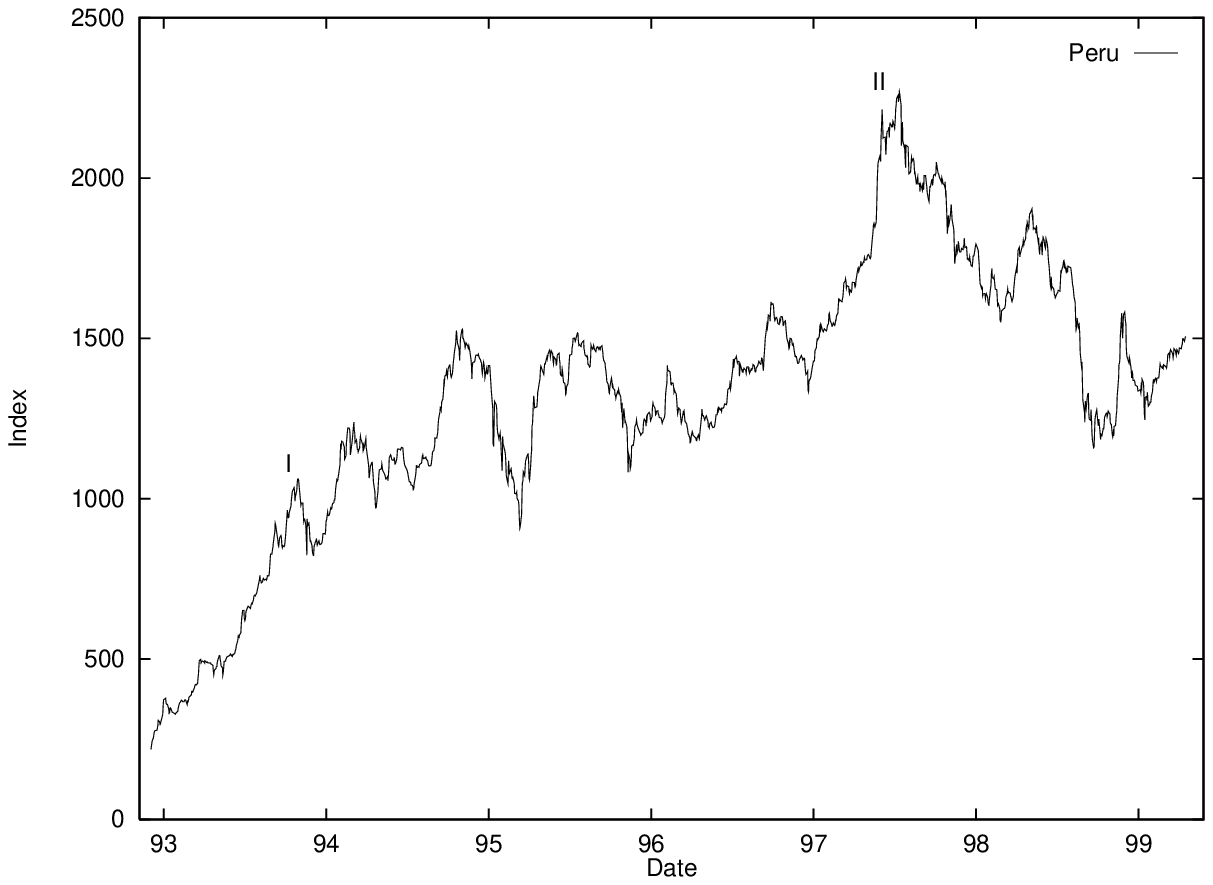,height=9cm,width=16cm}
\caption{\protect\label{peru} The Peruvian stock market index as a function
of date. 2 bubbles with a subsequent very large draw down can be identified.
The approximate dates are in chronological order late 93 (I) and mid-97 (II).}
\vspace{1cm}
\epsfig{file=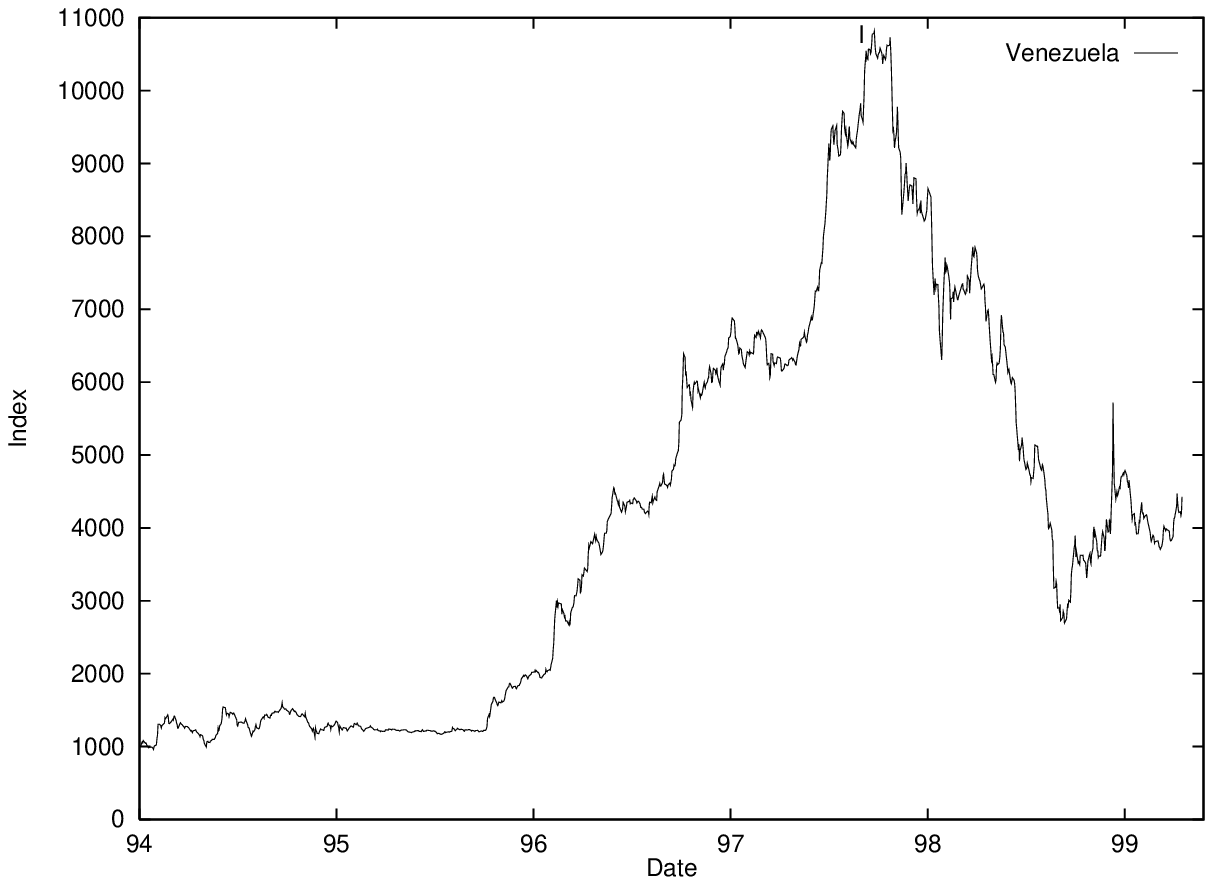,height=9cm,width=16cm}
\caption{\protect\label{venu}The Venezuelan stock market index as a function
of date. 1 bubble with a subsequent very large draw down can be identified.
The approximate date is mid-97 (I).}
\end{center}
\end{figure}

In figure \ref{arg} to \ref{venu}, the evolution of six Latin-American
stock market indices (Argentina, Brazil, Chile, Mexico, Peru and Venezuela)
is shown as a function of time from early in this decade to Feb. 1999.

We first define a bubble as a period of time going from a pronounced
minimum to a large maximum by a prolonged price acceleration, followed by
a crash or a large decrease represented by a bear-market. As for the WMFMs,
such a bubble is defined unambiguously by identifying its end with the date
$t_{max}$ where the highest value of the index is reached prior to the
crash/decrease. For the bubbles prior to the largest crashes on the WMFMs, the
beginning of a bubble is clearly identified as coinciding always with the date
of the lowest value of the index prior to the change in trend. However, this
identification is not as straightforward for the Latin-American and smaller
Asian indices analyzed here. Hence, in approximately half the cases, the date 
of the first data point used in defining the beginning of the bubble had to be
moved up and the bubble had to be truncated in order to obtain fits with
non-pathological values for $z$ and $\omega$. This may well be an artifact
stemming from the restrictions in the fitting imposed by using a single cosine
as the periodic function in eq. (\ref{lpeq}). We recall that the exponent
$z$ is expected to lie between zero and one and it should be not too close to
both zero and one: too small a $z$ implies a flat bubble with a very sudden
acceleration at the end. Too large a $z$ corresponds to an almost linear
non-accelerating bubble. The angular frequency $\omega$ of the log-periodic
oscillations must also not be too small or too large. If it is too small,
less than one oscillation occurs over the whole interval and the log-periodic
oscillation has little meaning. If it is too large, the oscillations are too
numerous and they start to fit the high-frequency noise.

\begin{figure}
\begin{center}
\parbox[l]{8cm}{\epsfig{file=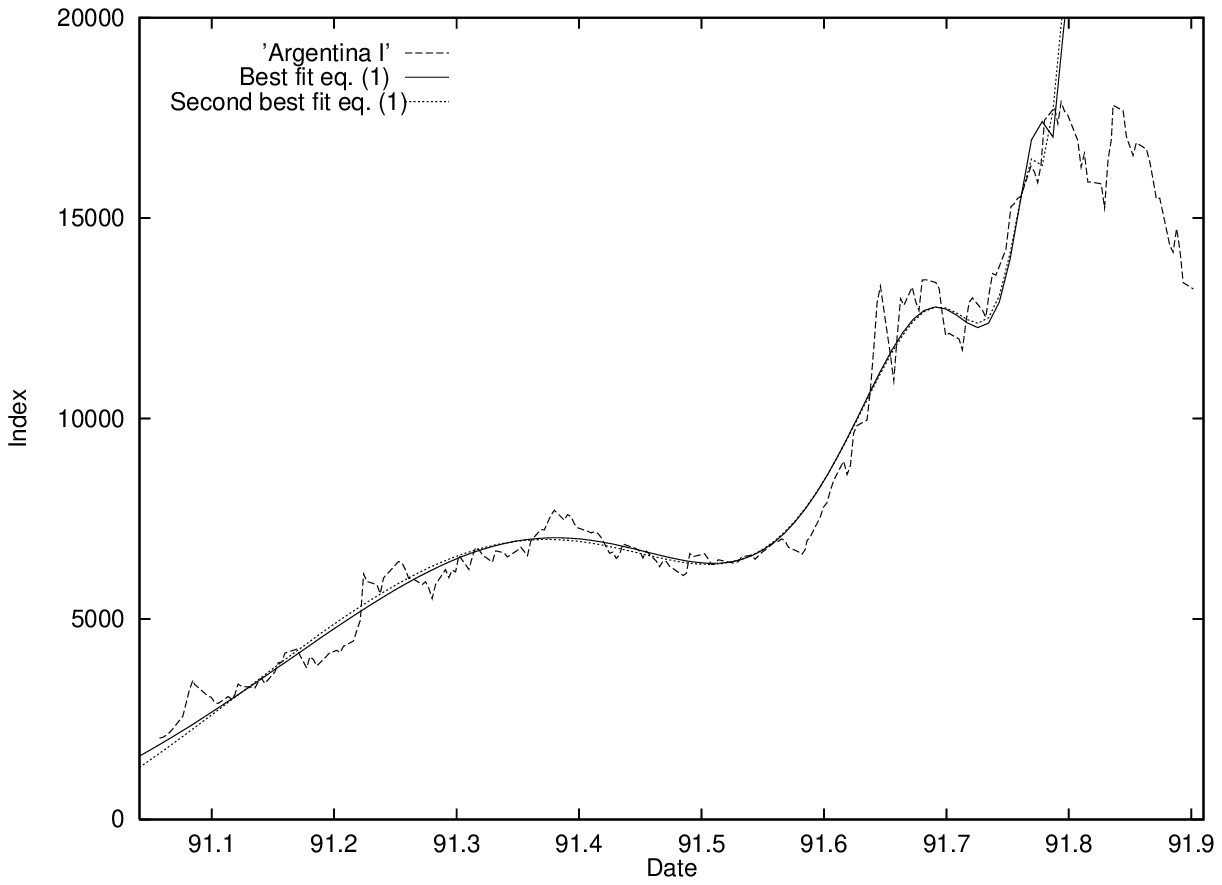,height=7cm,width=8cm} }
\hspace{5mm}
\parbox[r]{8cm}{\epsfig{file=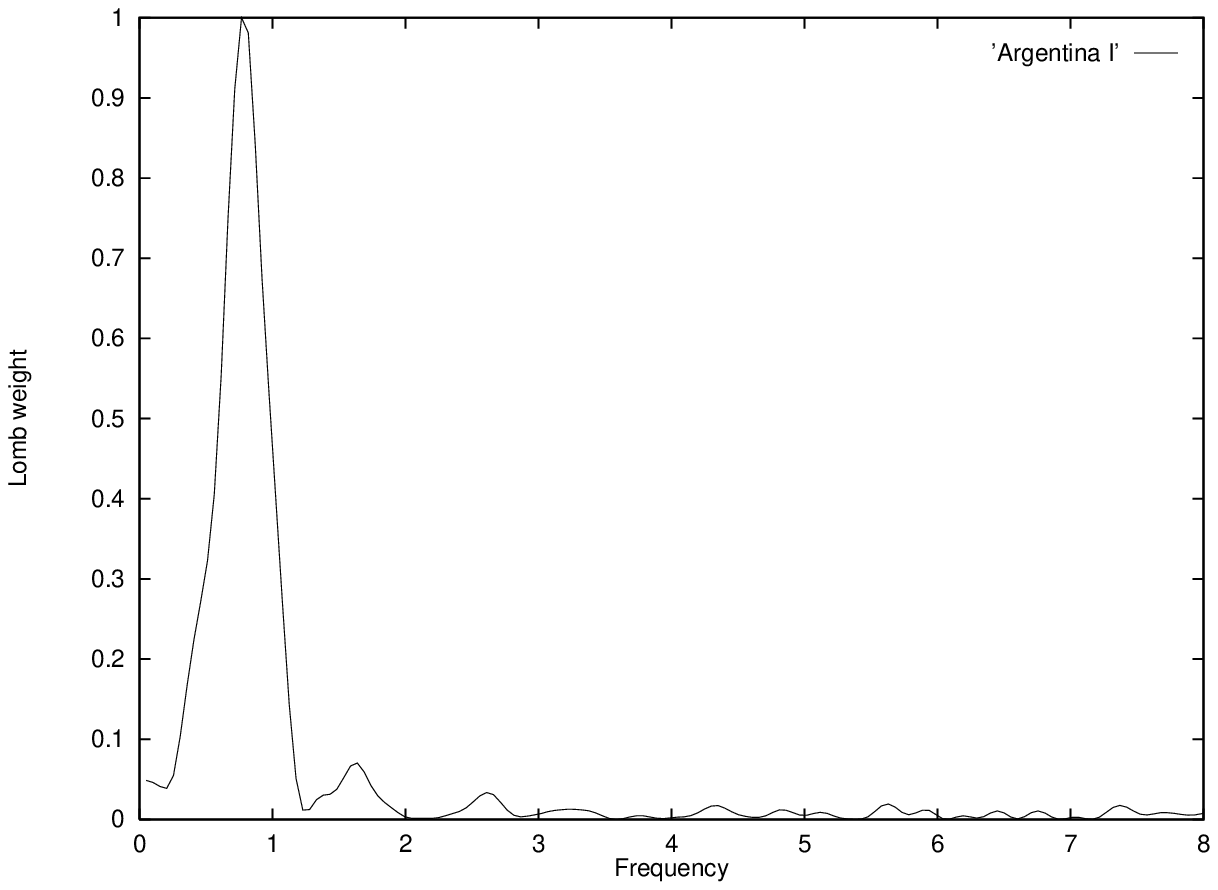,height=7cm,width=
8cm} }
\caption{\protect\label{argbub1} The Argentinian stock market bubble of 1991.
See table \protect\ref{lattab2} for the parameter values of the fits with
eq. (\protect\ref{lpeq}). Only the best fit is used in the Lomb periodogram.}

\vspace{5mm}

\parbox[l]{8cm}{\epsfig{file=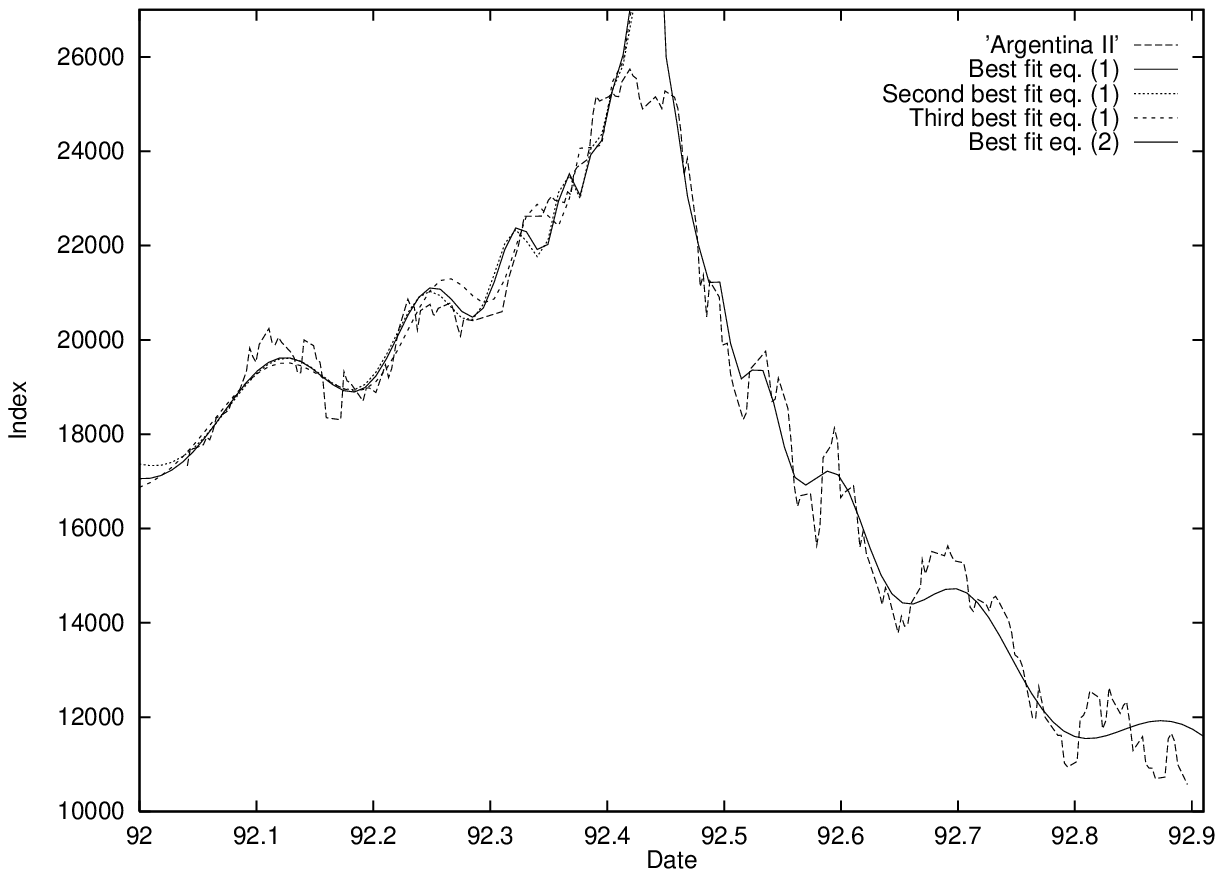,height=7cm,width=8cm} }
\hspace{5mm}
\parbox[r]{8cm}{\epsfig{file=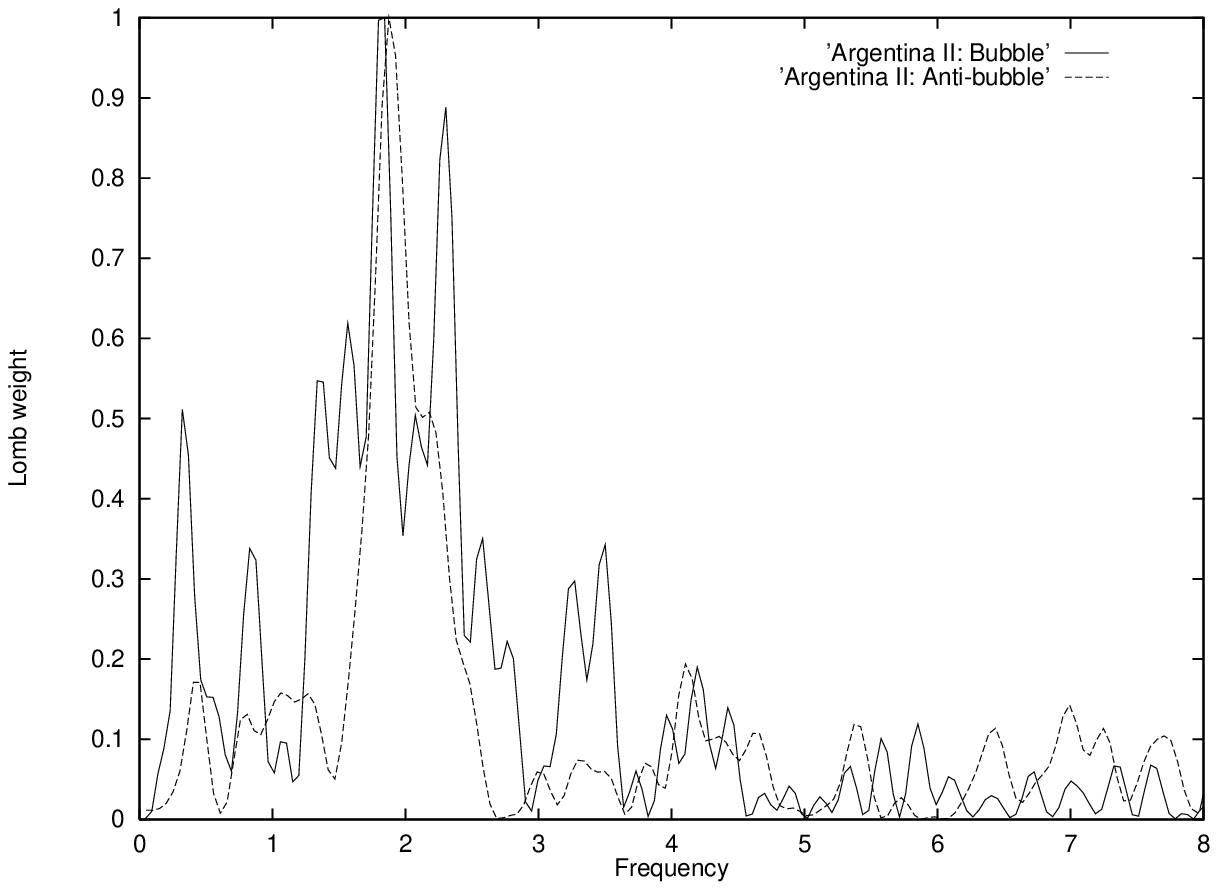,height=7cm,width=8cm}}
\caption{\protect\label{argbub2} The Argentinian stock market bubble and anti-
bubble of 1992.  See table \protect\ref{lattab2} for the parameter values of
the fit with eq. \protect\ref{lpeq} and table \protect\ref{lattab3} for eq.
(\protect\ref{lpdeceq}). Only the best fit is used in the Lomb periodograms.}
\end{center}
\end{figure}

\begin{figure}
\begin{center}
\parbox[l]{8cm}{\epsfig{file=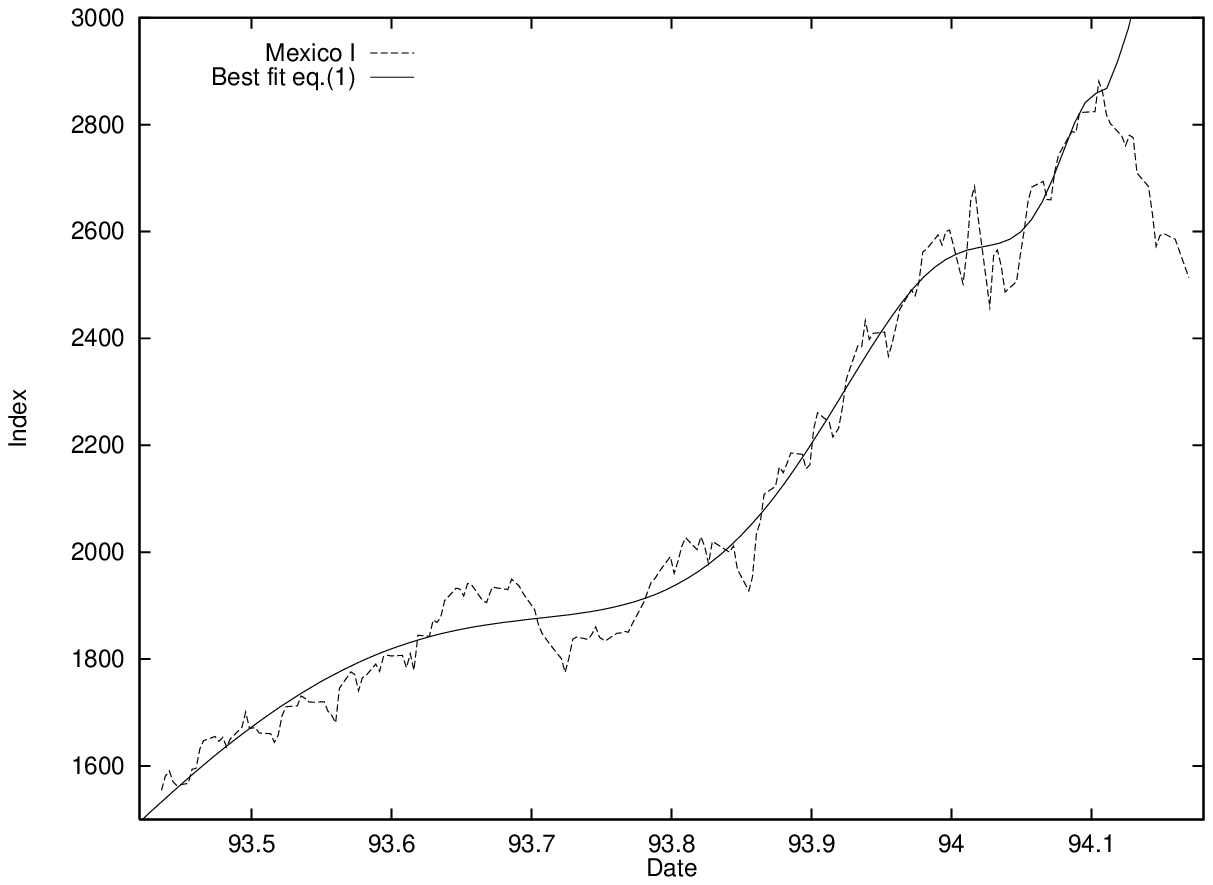,height=7cm,width=8cm} }
\hspace{5mm}
\parbox[r]{8cm}{\epsfig{file=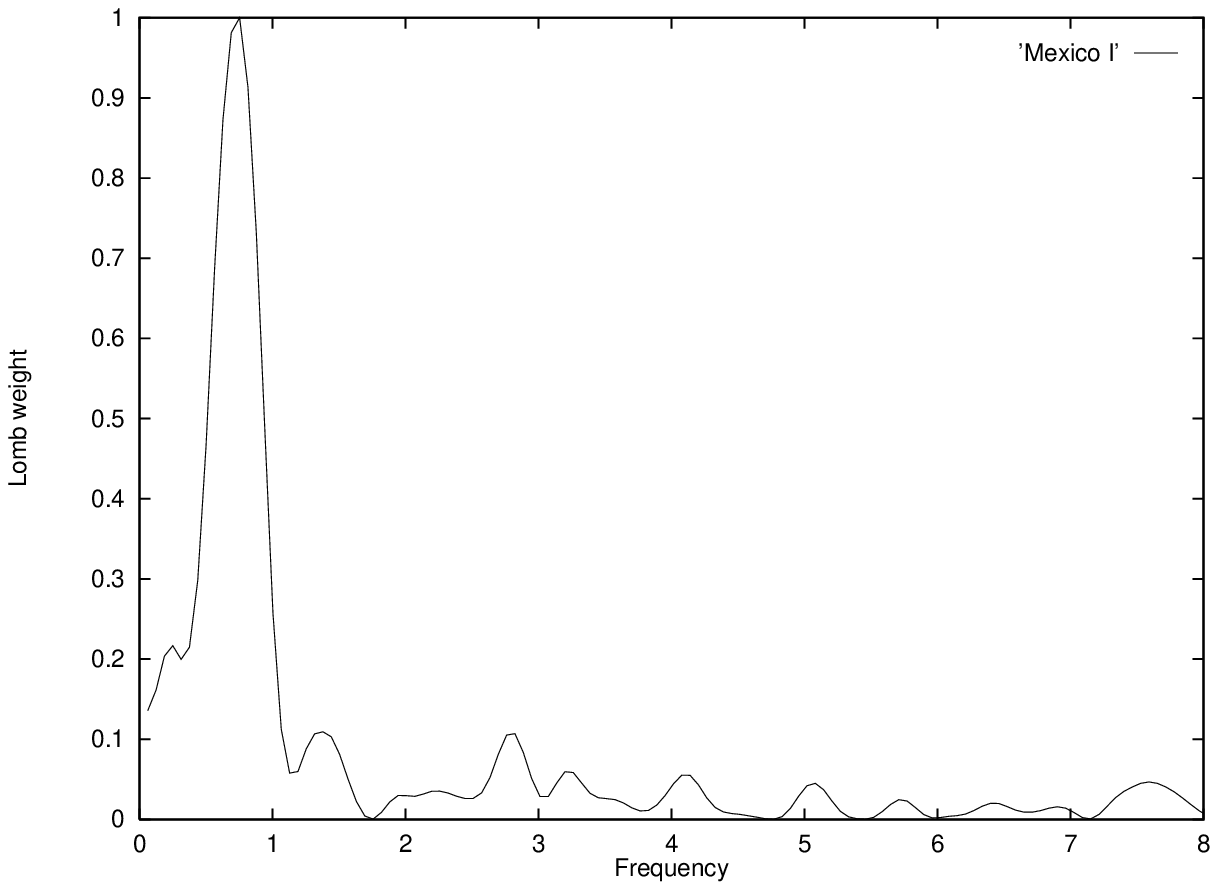,height=7cm,width=8cm} }
\caption{\protect\label{mexbub1} The Mexican bubble ending 1994.  See
table \protect\ref{lattab2} for the parameter values of the fit
with eq. (\protect\ref{lpeq}).}

\vspace{5mm}

\parbox[l]{8cm}{\epsfig{file=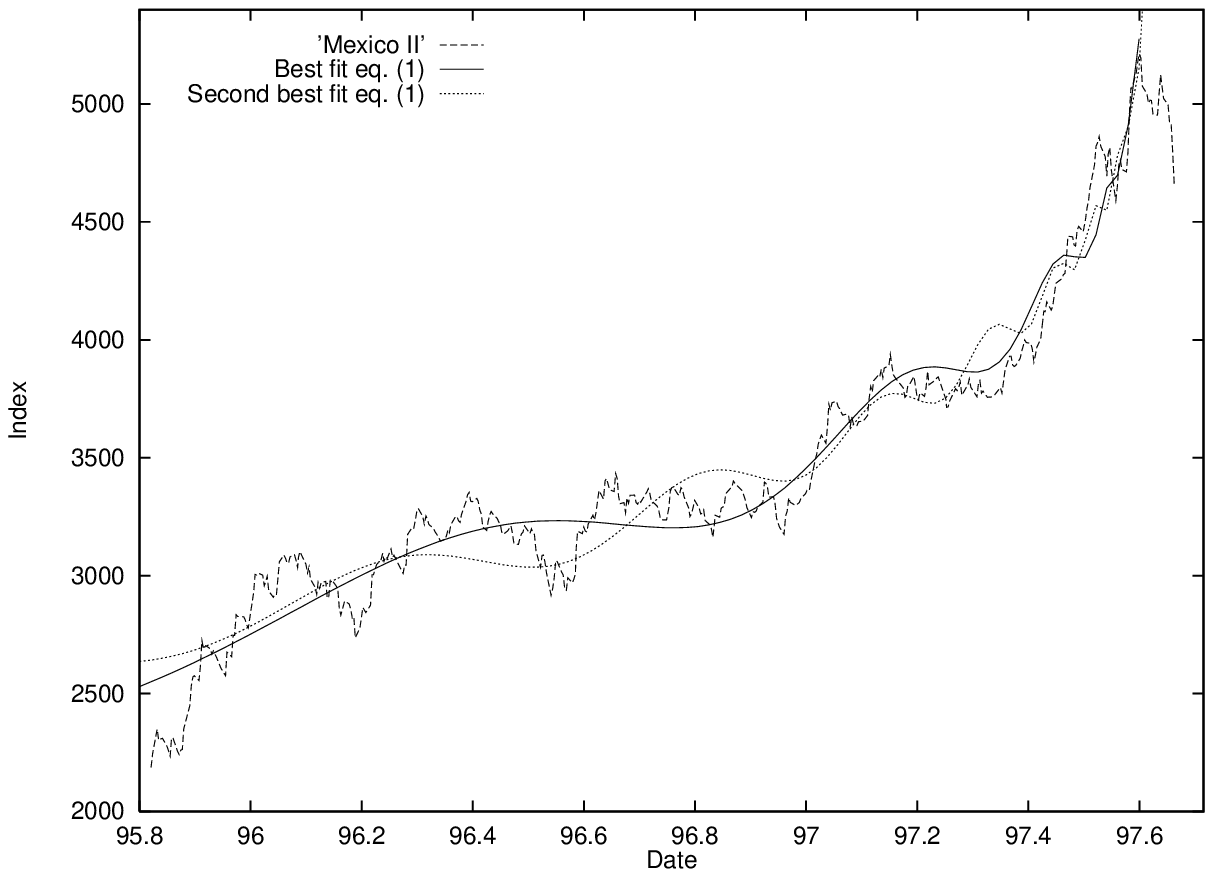,height=7cm,width=8cm} }
\hspace{5mm}
\parbox[r]{8cm}{\epsfig{file=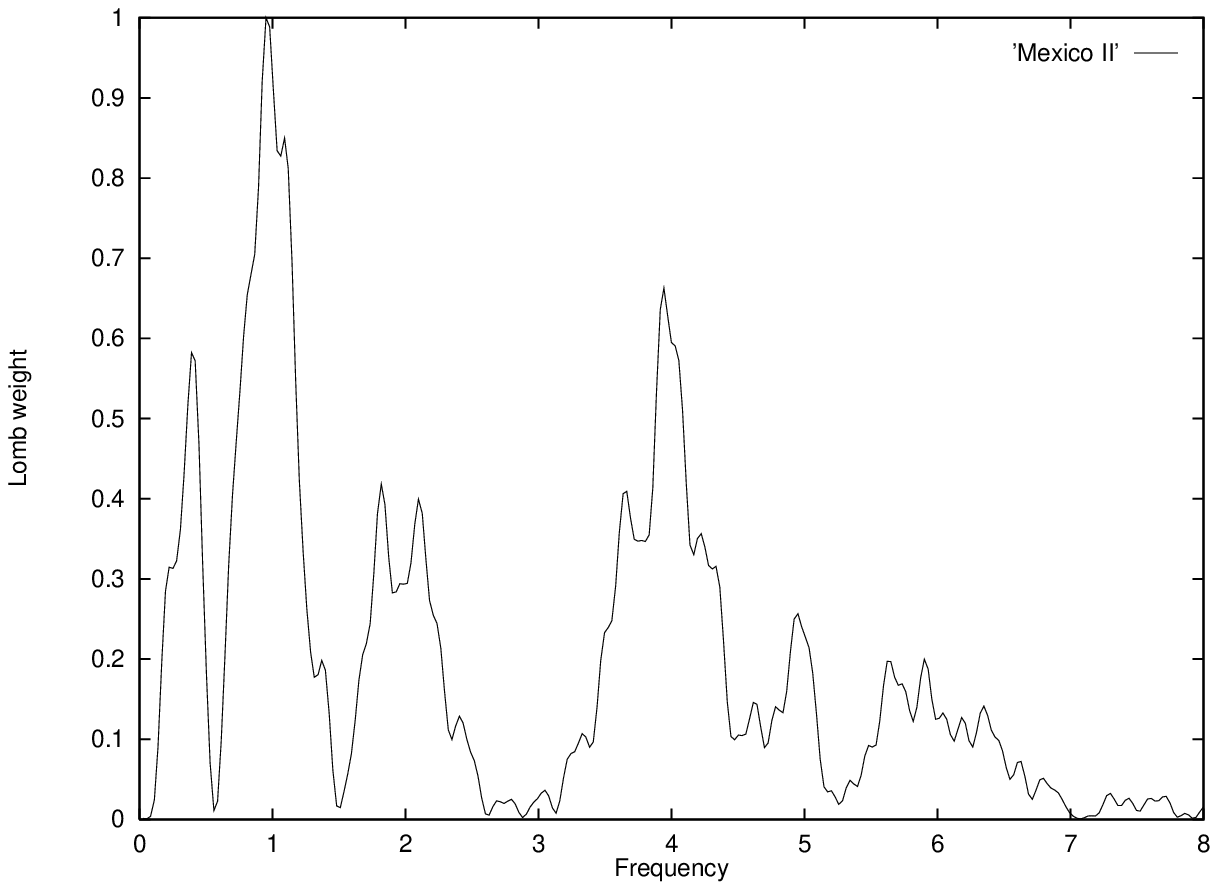,height=7cm,width=8cm} }
\caption{\protect\label{mexbub2} The Mexican bubble ending in 1997.
See table \protect\ref{lattab2} for the parameter values of the fit with
eq. (\protect\ref{lpeq}). Only the best fit is used in the Lomb periodogram.}
\end{center}
\end{figure}

From the six stock market indices, we have identified by eye four Argentinian
bubbles, one Brazilian bubble, two Chilean bubbles, two Mexican bubbles, two
Peruvian bubbles and a single Venezuelan bubble, with a subsequent large
crash/decrease, as shown in figures \ref{arg} to \ref{venu}. Before the reader 
starts to argue that our procedure is rather arbitrary and that many other 
bubbles can be seen on the figures, we stress that times scales considered 
should be comparable with those of the larger crashes analyzed in 
\cite{JS98.2,JLS,JSL} and not considerably less than one year. This has been 
achieved in most cases for the bubbles, whereas the life-time of the 
anti-bubbles seems to be shorter as a rule. Exceptions are the first and second
Argentinian bubbles, the second Chilean bubble and the first Mexican, as shown 
in figures \ref{argbub1},  \ref{argbub2}, \ref{chilbub2} and \ref{mexbub1}, 
where the fitted  interval is $\approx 0.7$ years except for the second 
Argentinian bubble, where only $\approx 0.4$ years could be fitted. On purpose,
we have restrained from analyzing log-periodic structures on smaller scales in 
order to obtain a good comparison with our previous analysis on WMFMs. That the
time-scales on which the bubbles has been identified in general are shorter
than for the WMFMs is as mentioned not very surprising.

\subsubsection{Results} \label{latresults}

\begin{figure}
\begin{center}
\parbox[l]{8cm}{\epsfig{file=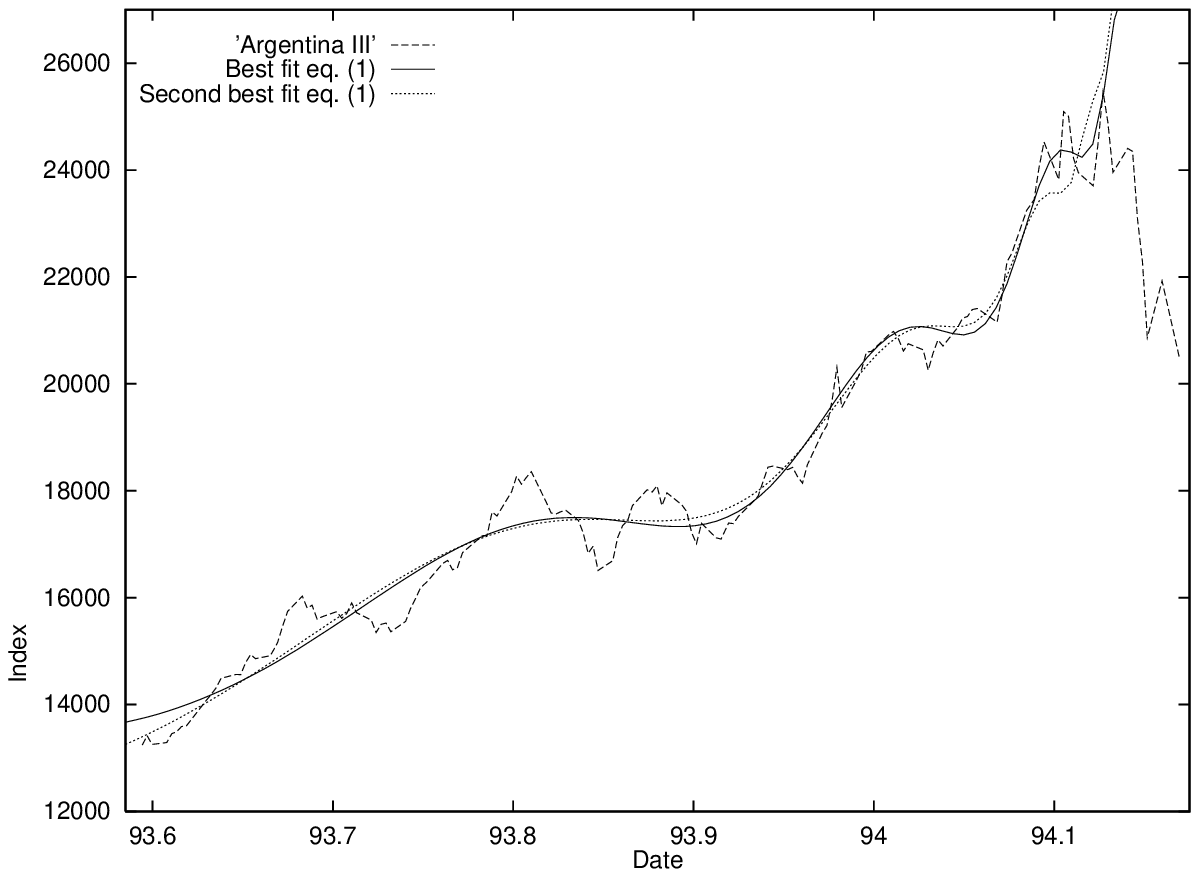,height=7cm,width=8cm} }
\hspace{5mm}
\parbox[r]{8cm}{\epsfig{file=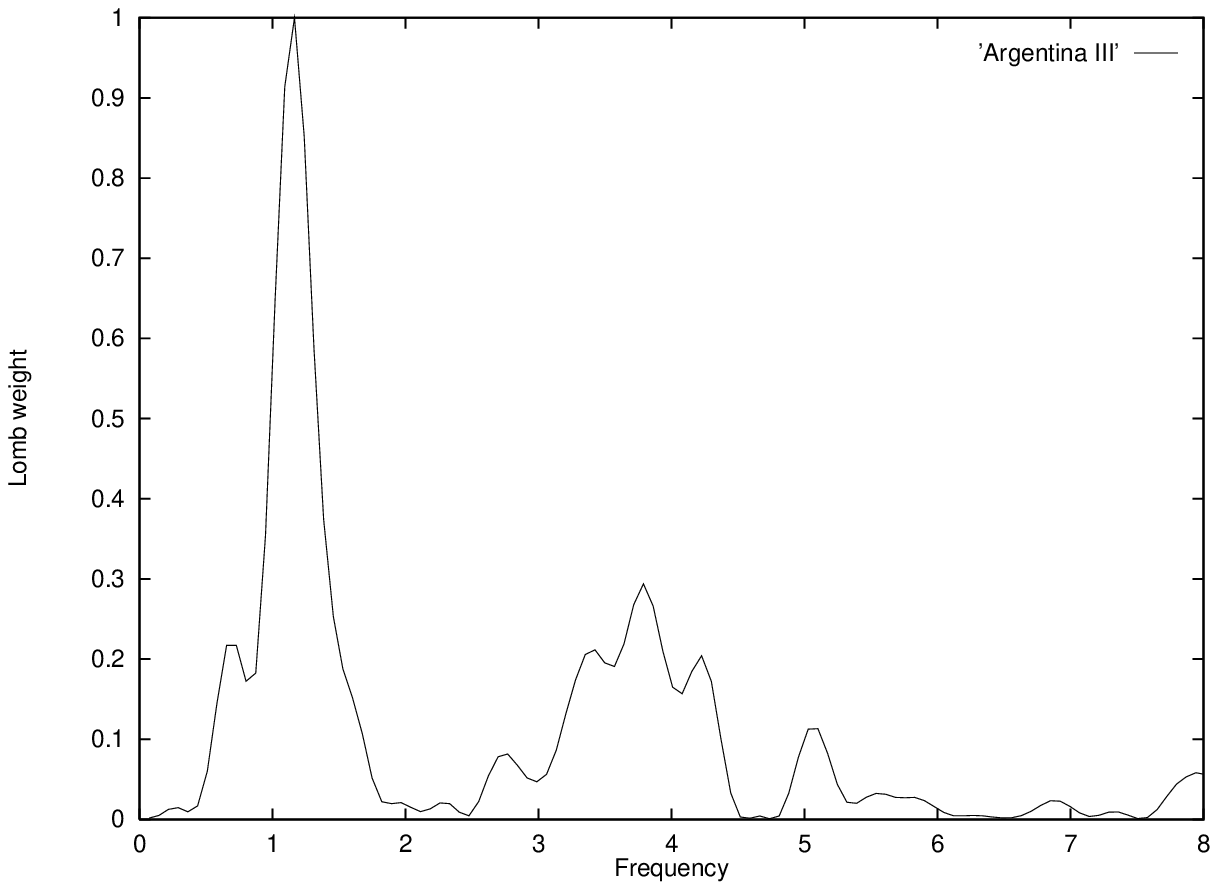,height=7cm,width
=8cm}}
\caption{\protect\label{argbub3} The Argentinian stock market bubble ending
in 1994.
See table \protect\ref{lattab2} for the parameter values of the fit with eq.
(\protect\ref{lpeq}). Only the best fit is used in the Lomb periodogram.}

\vspace{5mm}

\parbox[l]{8cm}{\epsfig{file=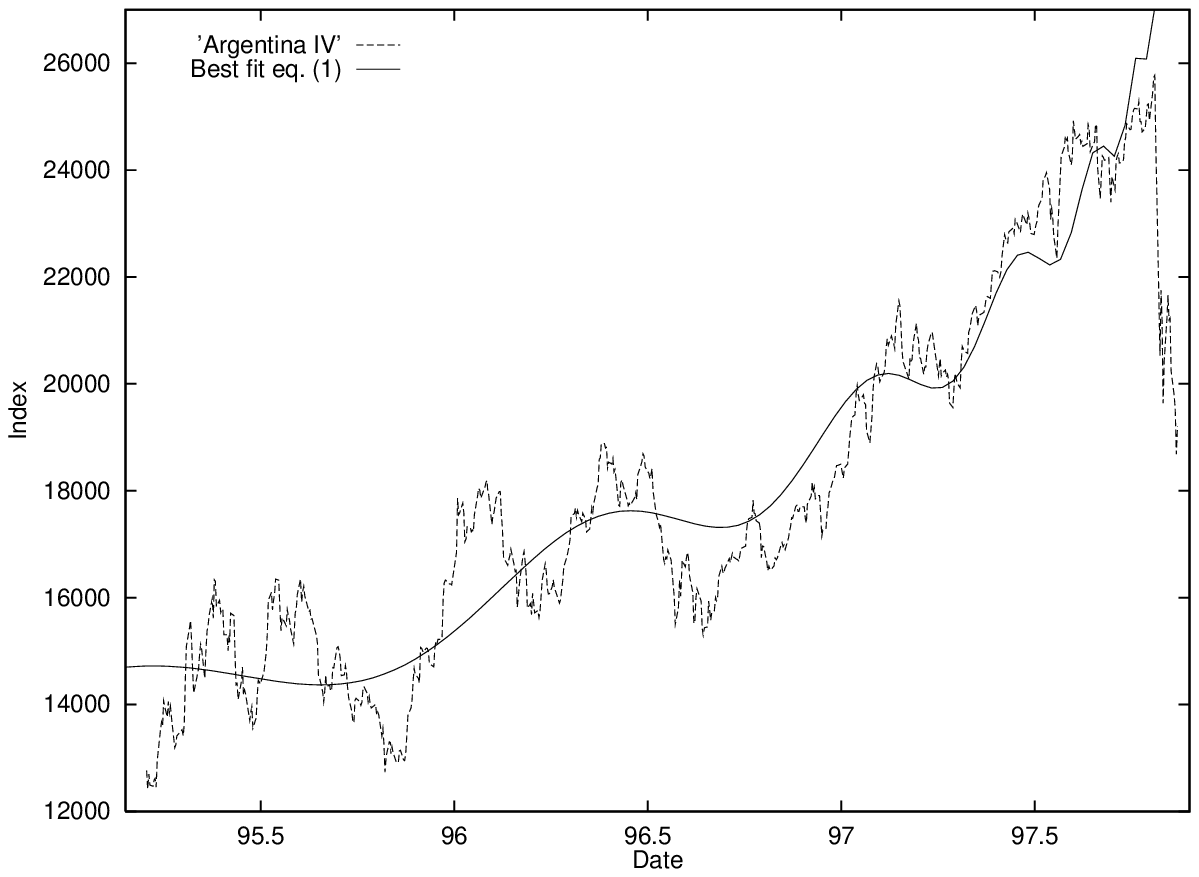,height=7cm,width=8cm} }
\hspace{5mm}
\parbox[r]{8cm}{\epsfig{file=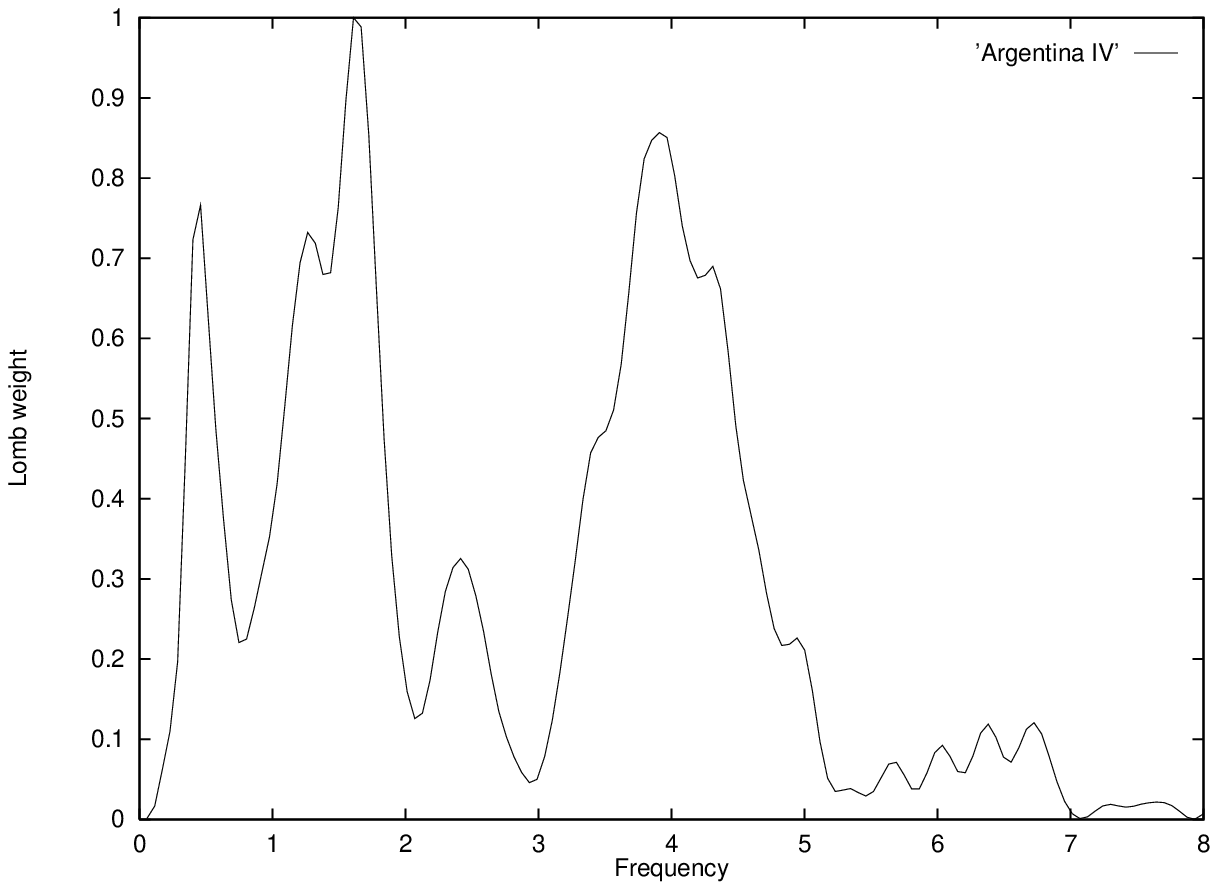,height=7cm,width=
8cm} }
\caption{\protect\label{argbub4} The Argentinian stock market bubble ending
in 1997. See table \protect\ref{lattab2} for the parameter values of the fit
with eq. (\protect\ref{lpeq}). Only the best fit is used in the Lomb
periodogram.}
\end{center}
\end{figure}

\begin{figure}
\begin{center}
\parbox[l]{8cm}{\epsfig{file=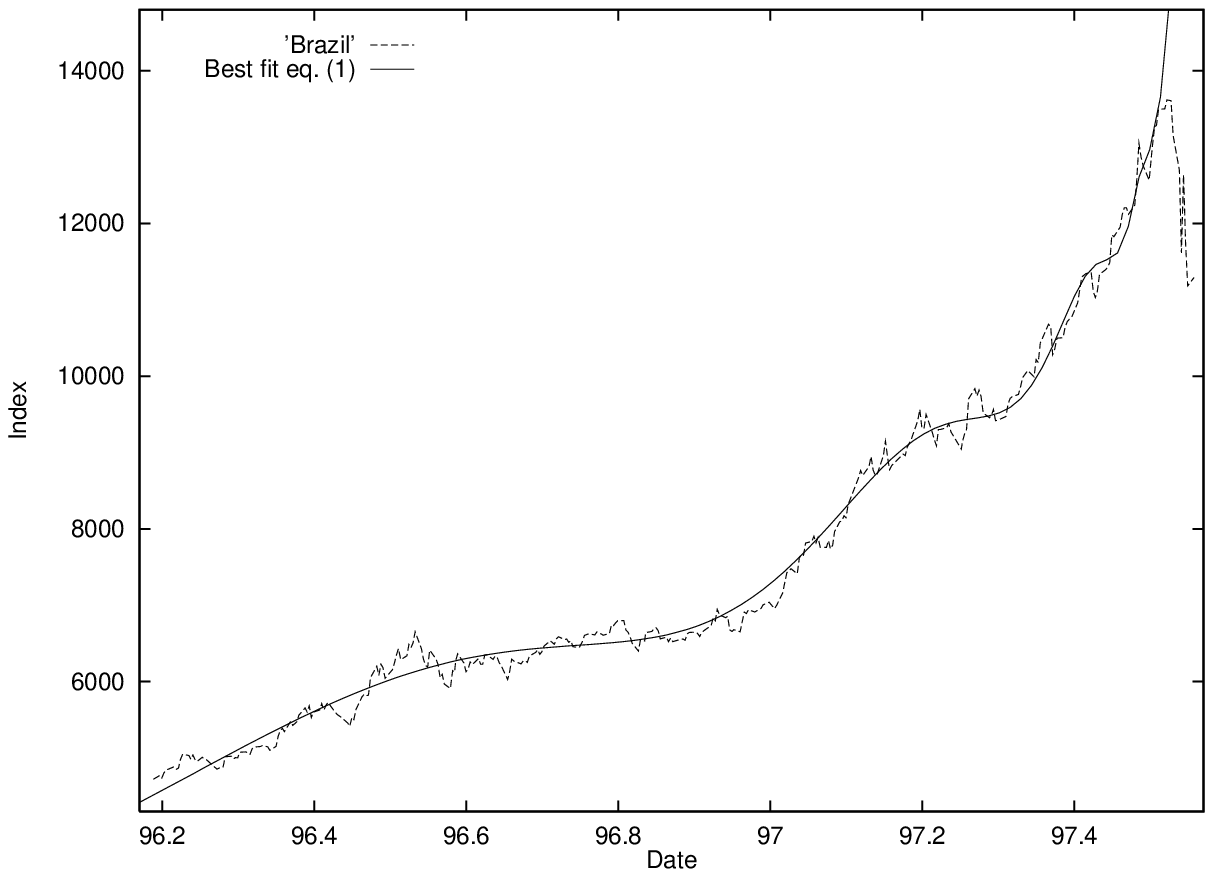,height=7cm,width=8cm}}
\hspace{5mm}
\parbox[r]{8cm}{\epsfig{file=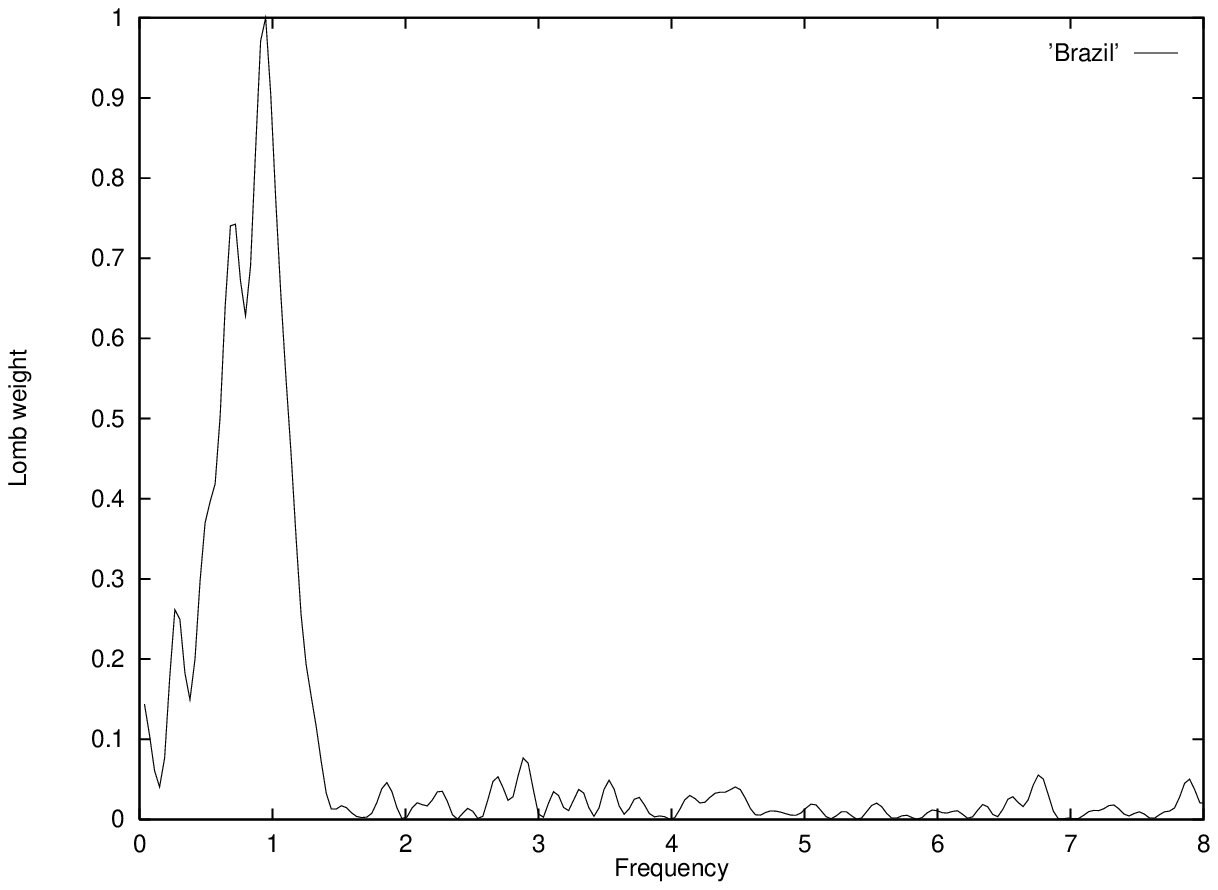,height=7cm,width=8
cm}}
\caption{\protect\label{brasbub} The Brazilian stock market bubble ending
in 1997. See table \protect\ref{lattab2} for the parameter values of the fit
with eq. (\protect\ref{lpeq}). Only the best fit is used in the Lomb
periodogram.}

\vspace{5mm}

\parbox[l]{8cm}{
\epsfig{file=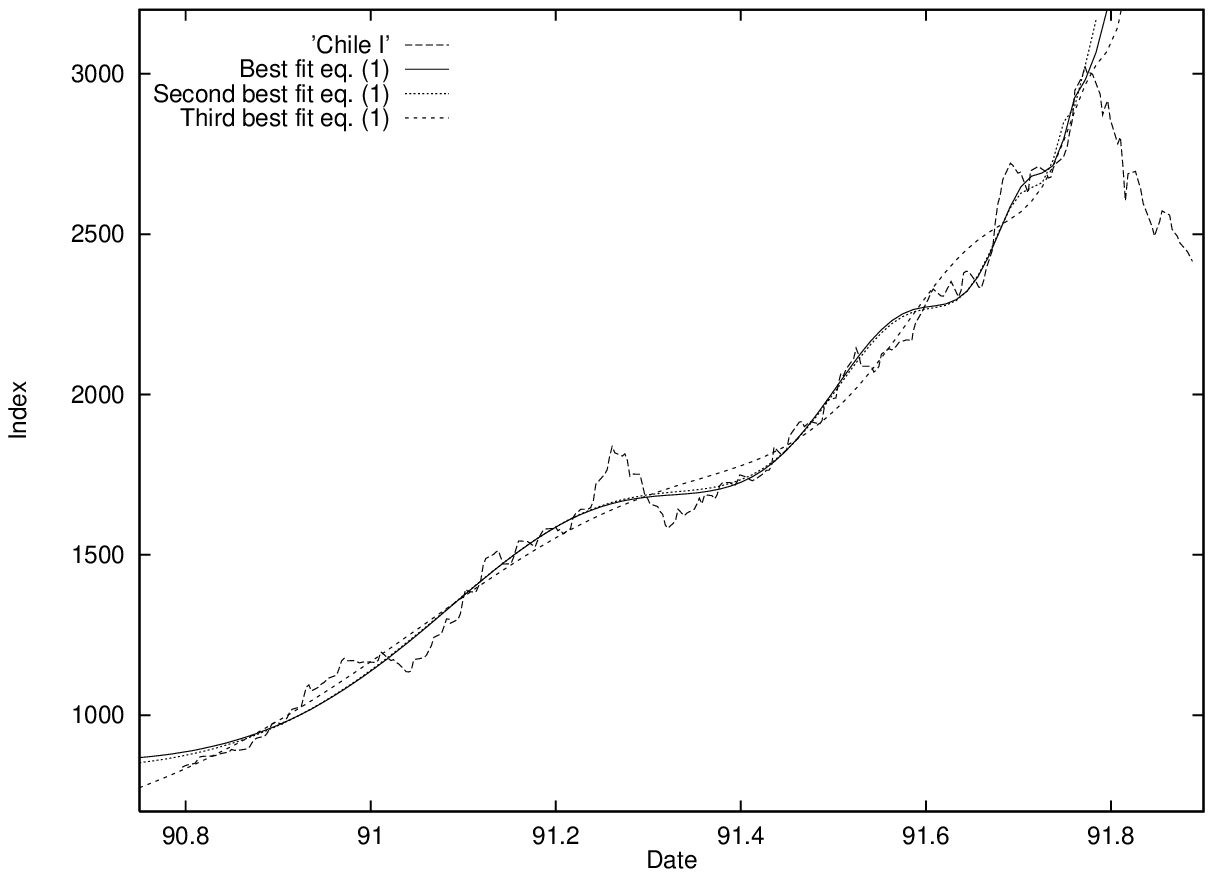,height=7cm,width=8cm}}
\hspace{5mm}
\parbox[r]{8cm}{\epsfig{file=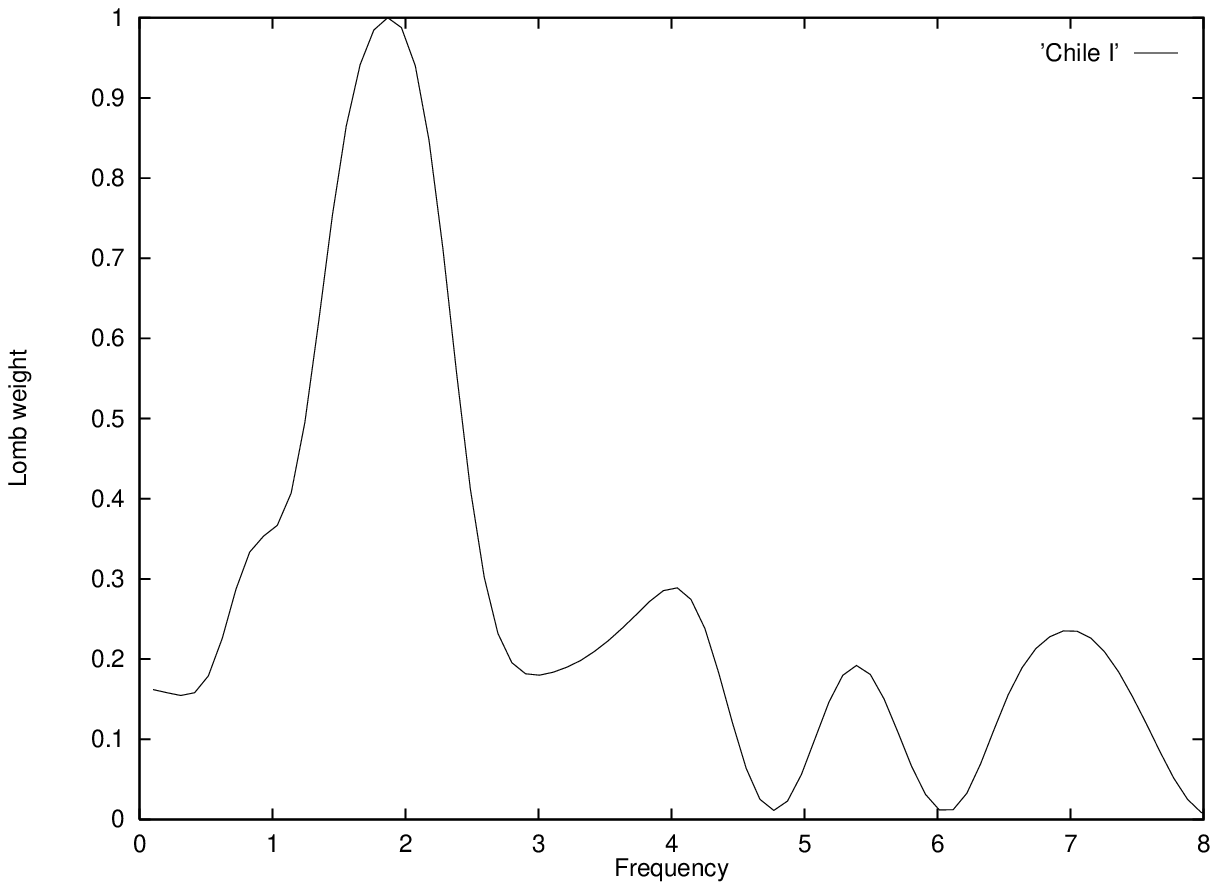,height=7cm,width=8cm}}
\caption{\protect\label{chilbub1} The Chilean bubble ending in 1991. See table
\protect\ref{lattab2} for the parameter values of the fit with eq.
(\protect\ref{lpeq}). Only the best fit is used in the Lomb periodogram.}
\end{center}
\end{figure}

\begin{figure}
\begin{center}
\parbox[l]{8cm}{\epsfig{file=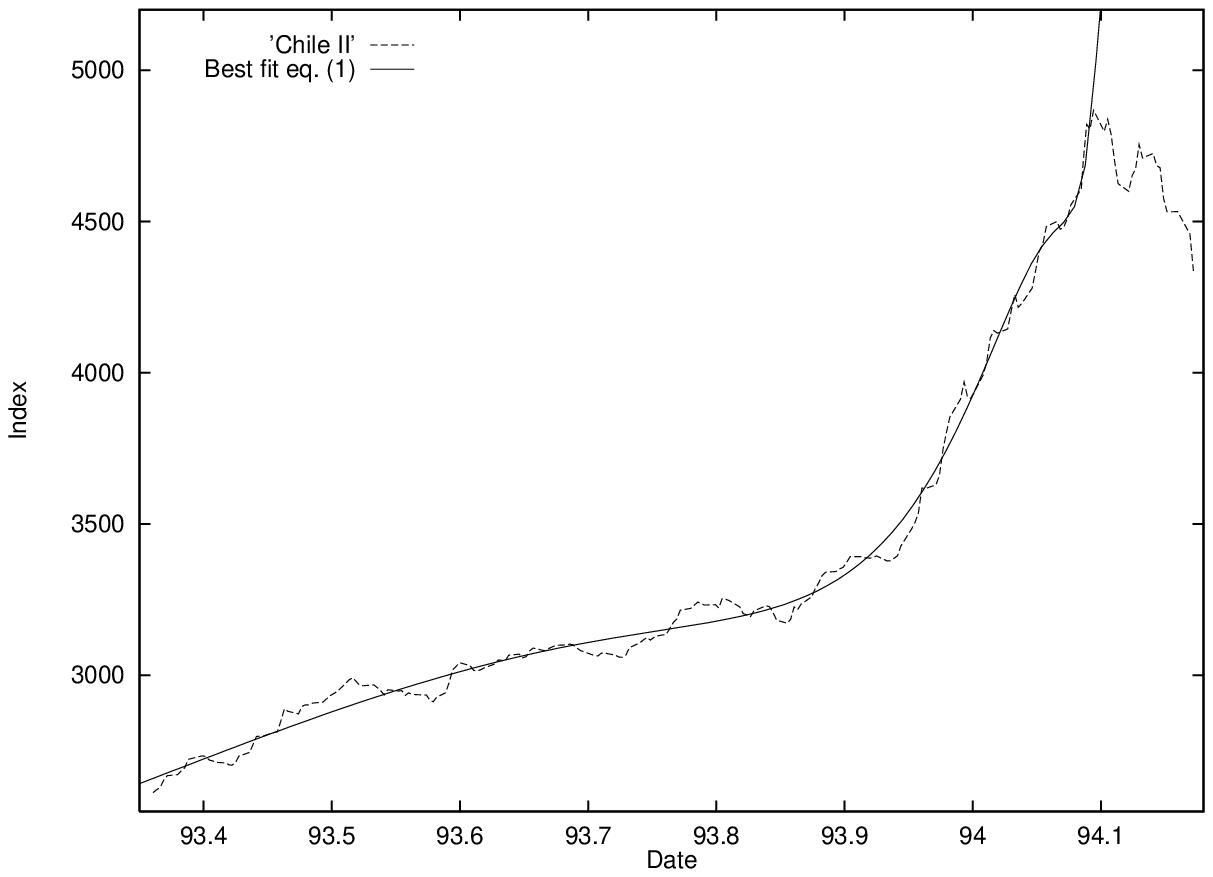,height=7cm,width=8cm} }
\hspace{5mm}
\parbox[r]{8cm}{\epsfig{file=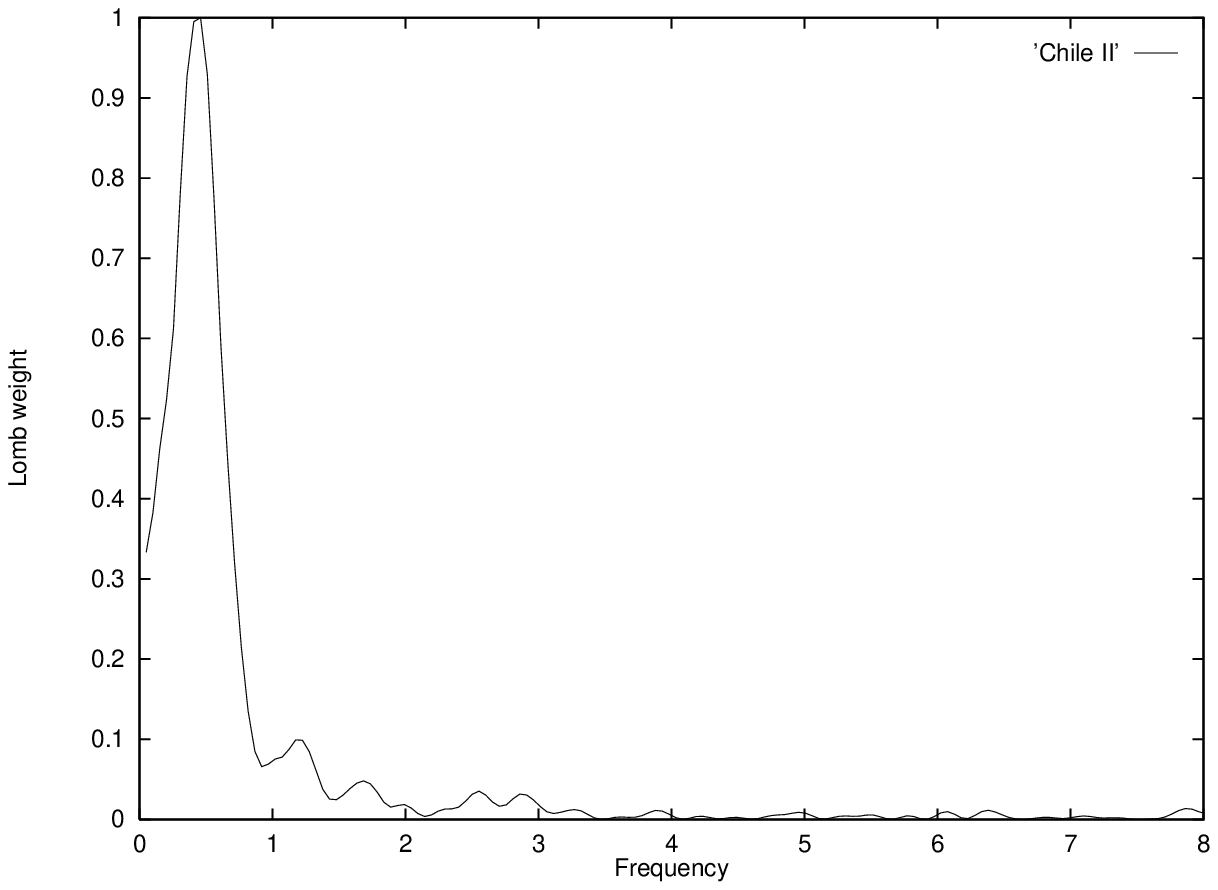,height=7cm,width=8cm} }
\caption{\protect\label{chilbub2} The Chilean bubble of 1993.  See
table \protect\ref{lattab2} for the parameter values of the fit
with eq. (\protect\ref{lpeq}).}

\vspace{5mm}

\parbox[l]{8cm}{\epsfig{file=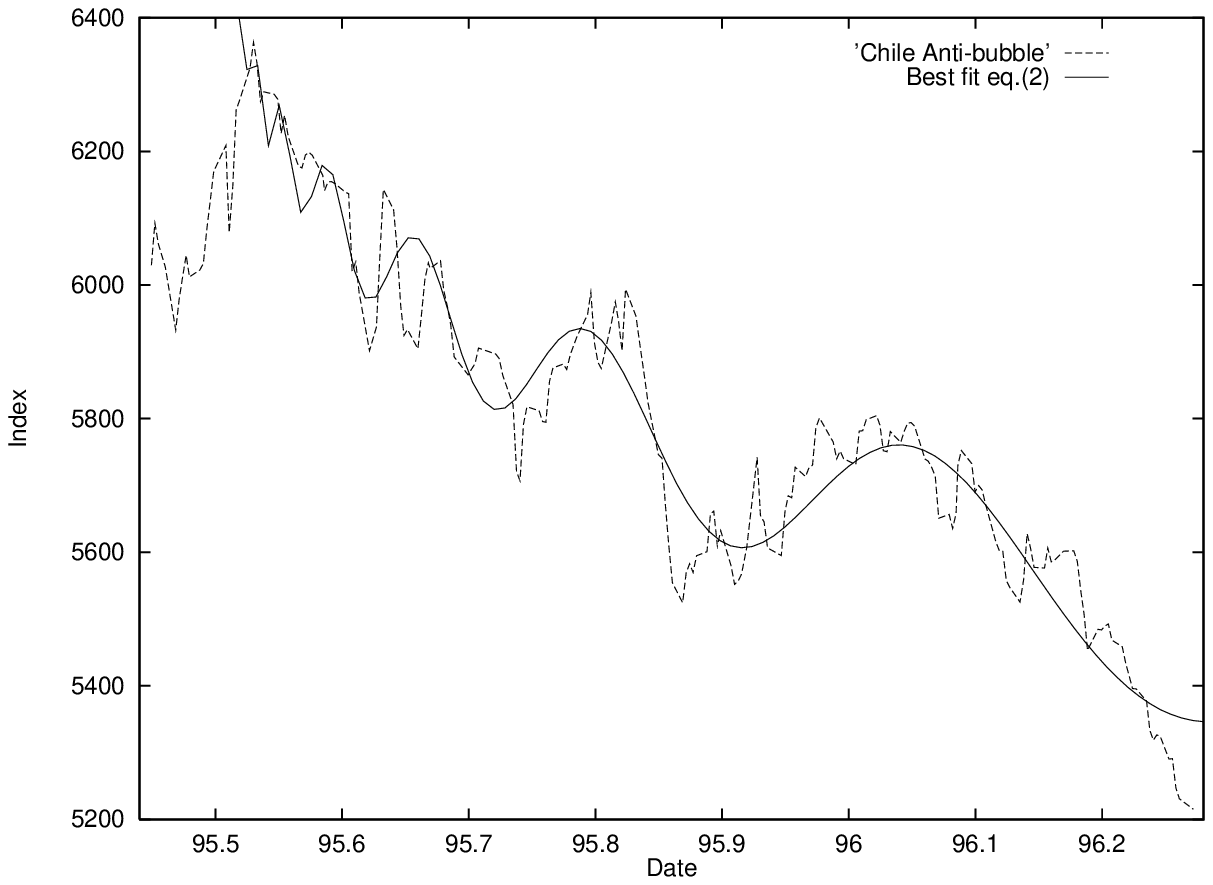,height=7cm,width=8cm} }
\hspace{5mm}
\parbox[r]{8cm}{\epsfig{file=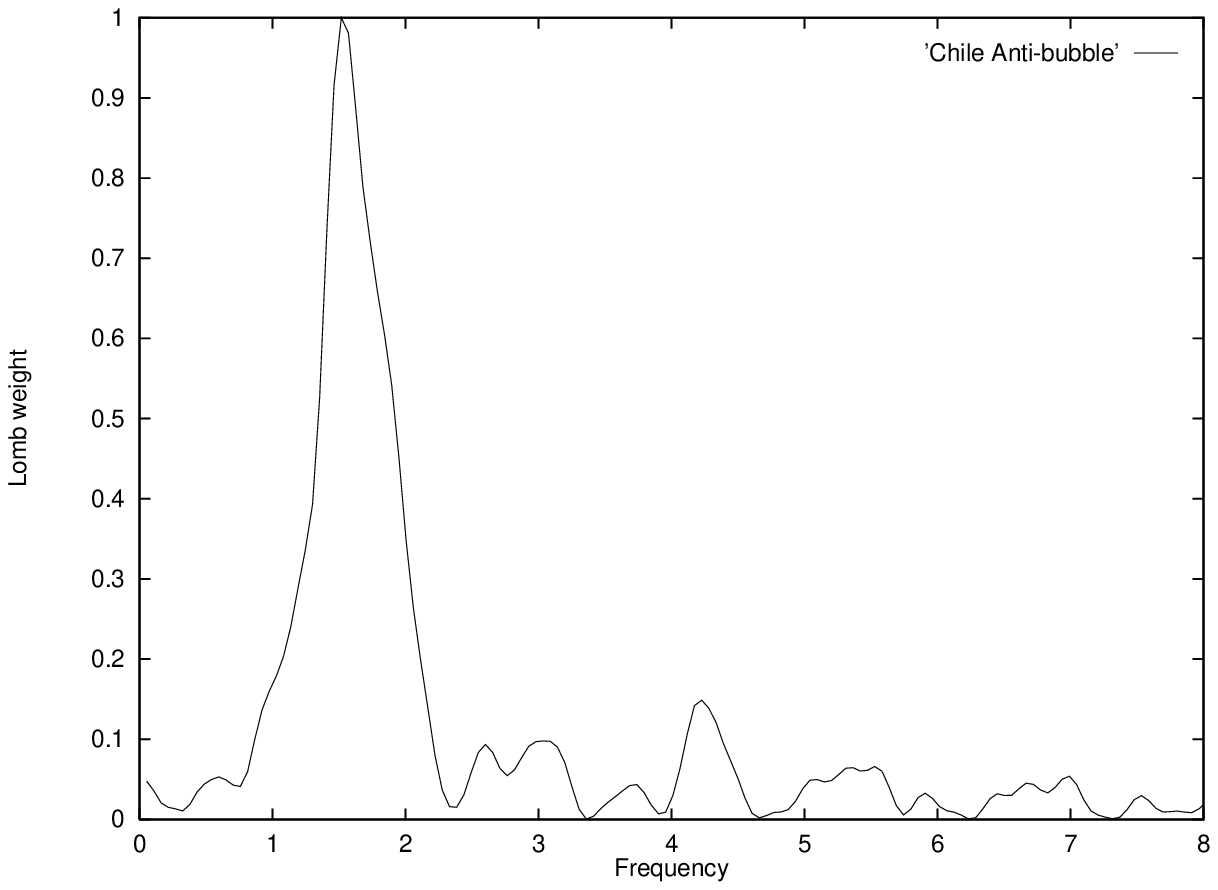,height=7cm,width=8cm} }
\caption{\protect\label{chilbub3} The Chilean anti-bubble beginning in 1995.
See table \protect\ref{lattab3} for the parameter values of the fit with
eq. (\protect\ref{lpdeceq}).}
\end{center}
\end{figure}

\begin{figure}
\begin{center}
\parbox[l]{8cm}{\epsfig{file=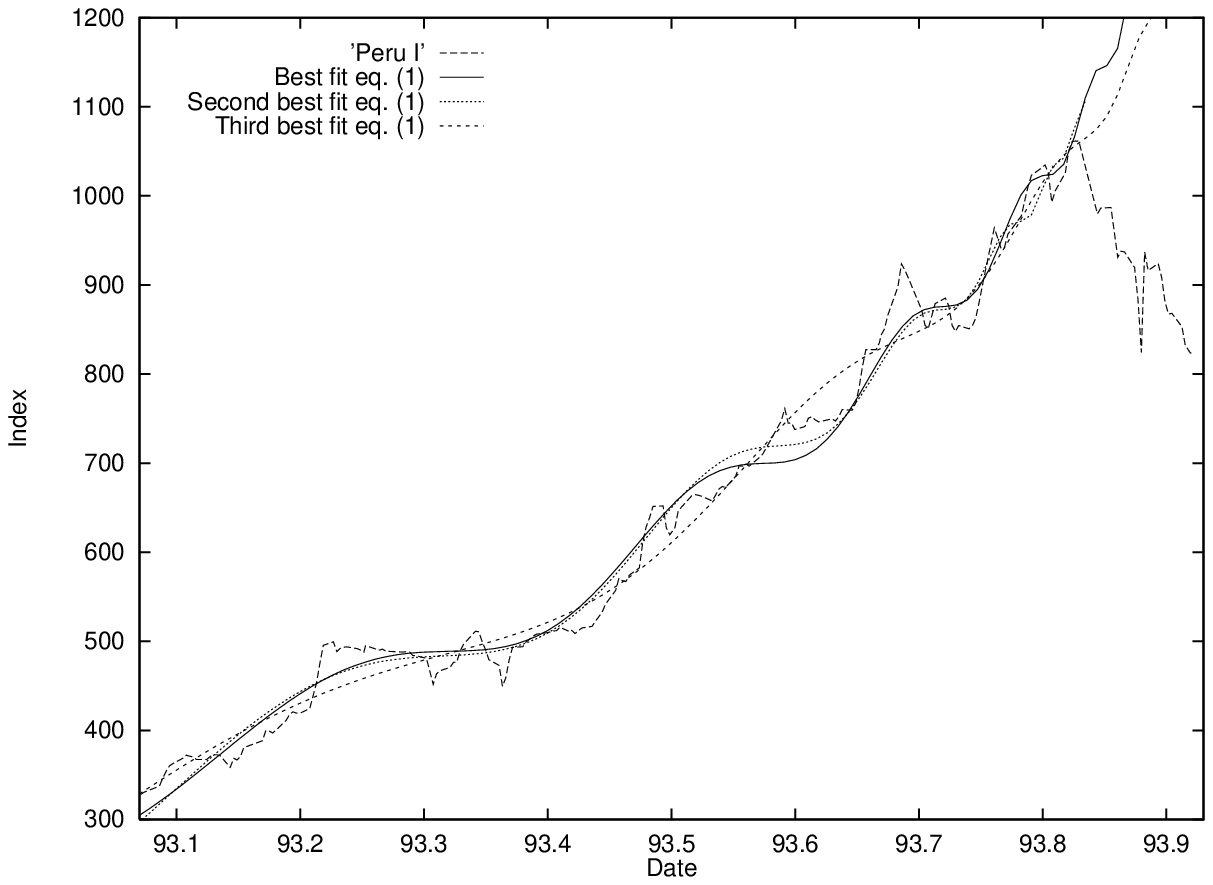,height=7cm,width=8cm} }
\hspace{5mm}
\parbox[r]{8cm}{\epsfig{file=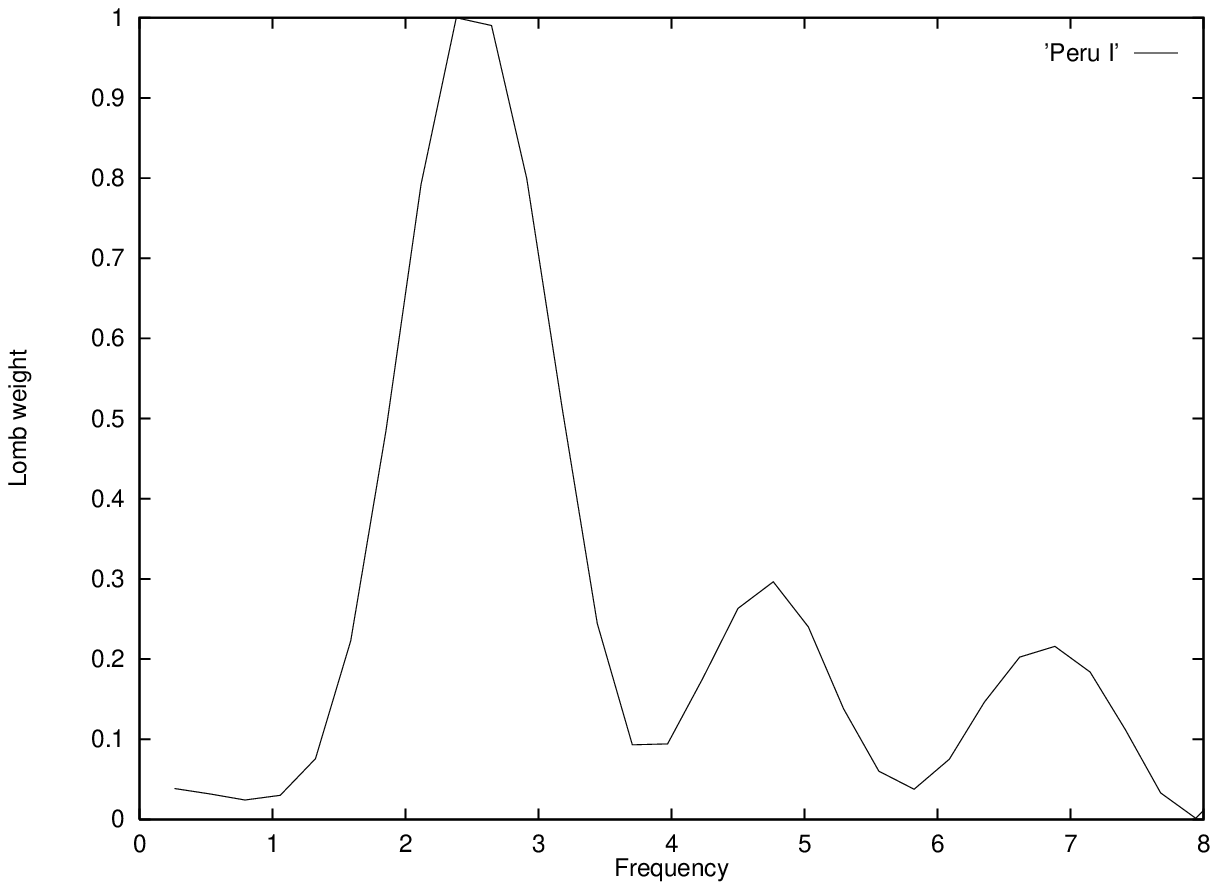,height=7cm,width=8cm} }
\caption{\protect\label{perubub} The Peruvian bubble of 1993.  See
table \protect\ref{lattab2} for the parameter values of the fit
with eq. (\protect\ref{lpeq}).}

\vspace{5mm}

\parbox[l]{8cm}{\epsfig{file=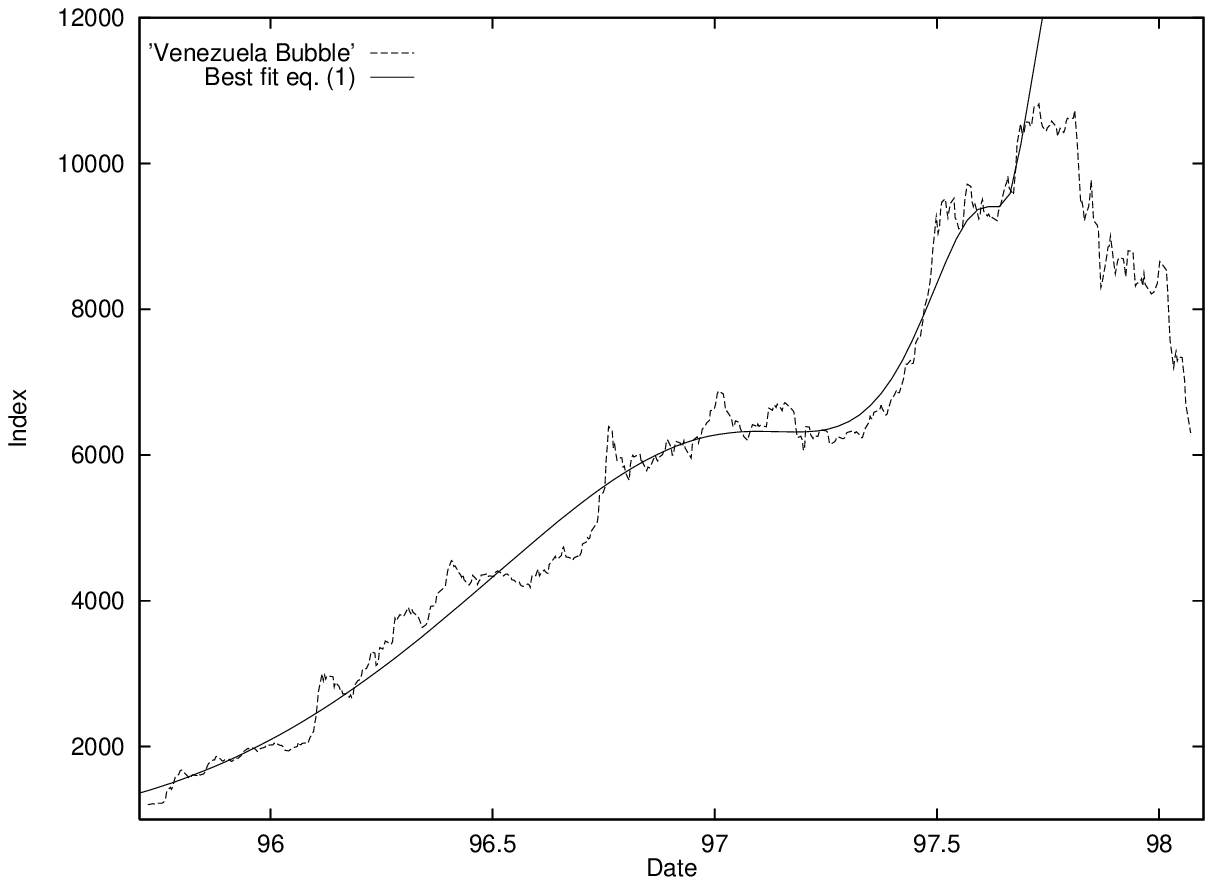,height=7cm,width=8cm} }
\hspace{5mm}
\parbox[r]{8cm}{\epsfig{file=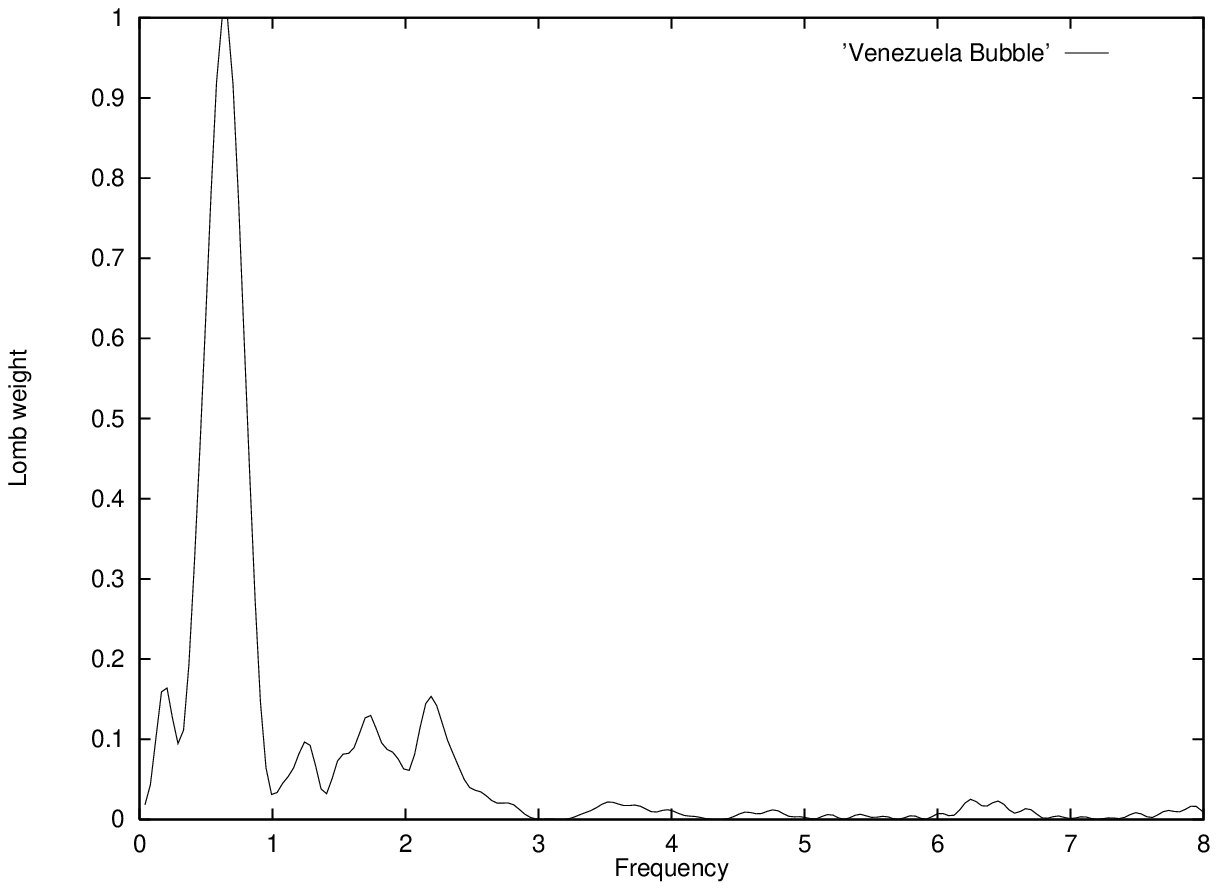,height=7cm,width=8cm} }
\caption{\protect\label{venbub1} The Venezuelan bubble ending in 1997.
See table \protect\ref{lattab2} for the parameter values of the fit with
eq. (\protect\ref{lpeq}).}
\end{center}
\end{figure}

\begin{figure}
\begin{center}
\parbox[l]{8cm}{\epsfig{file=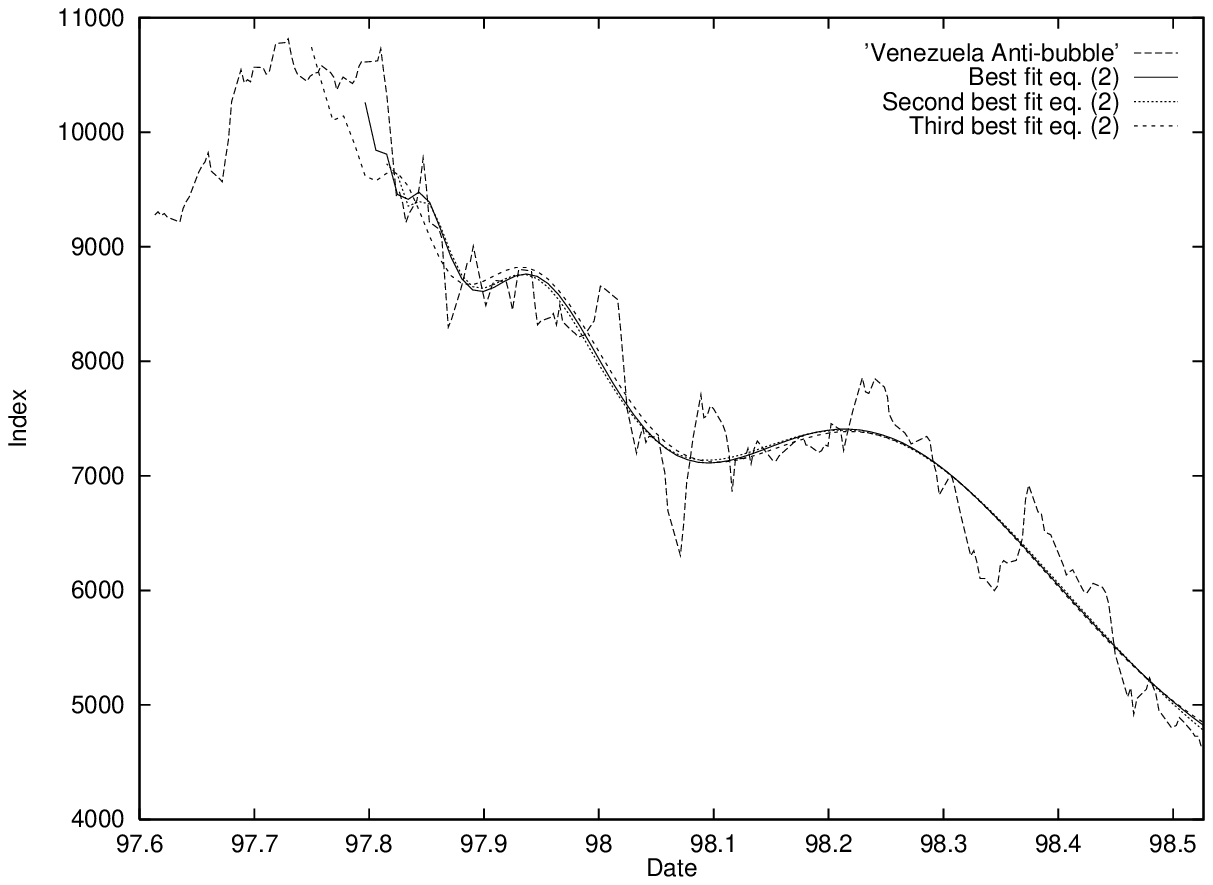,height=7cm,width=8cm} }
\hspace{5mm}
\parbox[r]{8cm}{\epsfig{file=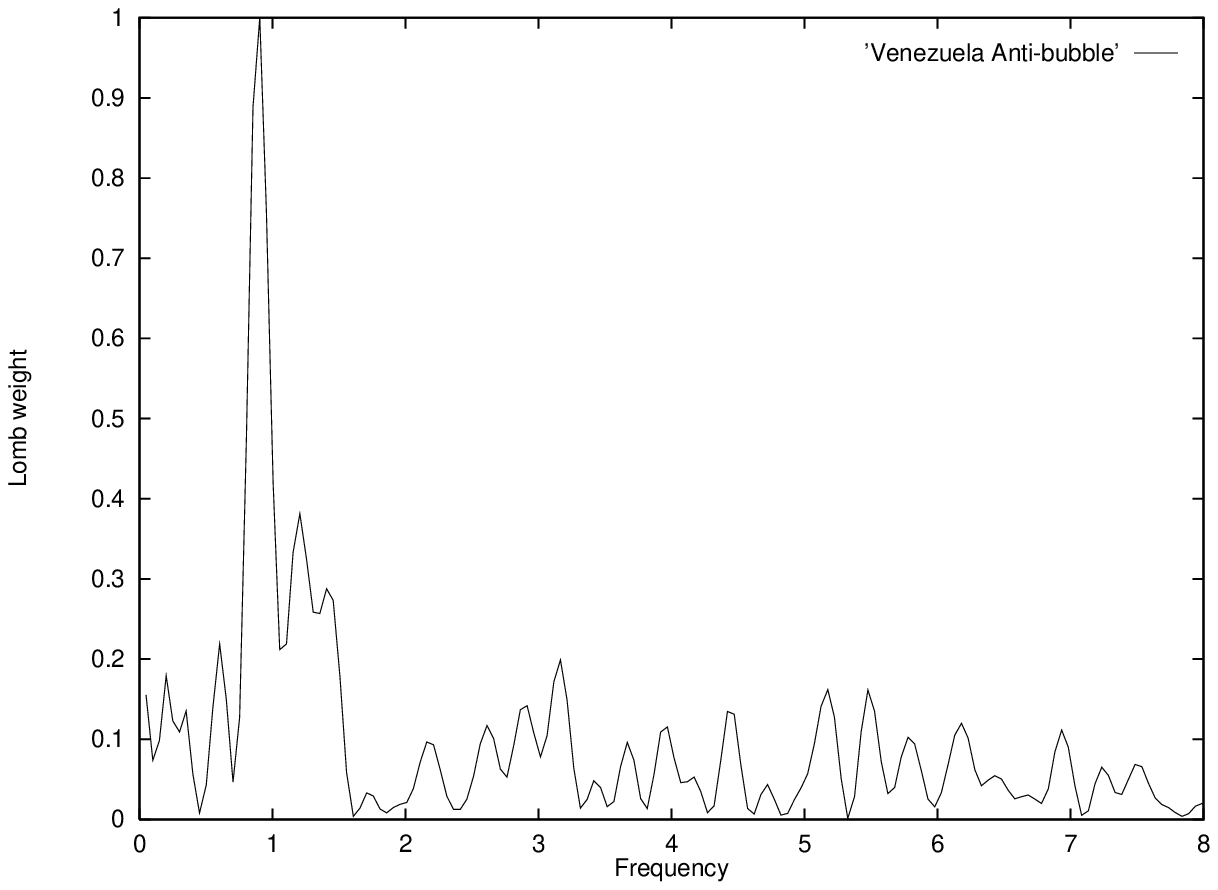,height=7cm,width=8cm} }
\caption{\protect\label{venbub2} The Venezuelan anti-bubble starting in 1997.
See table \protect\ref{lattab3} for the parameter values of the fit with eq.
(\protect\ref{lpdeceq}). Only the best fit is used in the Lomb periodogram.}
\end{center}
\end{figure}

In figures \ref{argbub1} to \ref{venbub2}, we see the fits of the bubbles
indicated in figures \ref{arg} to \ref{venu} as well as the spectral
Lomb periodogram \cite{Lomb} of the difference between the indices and the
pure power law defined as
\be \label{residue}
I\lp t\rp \rightarrow \frac{I\lp t\rp - \left[ A -
B(t_c-t)^{z}\right]}{C(t_c-t)^{z}}~ .
\ee
One exception is the second Peruvian bubble of which a numerically stable fit
could not be obtained due to an almost vertical raise at the very end of the
bubble. Using the logarithm of the index instead did only in part solve this
problem and we have not included this fit in the present paper.

If log-periodicity is present in the data as quantified by eq. (\ref{lpeq}), 
the residue defined by eq. (\ref{residue}) should be a pure cosine of $\omega 
\ln (t_c -t)$ and a spectral analysis of this variable should give a strong 
peak around $\omega$. For this, we use the Lomb spectral analysis, which 
corresponds to a harmonic analysis using a series of local fits of a cosine 
(with a phase) with some user chosen range of frequencies. The advantage of 
the Lomb periodogram over a Fast Fourier transform is that the points do not 
have to be equidistantly sampled, which is the generic case when dealing with 
power laws. For unevenly sampled data, the Lomb method is superior to 
FFT-methods because it weights data on a ``per point'' basis
instead of ``per time interval'' basis. Furthermore, the significance level
of any frequency component can be estimated quite accurately if the nature
of the noise is known.

It is clear from simply looking at the figures, that the overall quality of
these fits is rather good and both the acceleration and the accelerating
oscillations are rather well captured by eq. (\ref{lpeq}). We let the
reader directly appreciate the quality of the fits on the figures. We notice
that, notwithstanding their value, the fits does not have the same excellent
over-all quality as for those obtained for the WMFMs as well as for the
Russian stock market \cite{JSL}. A plausible interpretation is that we deal
here with relatively small markets in terms of capitalization and number of
investors, for which {\it finite size effects}, in the technical sense given
in Statistical Physics \cite{finitesize}, are expected and thus may blur out
the signal with systematic distortions and unwanted fluctuations. In this vein,
numerical simulations of all (with one single exception \cite{percostausor})
available microscopic stock market models have shown that simple regular
deterministic dynamics is obtained when the limit of a large effective number
$N$ of traders is taken while the stock market behavior seems realistically
random and complex when only a few hundred traders are simulated
\cite{StaufferlargeN1,StaufferlargeN2,StaufferlargeN3,StaufferlargeN4}.

In tables \ref{lattab1} and \ref{lattab2}, the parameters of the various fits
are given as well as the beginning and ending dates of the bubble and the size
of the crash/correction, defined as
\be \label{dropeq}
\mbox{drop \%} = \frac{I\lp t_{max}\rp - I\lp t_{min}\rp}{I\lp t_{max}\rp}~.
\ee
Here, $t_{min}$ is is defined as the date after the crash/correction
where the index achieves its lowest value before a clear novel market regime
is observed. The duration $t_{max} - t_{min}$ of the crash/correction is
found to range from a few days (a crash) to a few months (a change of regime).

From table \ref{lattab1}, we observe that the fluctuations in the parameters
values $z$ and $\omega$ obtained for the $11$ Latin-American crashes
are considerable. The lower and upper values for the exponent $z$ are $0.12$
and $0.62$, respectively. For $\omega$, the lower and upper values are
$2.9$ and $11.4$ corresponding to a range of $\lambda$'s in the interval
$1.8-8.8$. Removing the two largest values for $\lambda$ reduces the 
fluctuations to $2.8 \pm 1.1$, which is still much larger than the 
$2.5 \pm 0.3$ previously seen on WMFMs \cite{JS98.2}. Again, we attribute 
these larger fluctuations to finite-size effects.

Last, we note that three cases of anti-bubbles could be identified for the
Latin-American markets analysed here, see figures \ref{argbub2}, \ref{chilbub3}
and \ref{venbub2} and table \ref{lattab3}. Quite remarkably, the first and
the last are preceded by a bubble thus exhibiting a qualitative symmetry
around comparable $t_c$'s as defined in eq.'s (\ref{lpeq}) and (\ref{lpdeceq}).

\subsection{Asian markets}

\subsubsection{Identification of bubbles}

\begin{figure}
\begin{center}
\epsfig{file=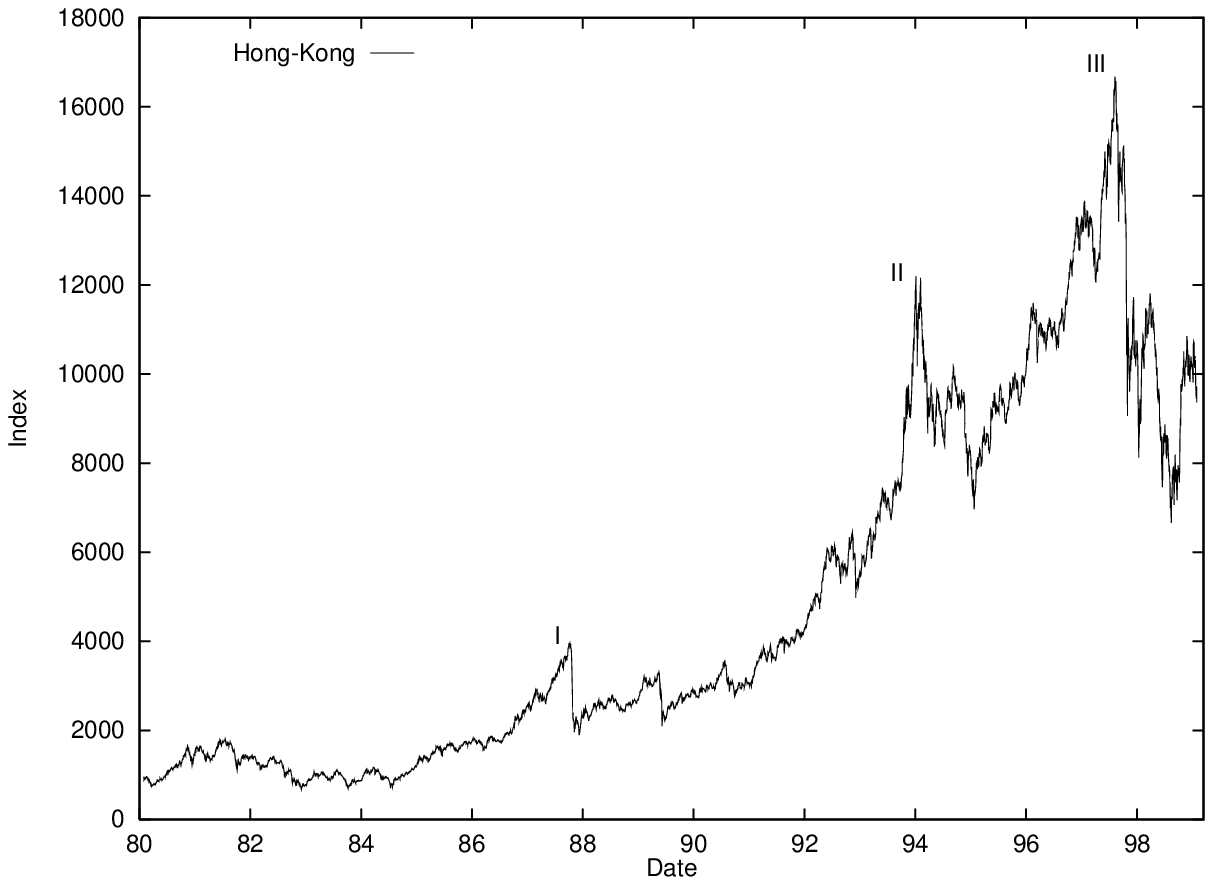,height=9cm,width=16cm}
\caption{\protect\label{hk} The Hong-Kong stock market index as a function
of date. 3 extended bubbles followed by large draw downs can be identified.
The approximate dates of the crashes are  Oct. 87 (I),  Jan 94 (II)
and Oct 97 (III).}

\vspace{5mm}

\epsfig{file=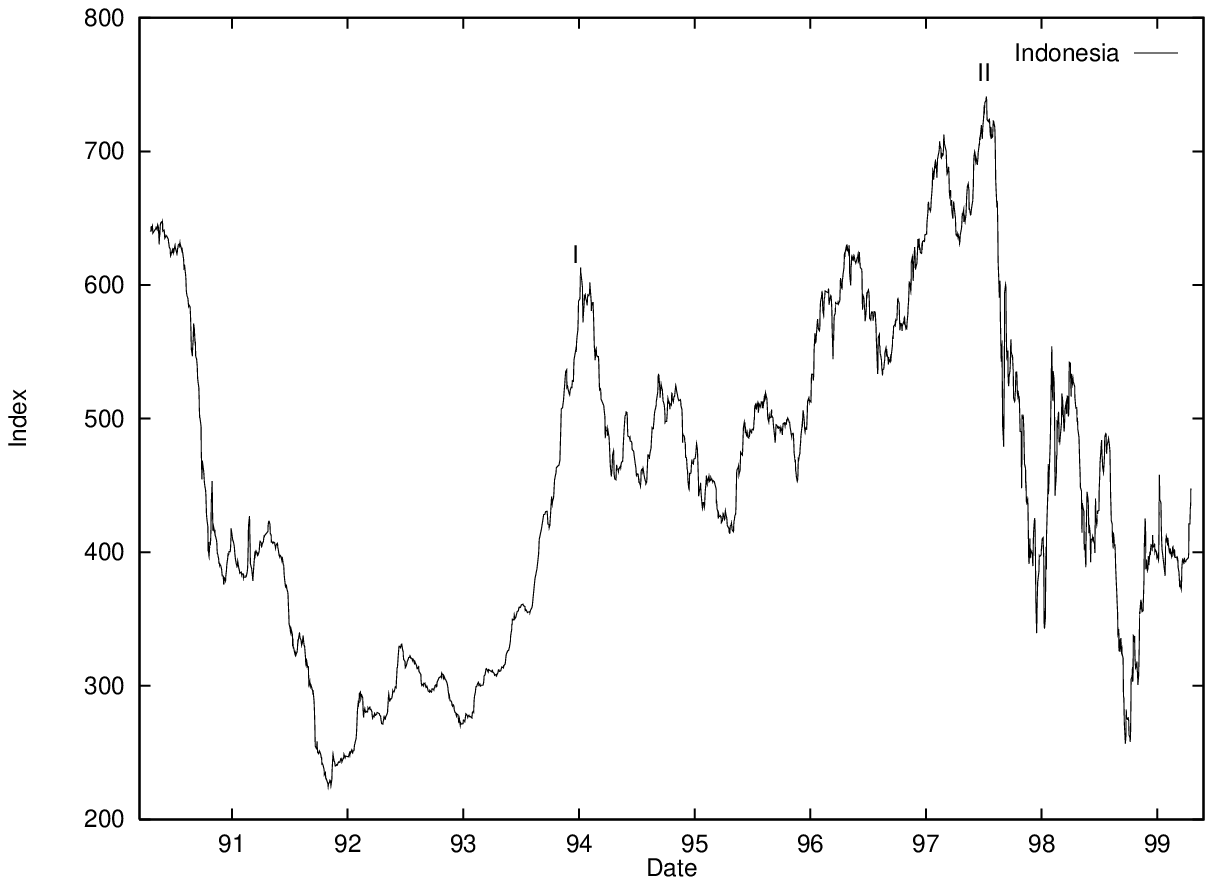,height=9cm,width=16cm}
\caption{\protect\label{indo}The Indonesian stock market index as a function of
date. 2 bubbles with a subsequent very large draw down can be identified. The
approximate dates for the draw downs are early 94 (I) and mid-97 (II), see
figure. Note that only the first bubble gives a reasonable fit with eq.
(\protect\ref{lpeq}).}
\end{center}
\end{figure}

\begin{figure}
\begin{center}
\epsfig{file=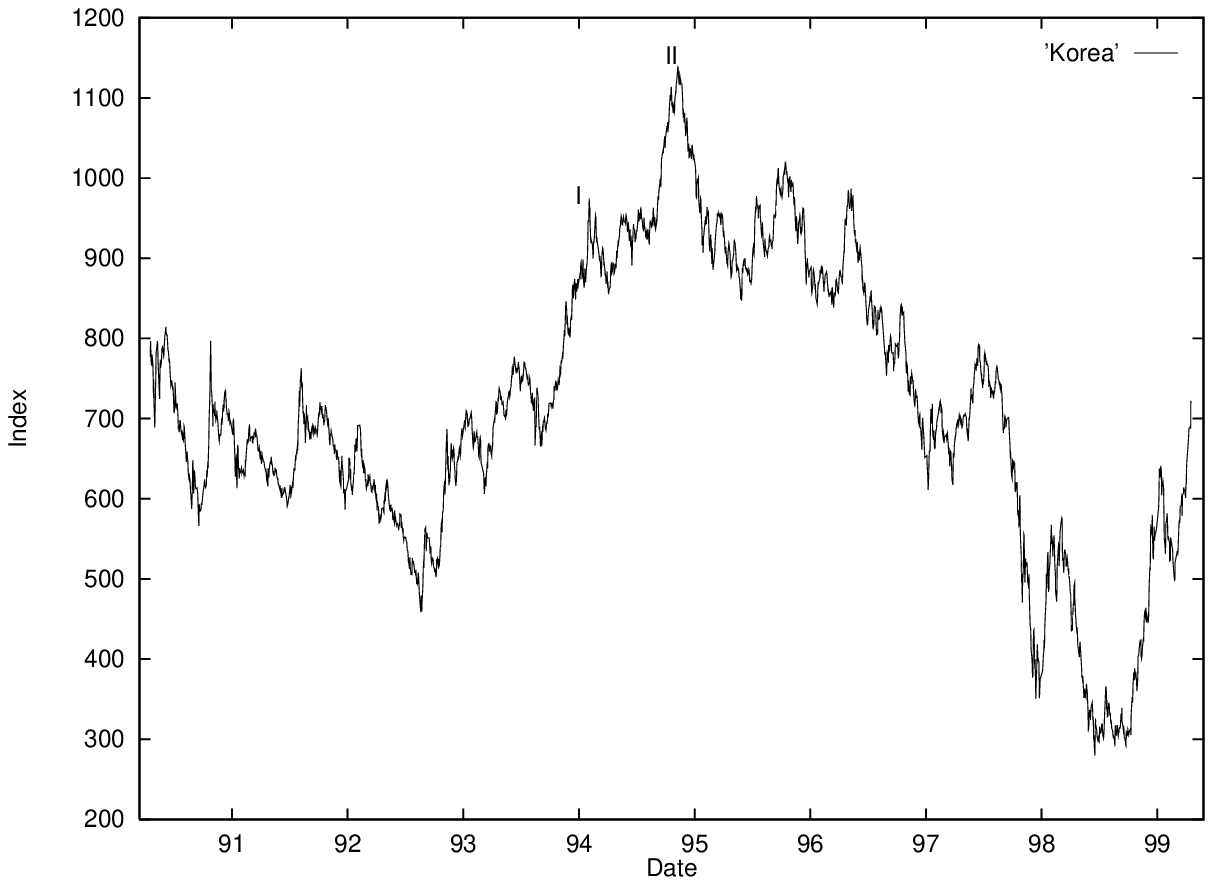,height=9cm,width=16cm}
\caption{\protect\label{kore} The Korean stock market index as a function
of date. 2 bubbles with a subsequent very large draw down can be identified.
The approximate dates are early 94 (I) and  late 94 (II). None
of the two bubbles gave a reasonable fit with  eq. (\protect\ref{lpeq}).}

\vspace{5mm}

\epsfig{file=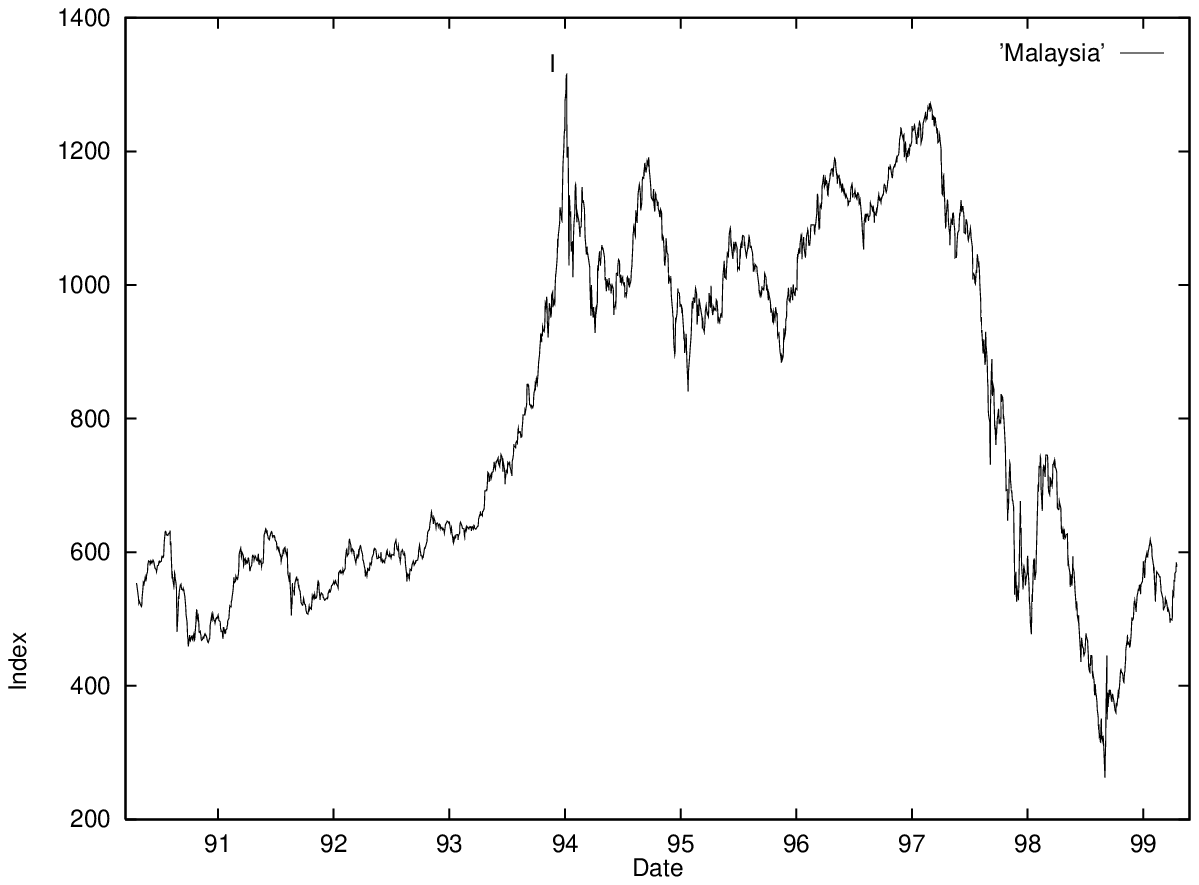,height=9cm,width=16cm}
\caption{\protect\label{mal} The Malaysian stock market index as a function
of date. 1 extended bubble with a subsequent very large draw down can be
identified. The approximate date for the draw down is early 94, see figure.}
\end{center}
\end{figure}

\begin{figure}
\begin{center}
\epsfig{file=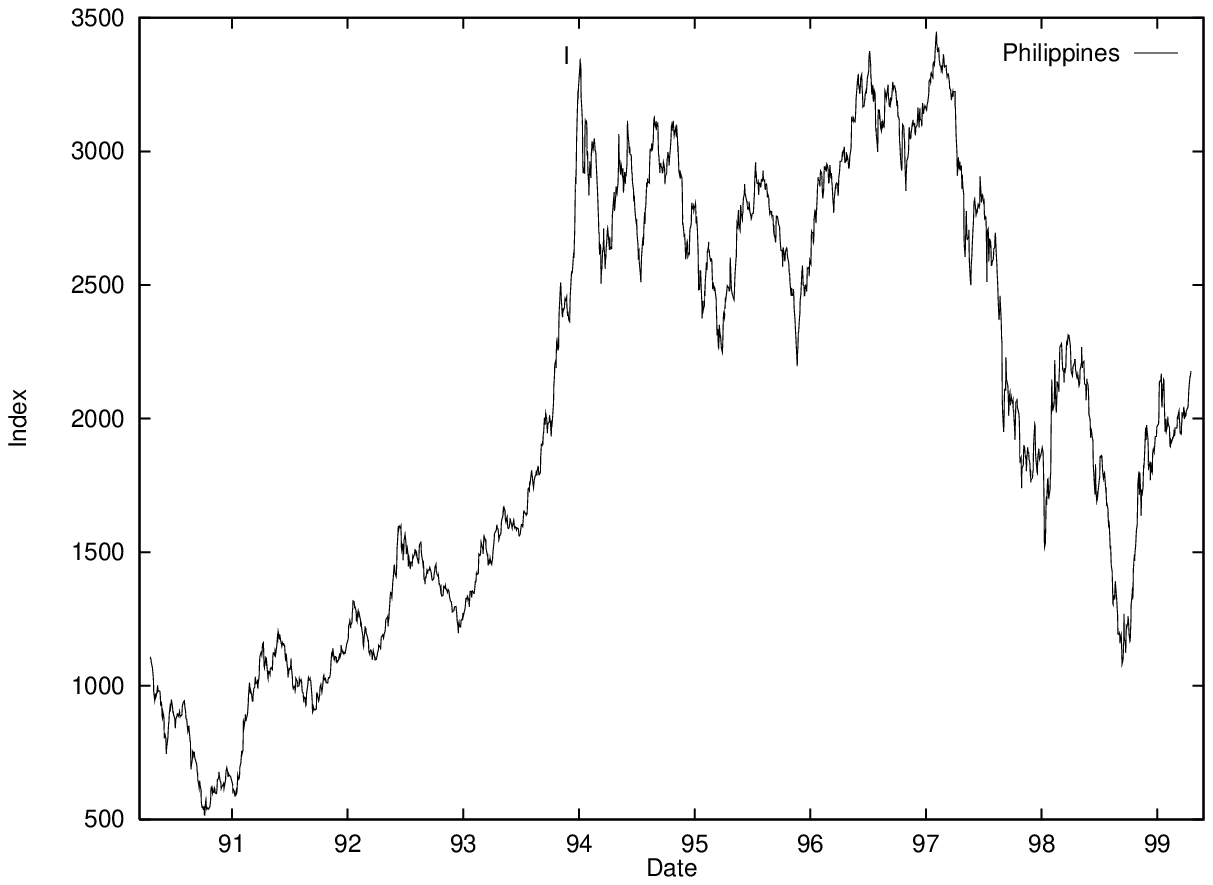,height=9cm,width=16cm}
\caption{\protect\label{phil} The Philippines stock market index as a function
of date. 1 bubble with a subsequent very large draw down can be identified.
The approximate date for the draw down is early 94, see figure.}

\vspace{5mm}

\epsfig{file=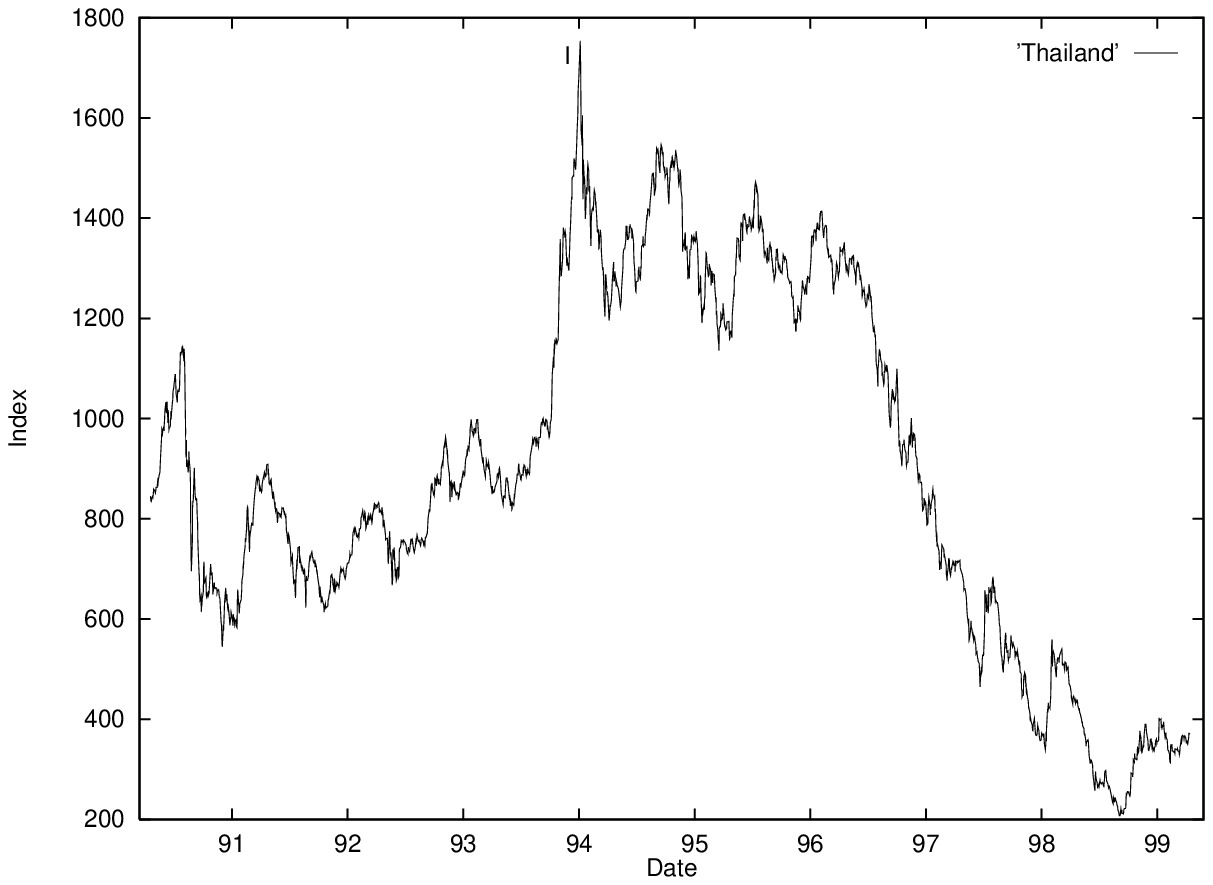,height=9cm,width=16cm}
\caption{\protect\label{thai} The Thai stock market index as a function
of date. 1 bubble with a subsequent very large draw down can be identified.
The approximate date for the draw down is early 94, see figure.}
\end{center}
\end{figure}

In figures \ref{hk} to \ref{thai}, the evolution of six Asian stock market
indices (Hong-Kong, Indonesia, Korea, Malaysia, Philippines and Thailand) is
shown as a function of time from 1990 to Feb. 1999 except for Hong-Kong, which
goes back to 1980.

From the six stock market indices, we have identified three bubbles on the
Hong-Kong stock market, two on the Indonesian, two on the Korean and one on
the Malaysian, Philippine and Thai stock markets, respectively, with subsequent
crashes/decreases that could be identified by eye, as indicated in figures
\ref{hk} to \ref{thai}. Of these, the two Korean bubbles and the second
Indonesian could not be quantified using eq. (\ref{lpeq}). Of the
remaining seven which could, all except the Hong-Kong crashes of Oct. 1987 and
Oct. 1997 belonged to the same period ending in Jan. 1994 as also found for
the Latin-American markets analysed in section \ref{latresults} with the
exception of Venezuela. As we shall see in section \ref{antibub}, this
globally
coordinated crash on emerging markets triggered a correlated anti-bubble on
the smaller Western stock markets.

\subsubsection{Results} \label{asiresults}

\begin{figure}
\begin{center}
\parbox[l]{8cm}{\epsfig{file=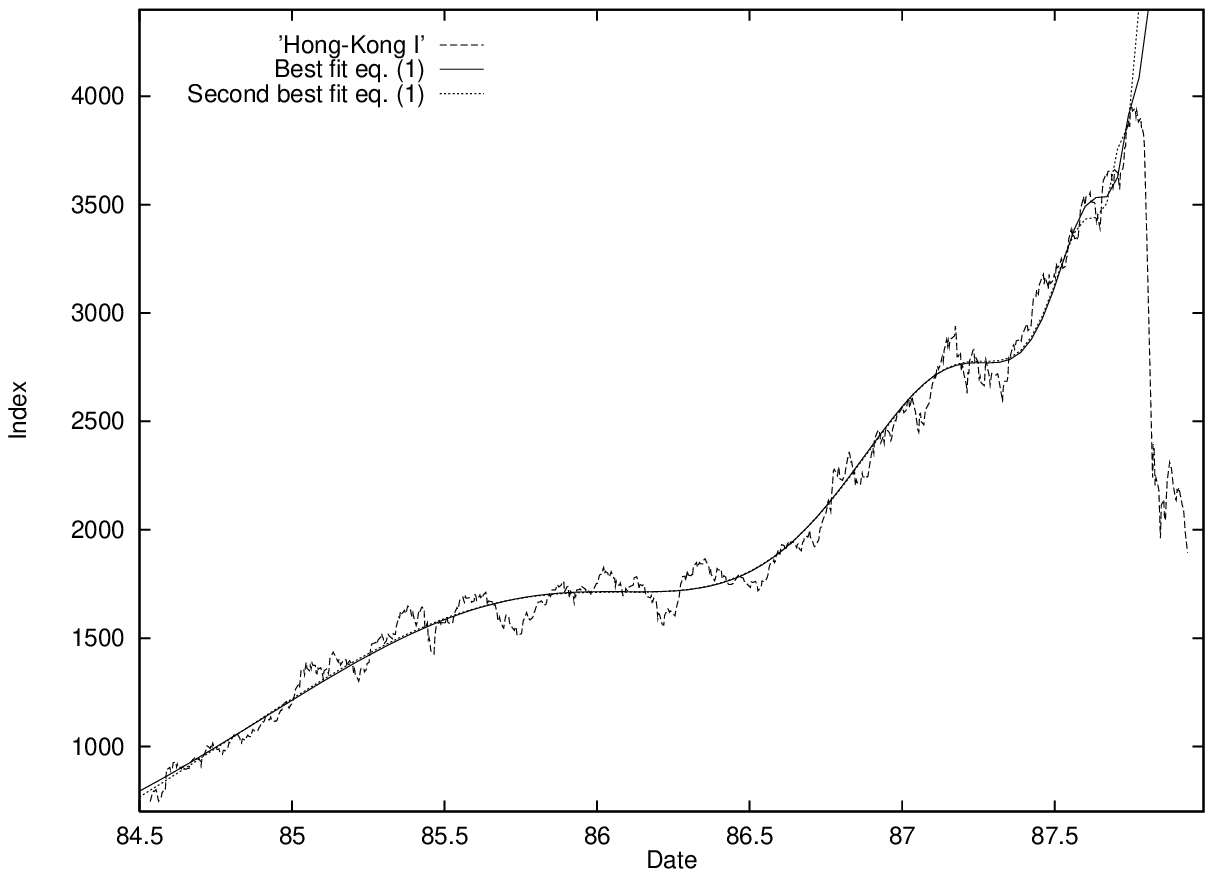,height=7cm,width=8cm} }
\hspace{5mm}
\parbox[r]{8cm}{\epsfig{file=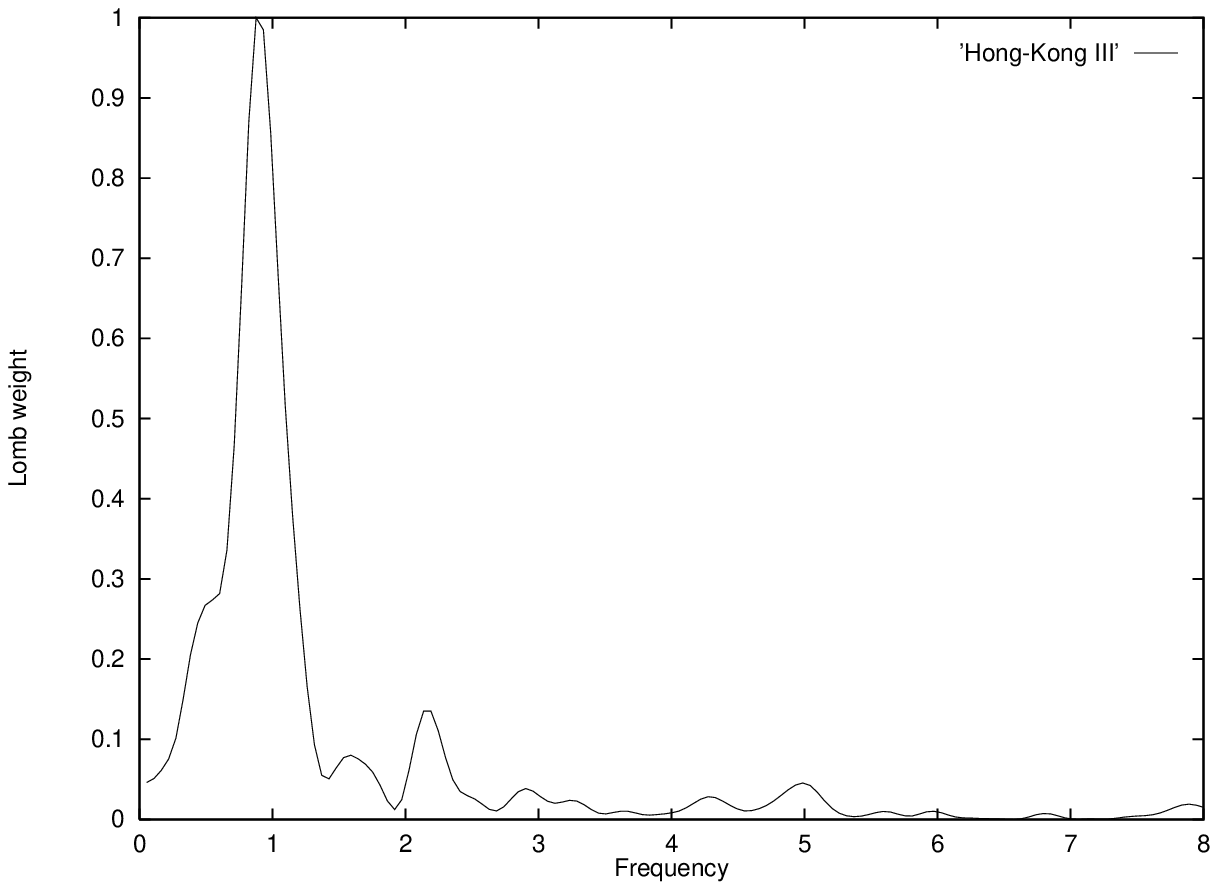,height=7cm,width=8cm}}
\caption{\protect\label{hkbub1} Hong Kong stock market bubble ending with the
crash of Oct. 87. See table \protect\ref{asitab2} for the parameter
values of the fit with eq. (\protect\ref{lpeq}). Only the best fit is used in
the Lomb periodogram.}

\vspace{5mm}

\parbox[l]{8cm}{\epsfig{file=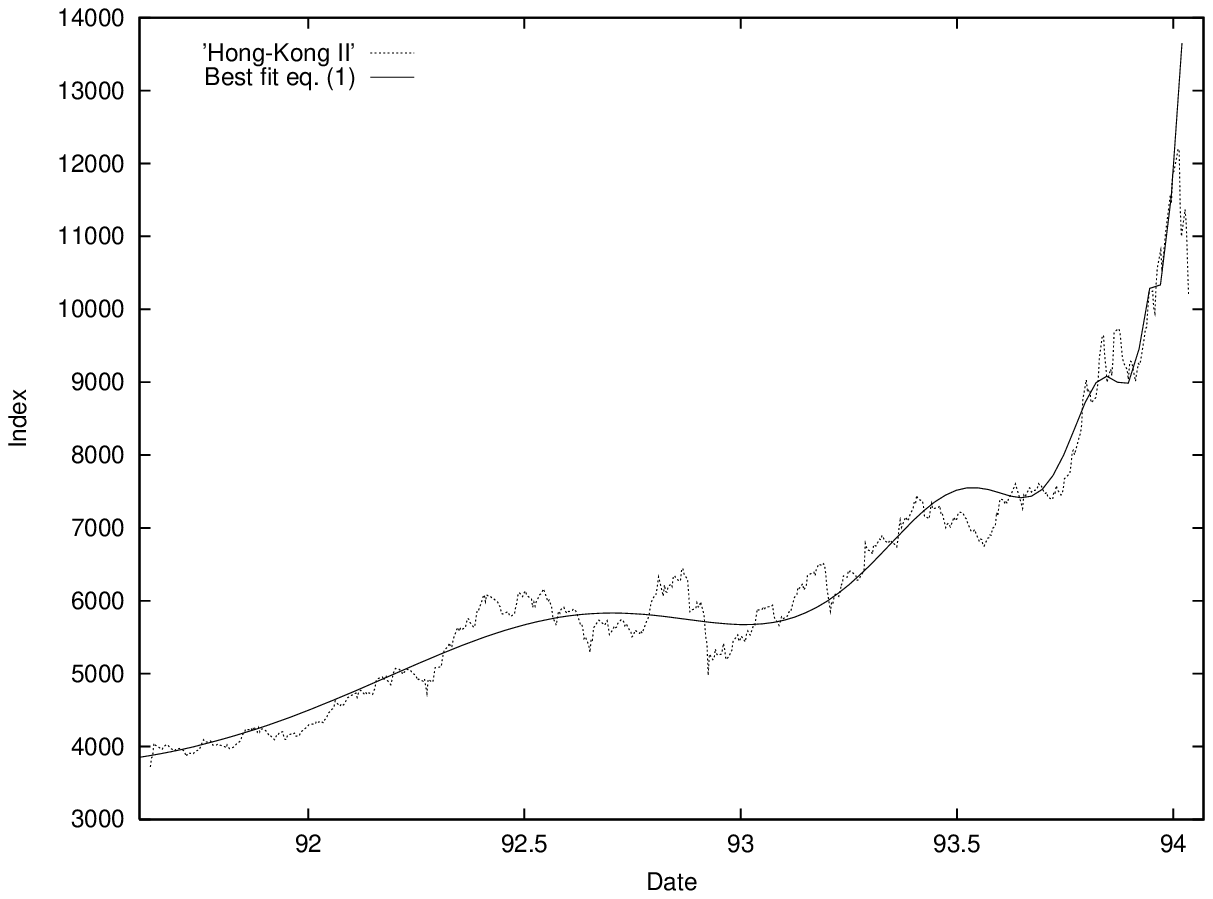,height=7cm,width=8cm}}
\hspace{5mm}
\parbox[r]{8cm}{\epsfig{file=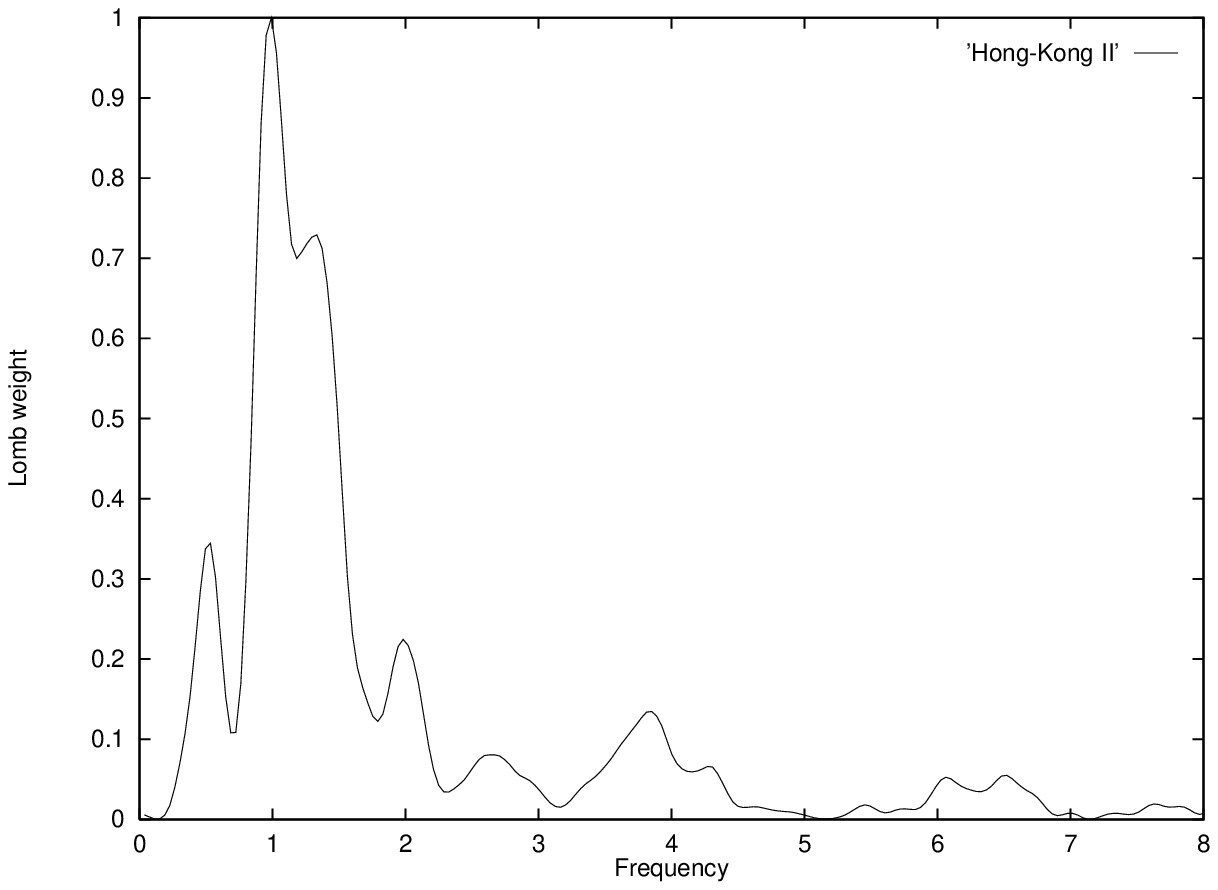,height=7cm,width=8cm}}
\caption{\protect\label{hkbub2} Hong Kong stock market bubble ending with the
crash of Jan. 94. See table \protect\ref{asitab2} for the parameter
values of the fit with eq. (\protect\ref{lpeq}).}
\end{center}
\end{figure}

\begin{figure}
\begin{center}
\parbox[l]{8cm}{\epsfig{file=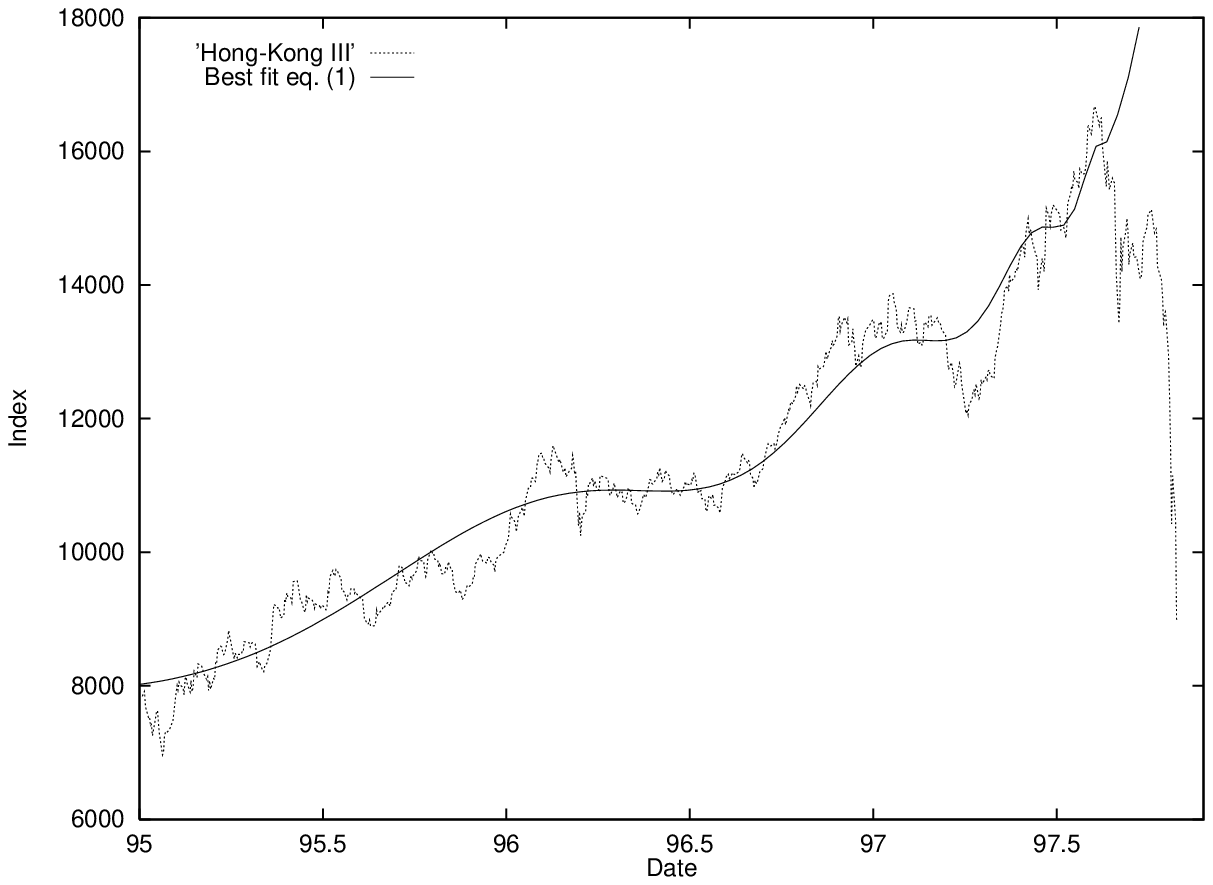,height=7cm,width=8cm}}
\hspace{5mm}
\parbox[r]{8cm}{\epsfig{file=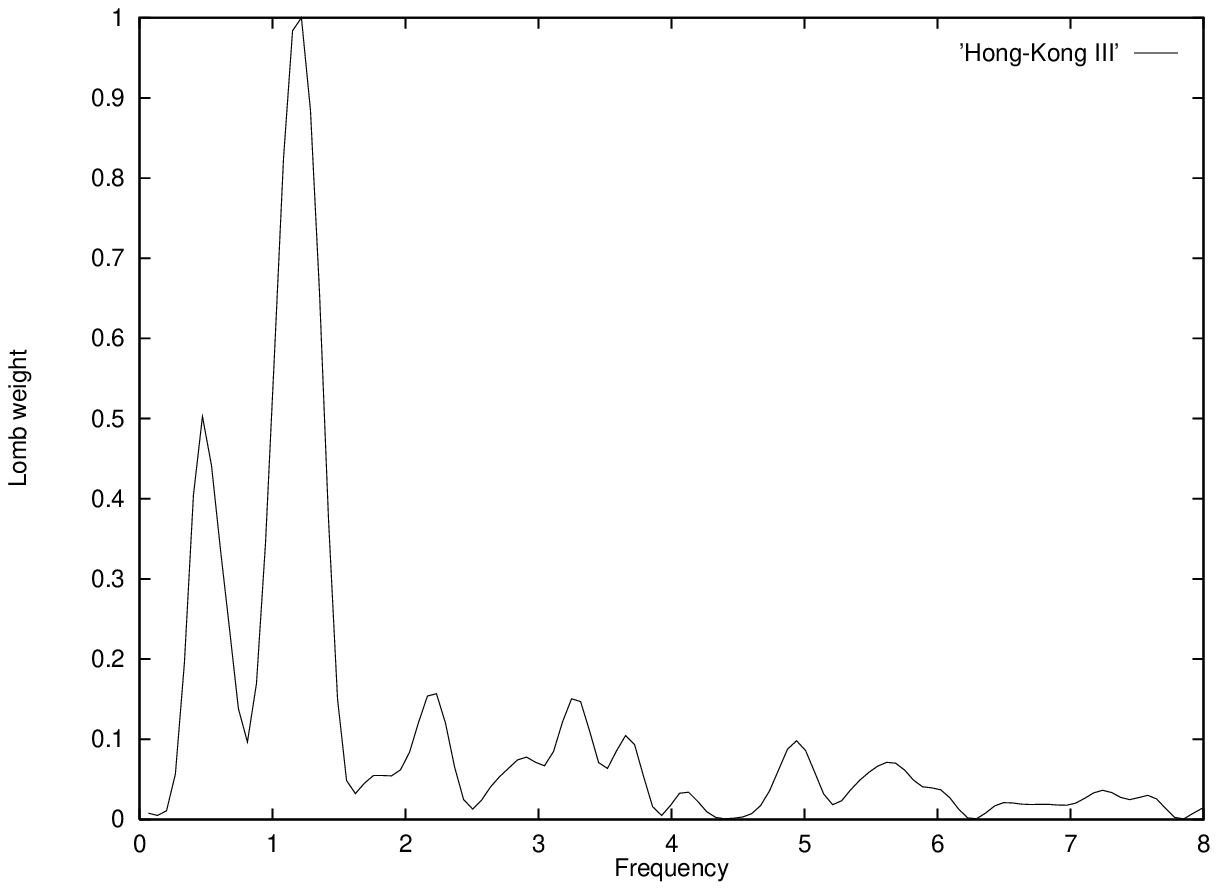,height=7cm,width=8cm}}
\caption{\protect\label{hkbub3} Hong Kong stock market bubble ending with the
crash of Oct. 97. See table \protect\ref{asitab2} for the parameter
values of the fit with eq. (\protect\ref{lpeq}).}

\vspace{5mm}

\parbox[l]{8cm}{\epsfig{file=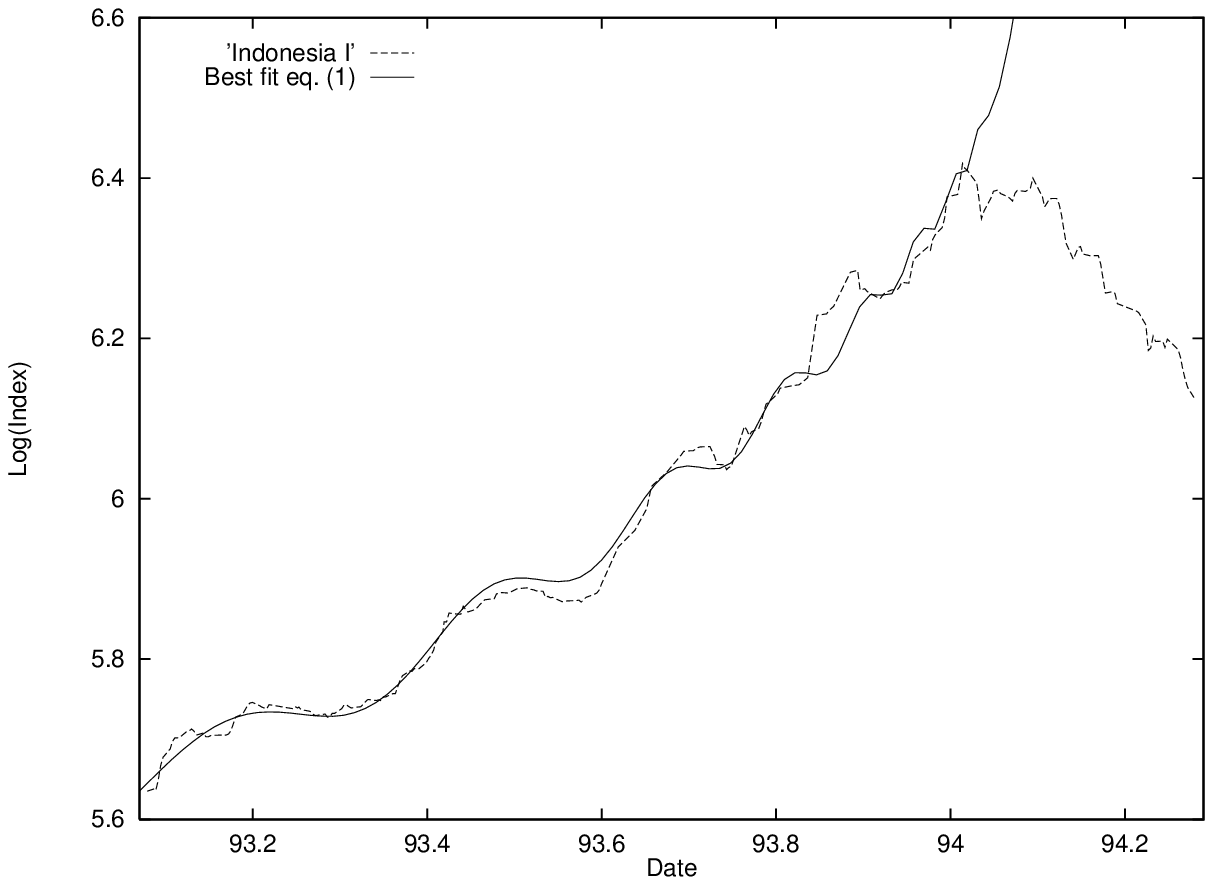,height=7cm,width=8cm}}
\hspace{5mm}
\parbox[r]{8cm}{\epsfig{file=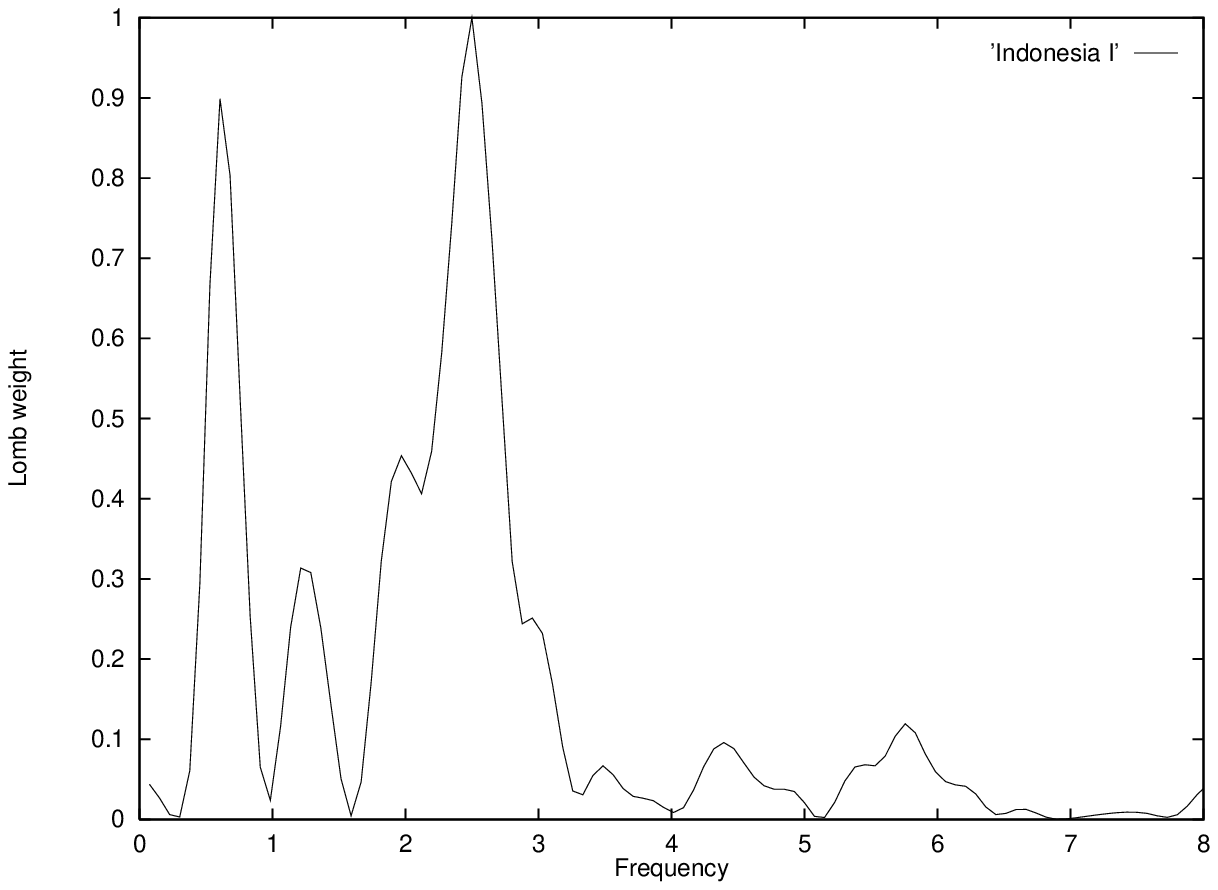,height=7cm,width=8cm}}
\caption{\protect\label{indobub} Indonesian stock market bubble ending in
Jan. 1994. See table \protect\ref{asitab2} for the parameter values of the fit
with eq. (\protect\ref{lpeq}).}
\end{center}
\end{figure}

\begin{figure}
\begin{center}
\parbox[l]{8cm}{\epsfig{file=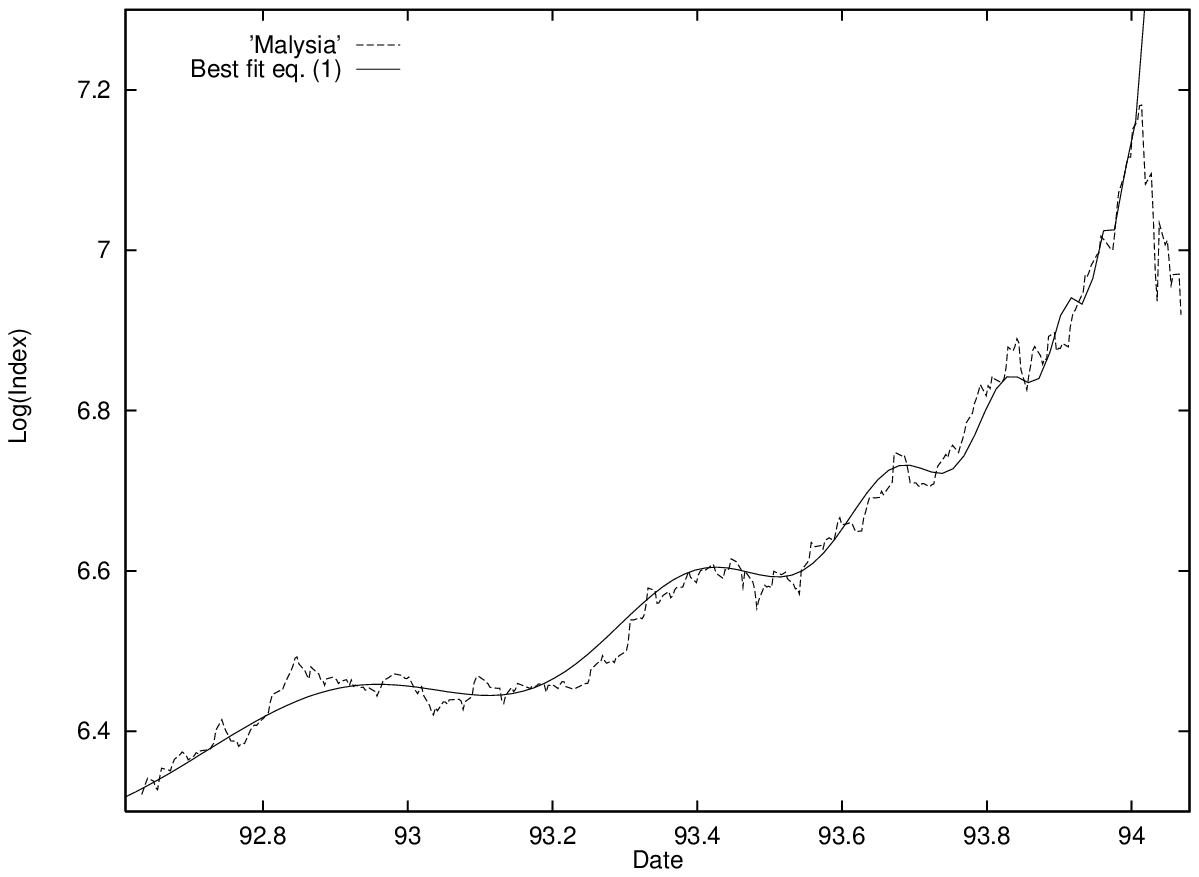,height=7cm,width=8cm}}
\hspace{5mm}
\parbox[r]{8cm}{\epsfig{file=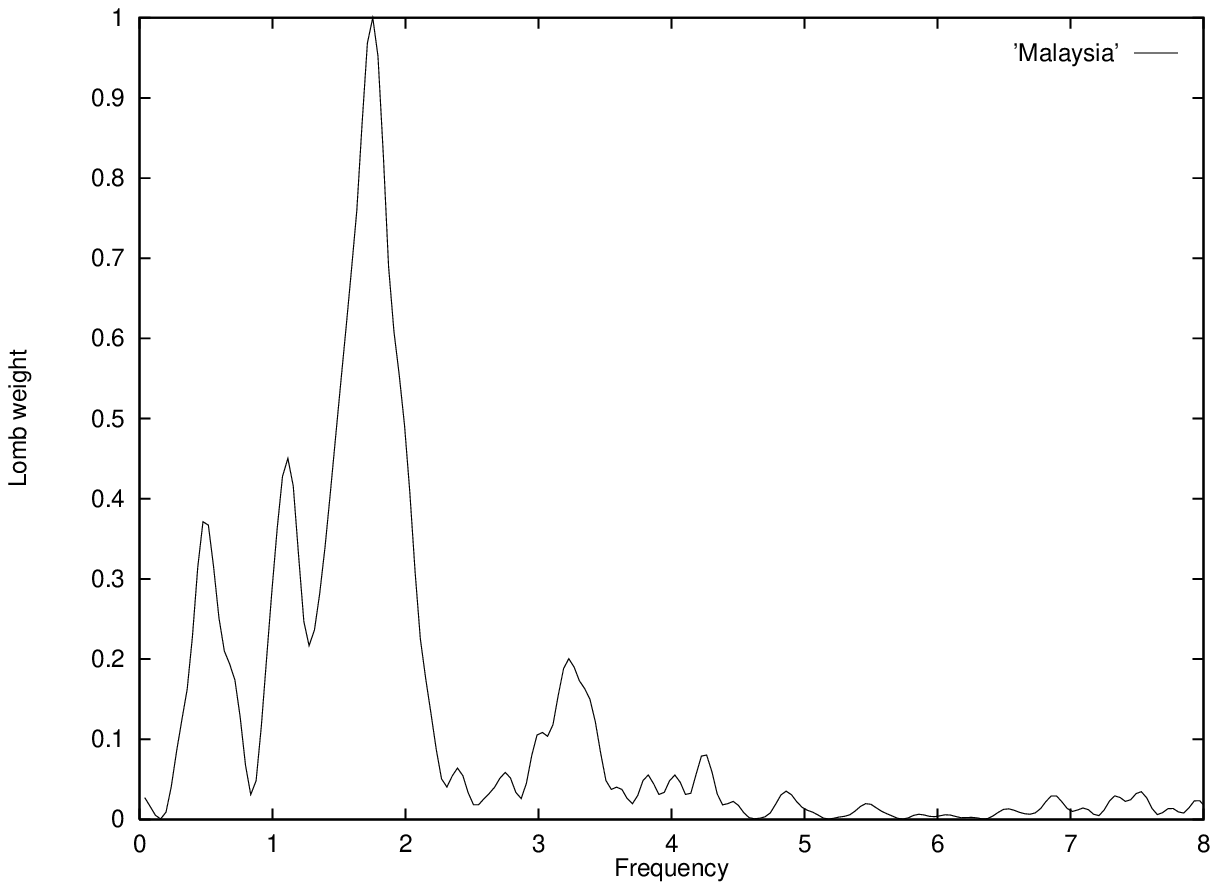,height=7cm,width=8cm}}
\caption{\protect\label{malbub} Malaysian stock market bubble ending with the
crash of Jan. 94. See table \protect\ref{asitab2} for the parameter
values of the fit with eq. (\protect\ref{lpeq}).}

\vspace{5mm}

\parbox[l]{8cm}{\epsfig{file=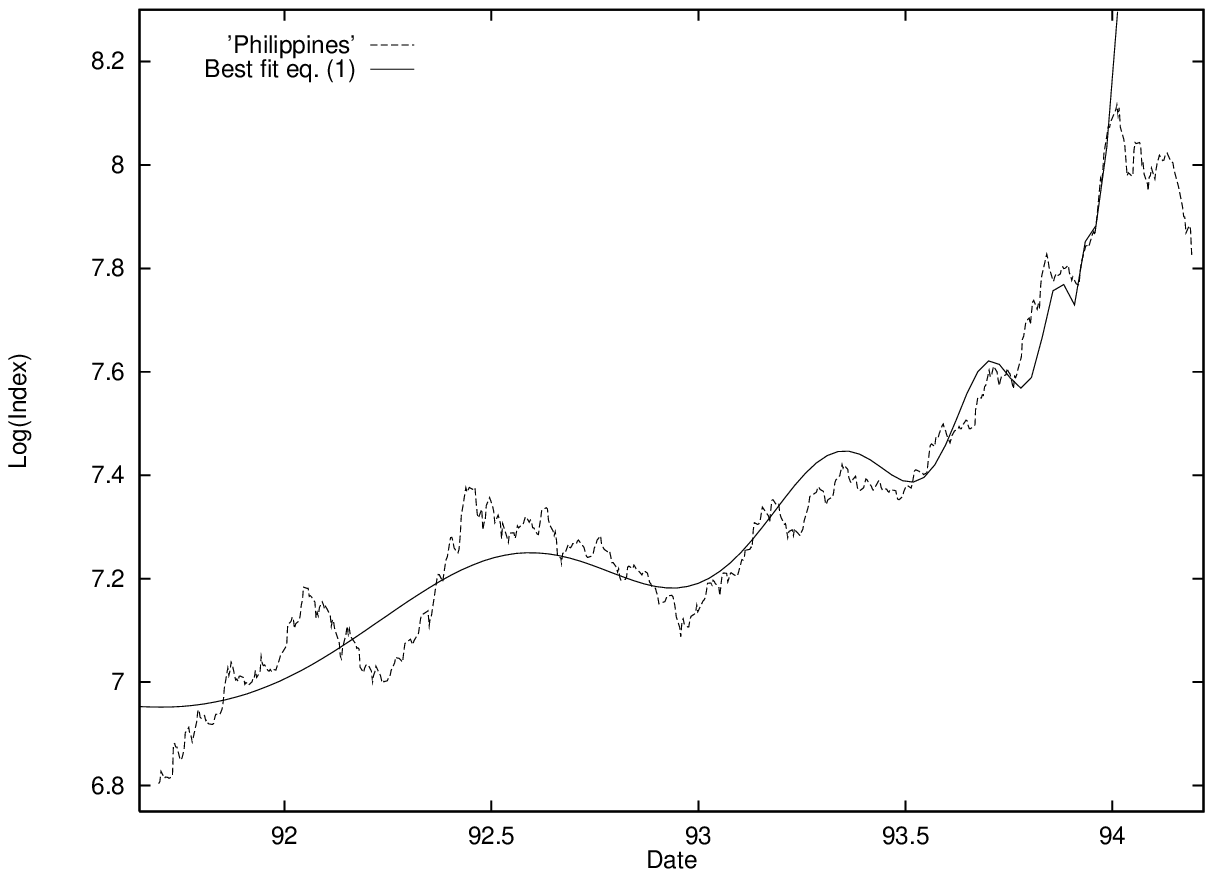,height=7cm,width=8cm}}
\hspace{5mm}
\parbox[r]{8cm}{\epsfig{file=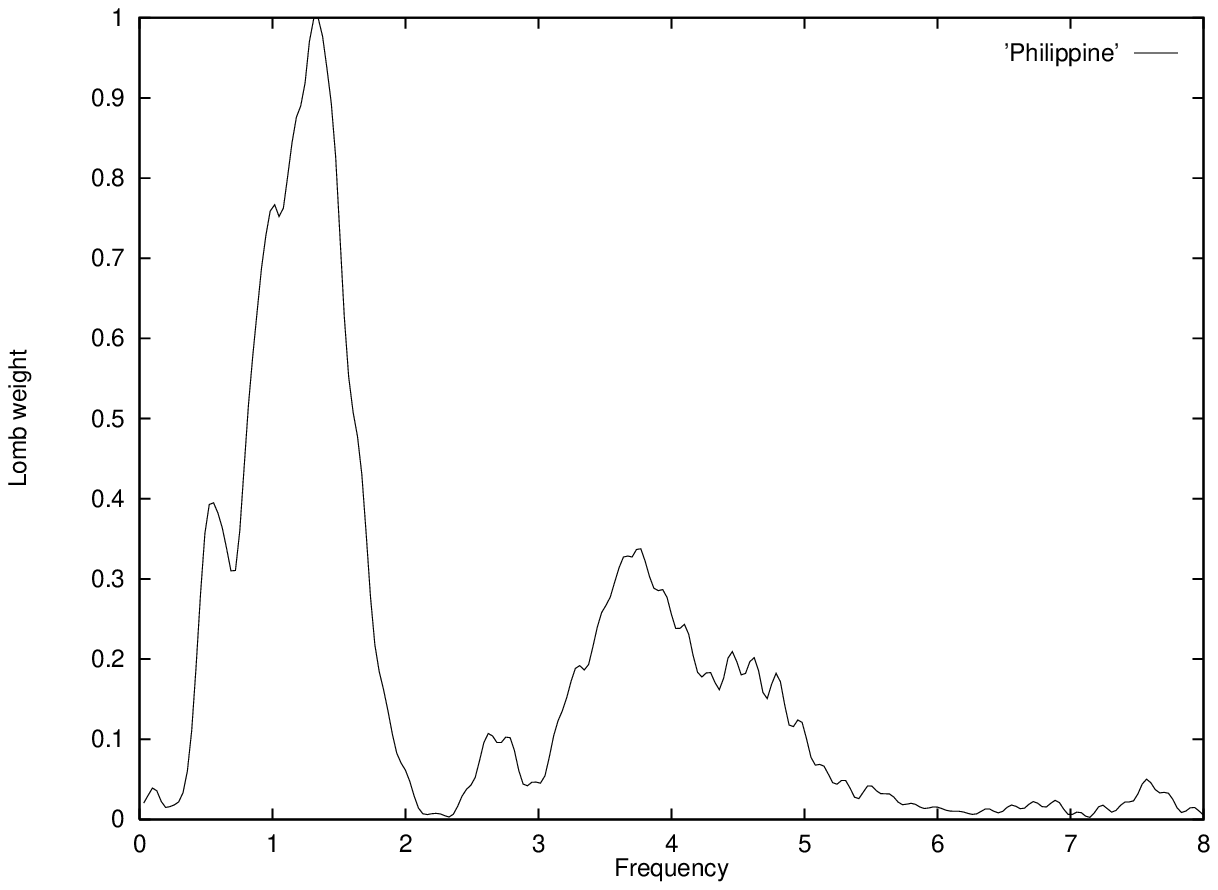,height=7cm,width=8cm}}
\caption{\protect\label{philbub} Philippine stock market bubble ending in
Jan. 1994. See table \protect\ref{asitab2} for the parameter values of the fit
with eq. (\protect\ref{lpeq}).}

\end{center}
\end{figure}

\begin{figure}
\begin{center}
\parbox[l]{8cm}{\epsfig{file=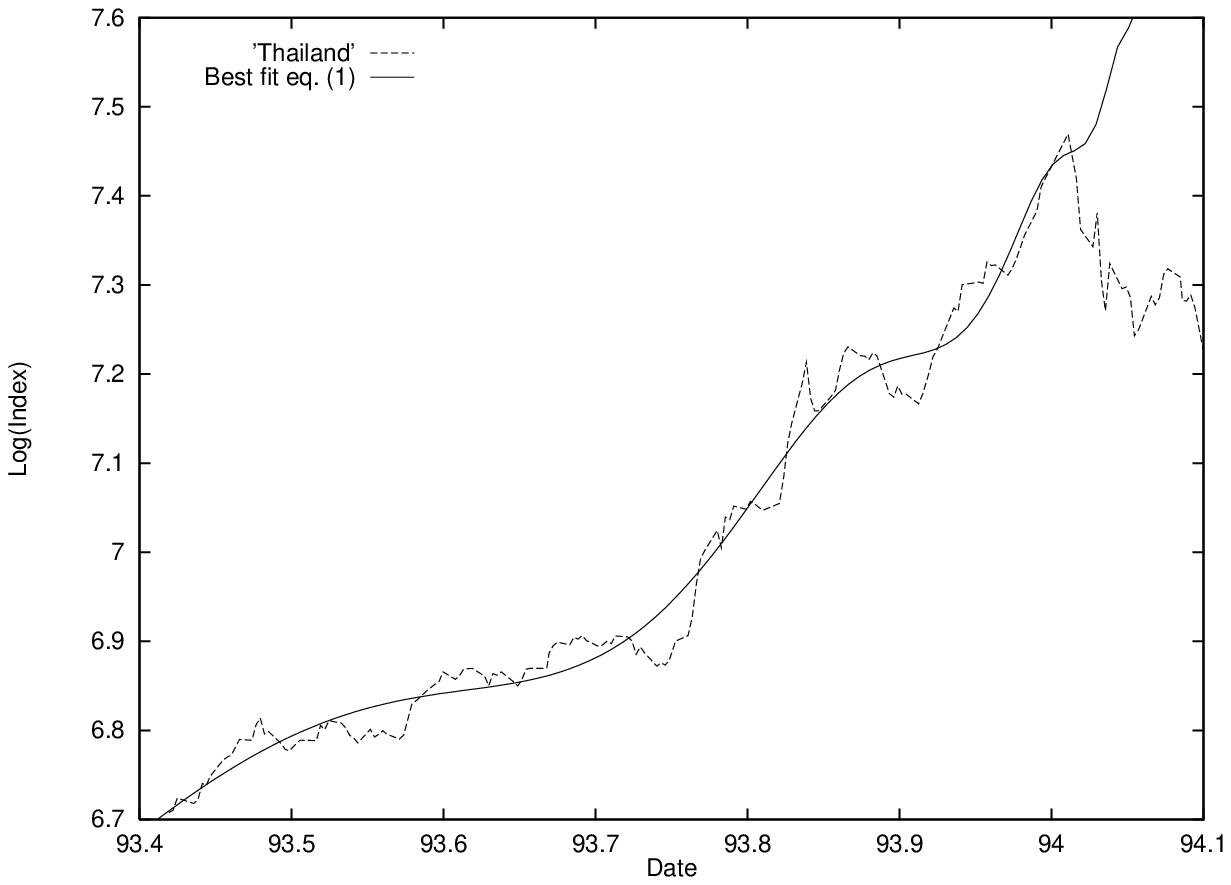,height=7cm,width=8cm}}
\hspace{5mm}
\parbox[r]{8cm}{\epsfig{file=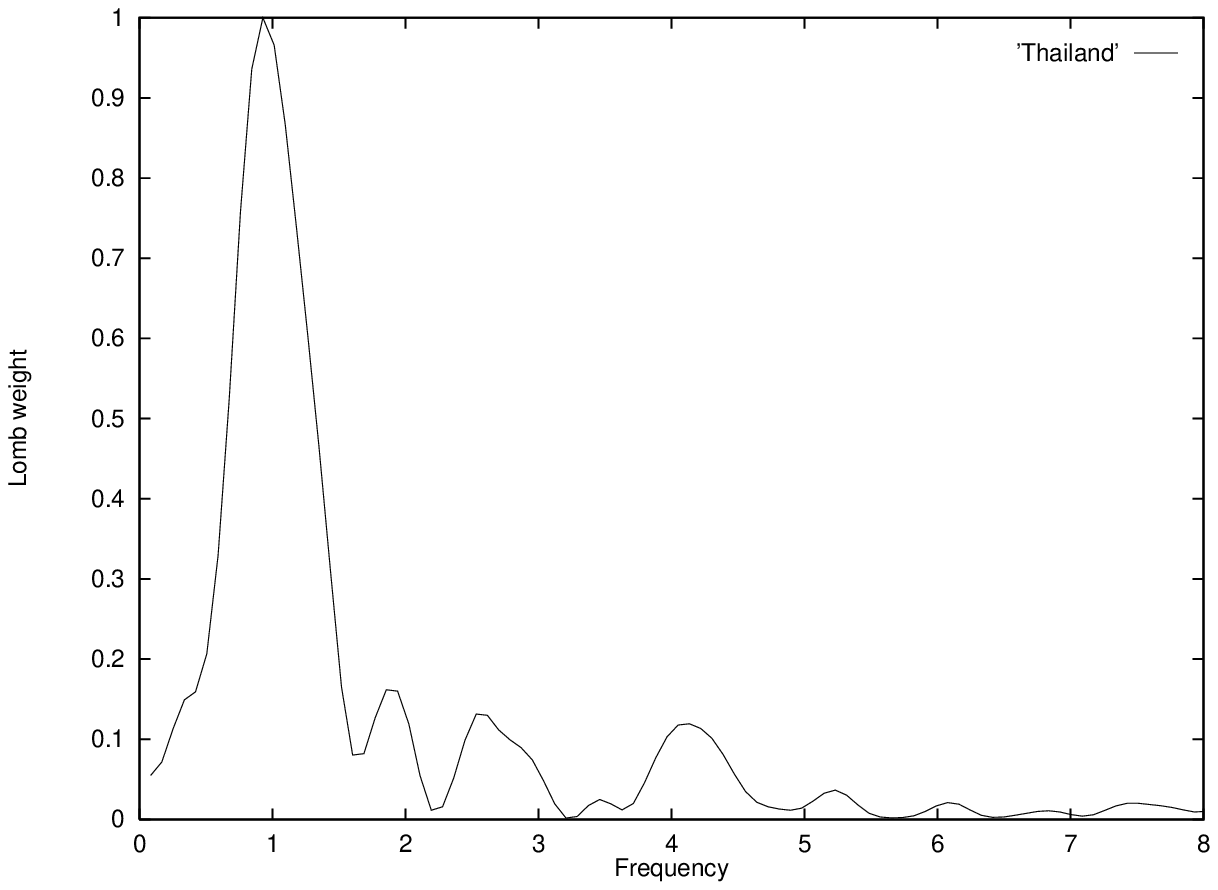,height=7cm,width=8cm}}
\caption{\protect\label{thaibub} Thai stock market bubble ending with the
crash of Jan. 94. See table \protect\ref{asitab2} for the parameter
values of the fit with eq. (\protect\ref{lpeq}).}
\end{center}
\end{figure}

In figures \ref{hkbub1} to \ref{thaibub}, we see the fits of the bubbles
indicated in figures \ref{hk} to \ref{thai} as well as the spectral
Lomb periodogram of the difference between the indices and the
pure power law with the exceptions of the Korean stock market and the second
Indonesian bubble, which could not quantified by eq. (\ref{lpeq}).

In tables \ref{asitab1} and \ref{asitab2}, the parameters of the various fits
are given as well as the beginning and ending dates of the bubble and the size
of the crash/correction. We again see somewhat larger fluctuations in the
values for the exponent $z$ and the log-angular frequency $\omega$ compared
to the WMFMs as for the Latin-American markets. However, except for the
Indonesian and Korean bubbles, the results are surprisingly consistent with
what
has been obtained for the WMFMs as well as for the Latin-American markets.

\section{Correlations across Markets} \label{antibub}

\begin{figure}
\begin{center}
\epsfig{file=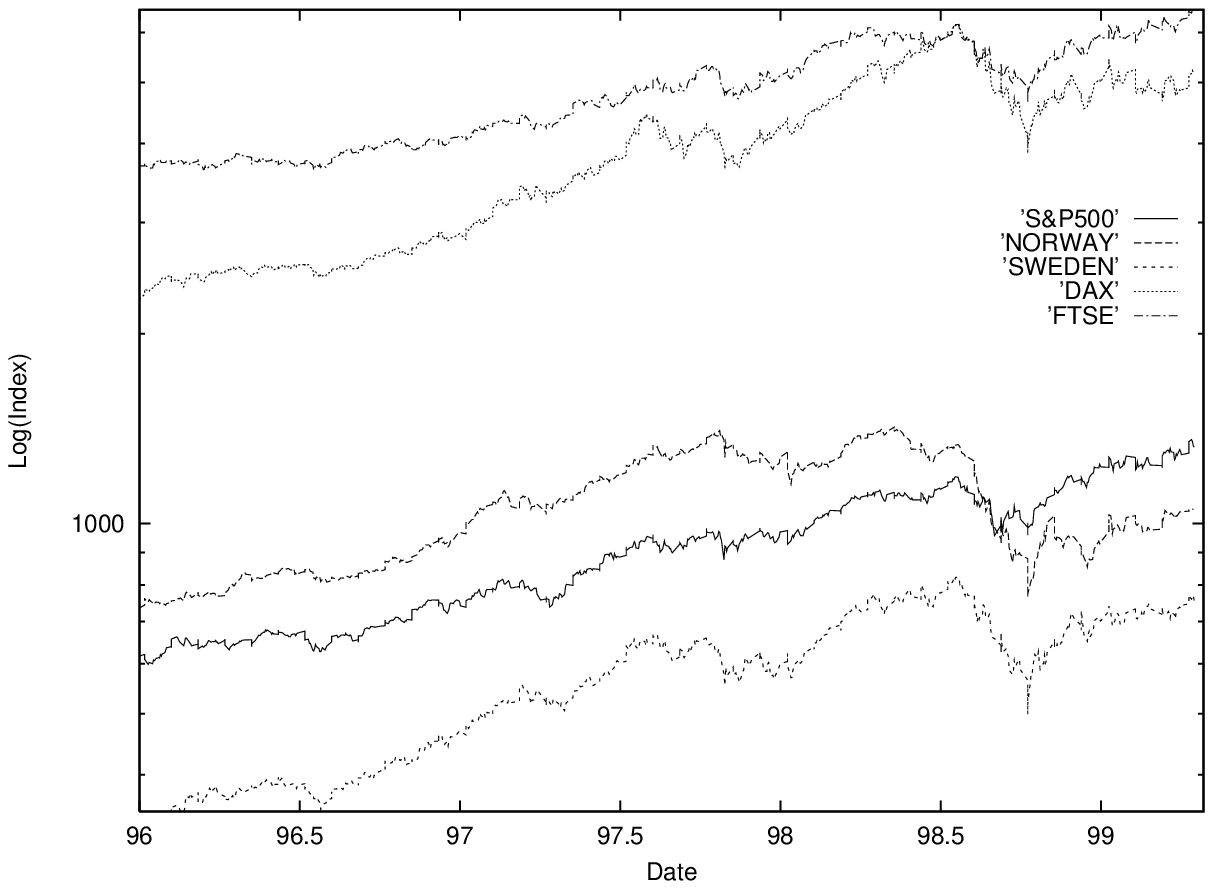,height=9cm,width=15cm}
\caption{\protect\label{weurosp96}The US (S\&P500), DAX (Franfurt), FTSE
(London), Norwegian and Swedish stock market indices as a function of time. The
``dips and peaks'' shown for the smaller markets seems to be better correlated
temporally with the S\&P500 in the end of the time interval
than in the beginning.}

\vspace{5mm}

\epsfig{file=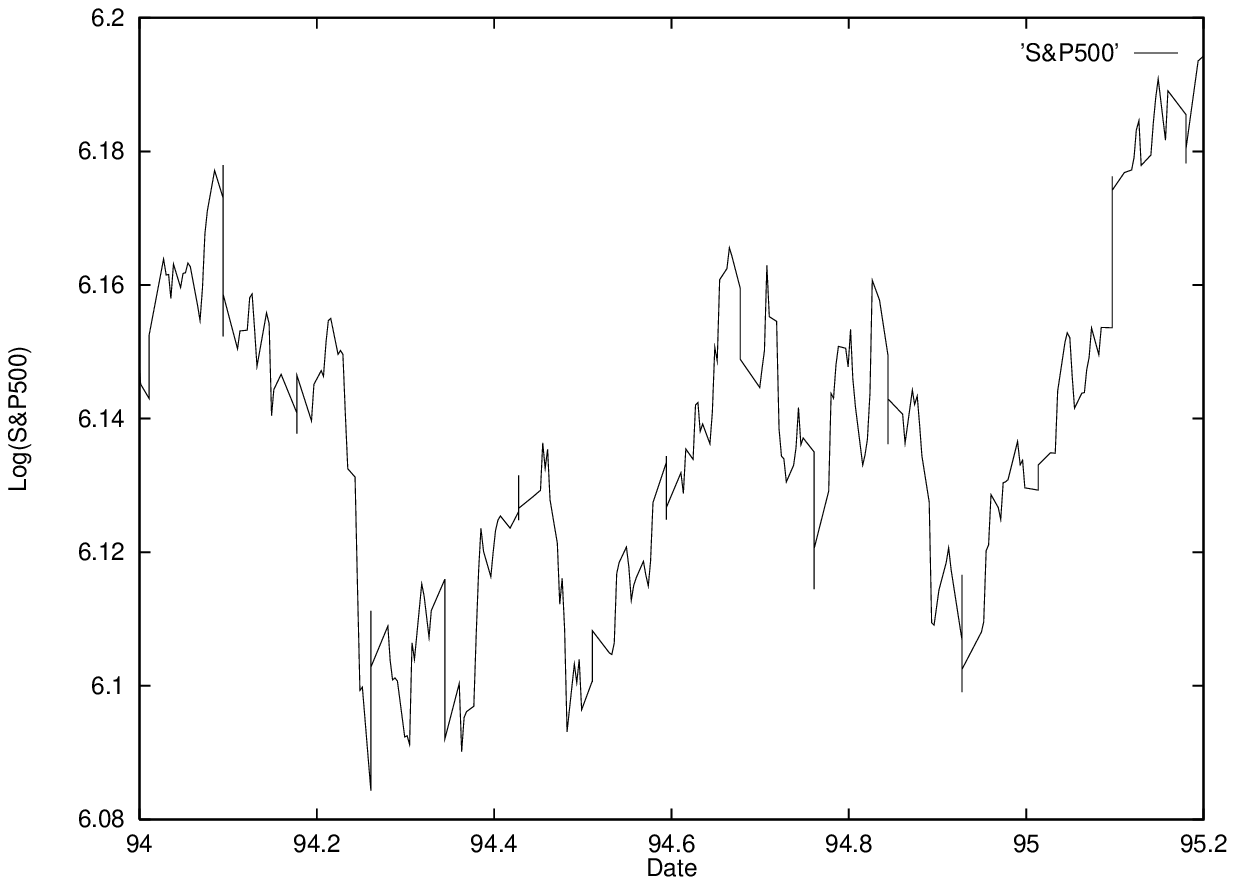,height=9cm,width=15cm}
\caption{\protect\label{sp} The S\&P500 just before and $\approx 1$ year
after the early 1994 financial crises on emerging markets.}
\end{center}
\end{figure}

It is well-known that the Oct. 1987 crash was an international event,
occurring within a few days in all major stock markets \cite{krach87}.
It is also often noted that smaller West-European stock markets  as well
as other markets around the world are influenced by dominating trends on
Wall Street. This correlation seems to have increases over the years as can
be seen with the naked eye by comparing the start and the end intervals of
figure \ref{weurosp96} (showing several market indices prior and after the
Aug. 1998 turmoil). We observe a clear qualitative strengthening of the
correlations between Wall Street and the smaller Western stock markets
in this decade. Specifically, identifying ``spikes'' in either direction
for the two end intervals of figure \ref{weurosp96}, a stronger correspondence
between changes in the various indices is clearly observed in the later period
compared to the former.
We stress that the suggested dependence between these markets would not
necessarily be detected by standard correlation measures, which are averages
over long period of times and detect only a part of possible dependence
structures. What we unravel here corresponds to ``phasing-up'' between markets
at special times of large moves and/or large volatilities.

\begin{figure}
\begin{center}
\parbox[l]{8cm}{\epsfig{file=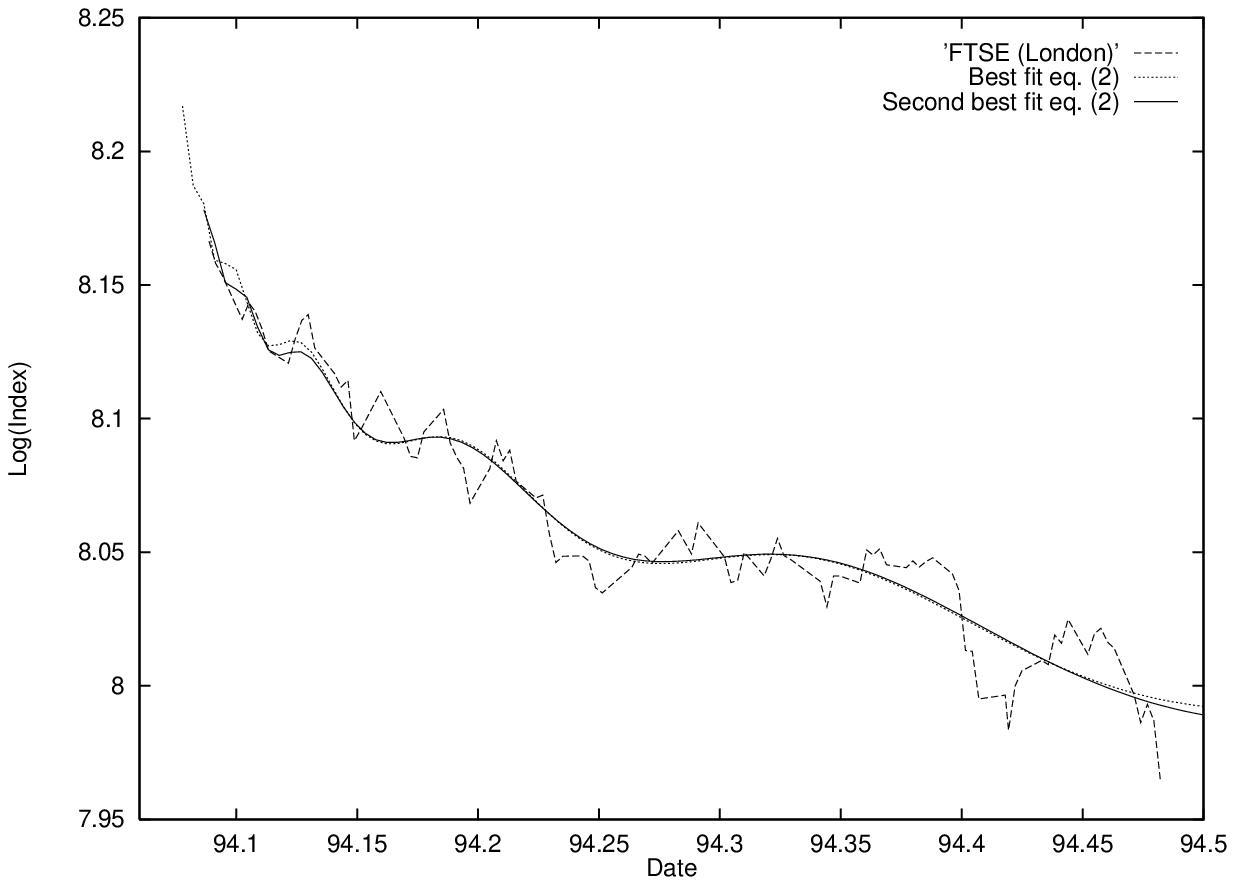,height=7cm,width=8cm}}
\hspace{5mm}
\parbox[r]{8cm}{
\epsfig{file=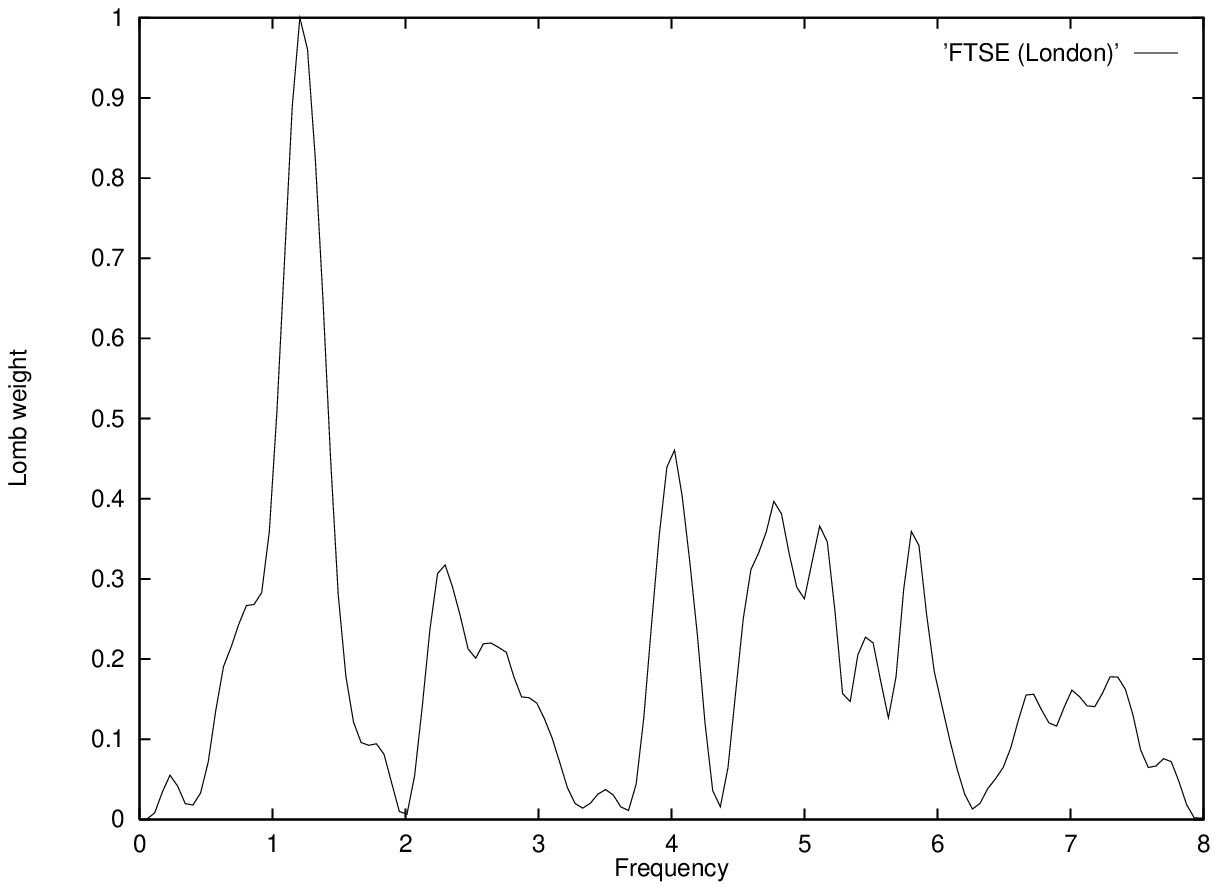,height=7cm,width=8cm}}
\caption{\protect\label{gb} FTSE (London). The 2 lines are the best and
the second best fit with eq. (\protect\ref{lpdeceq}). See table
\protect\ref{antitab2} for the parameter values of the fits.  Only the best
fit is used in the Lomb periodogram.}

\vspace{5mm}

\parbox[l]{8cm}{\epsfig{file=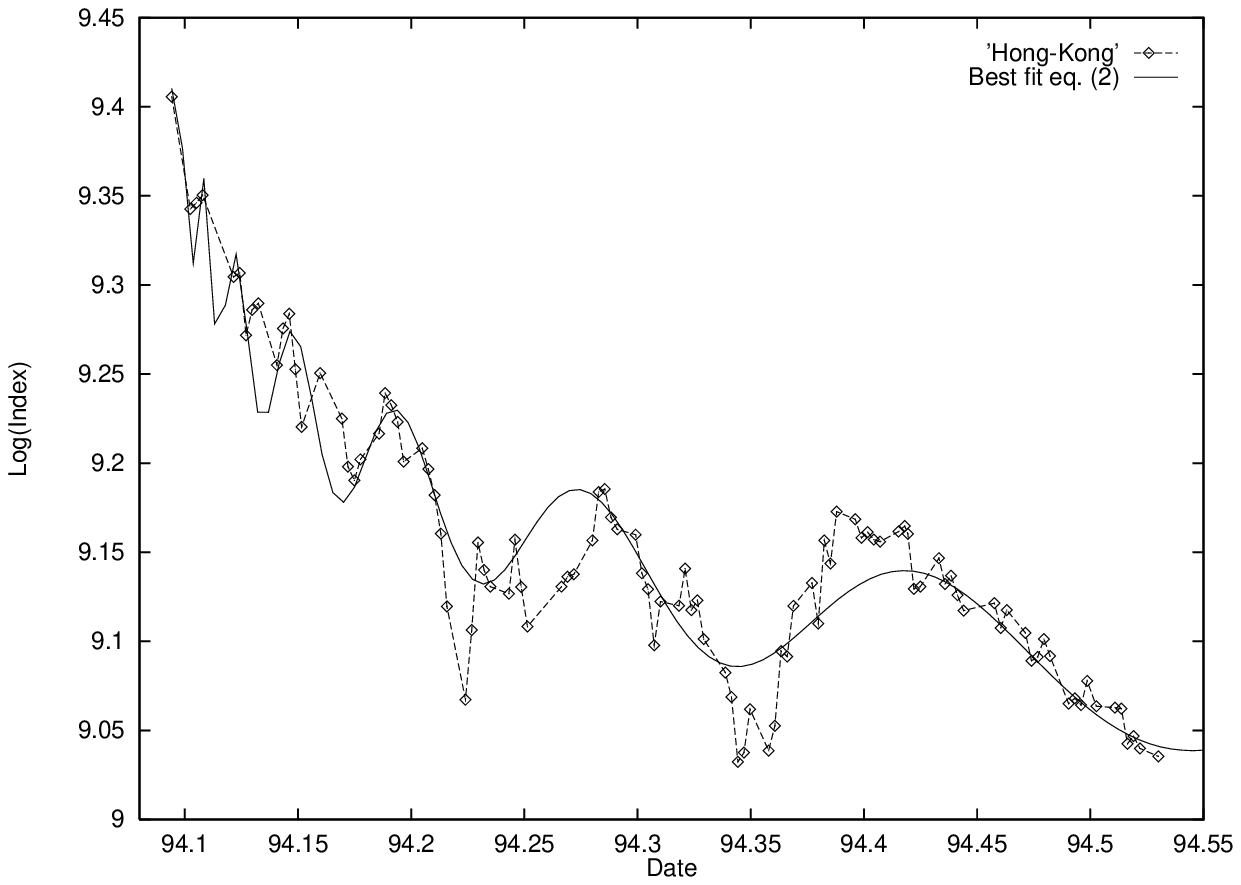,height=7cm,width=8cm}}
\hspace{5mm}
\parbox[r]{8cm}{\epsfig{file=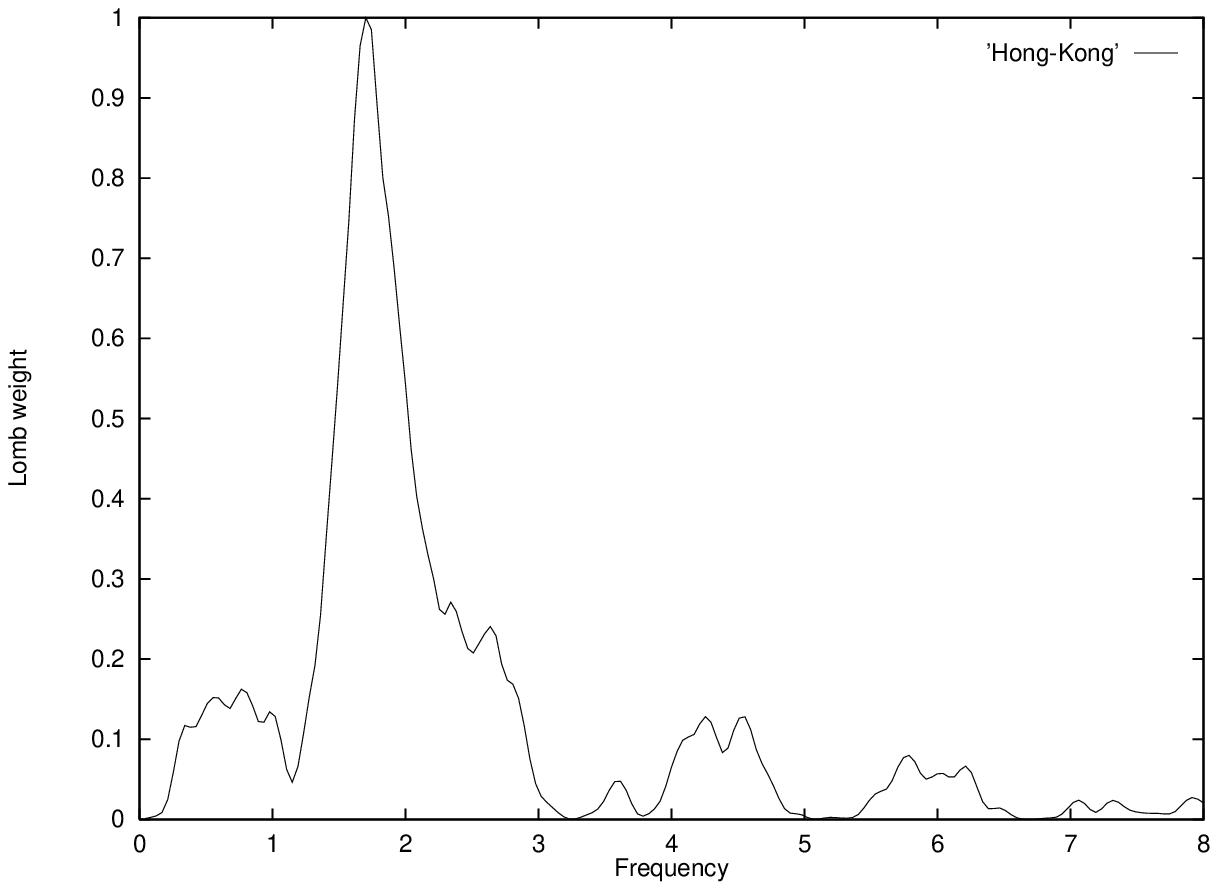,height=7cm,width=8cm} }
\caption{\protect\label{antihk} Hong-Kong. The line is the best
fit with eq. (\protect\ref{lpdeceq}). See table \protect\ref{antitab1}
for the parameter values of the fits. Note the small 
value for the exponent
$z$. This is presumably due to the under-sampling of the data
in the very first part of the data set.}

\end{center}
\end{figure}

\begin{figure}
\begin{center}
\parbox[l]{8cm}{
\epsfig{file=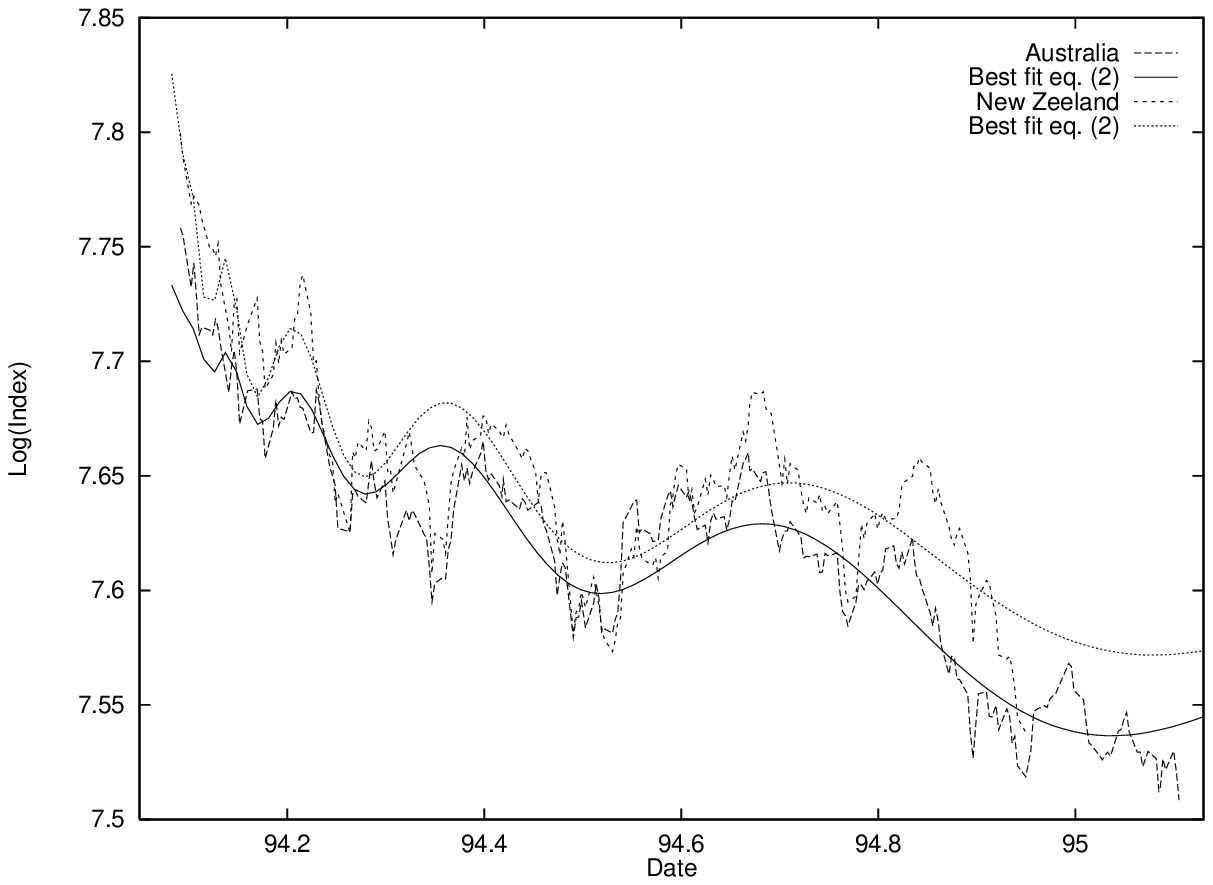,height=7cm,width=8cm}}
\hspace{5mm}
\parbox[r]{8cm}{
\epsfig{file=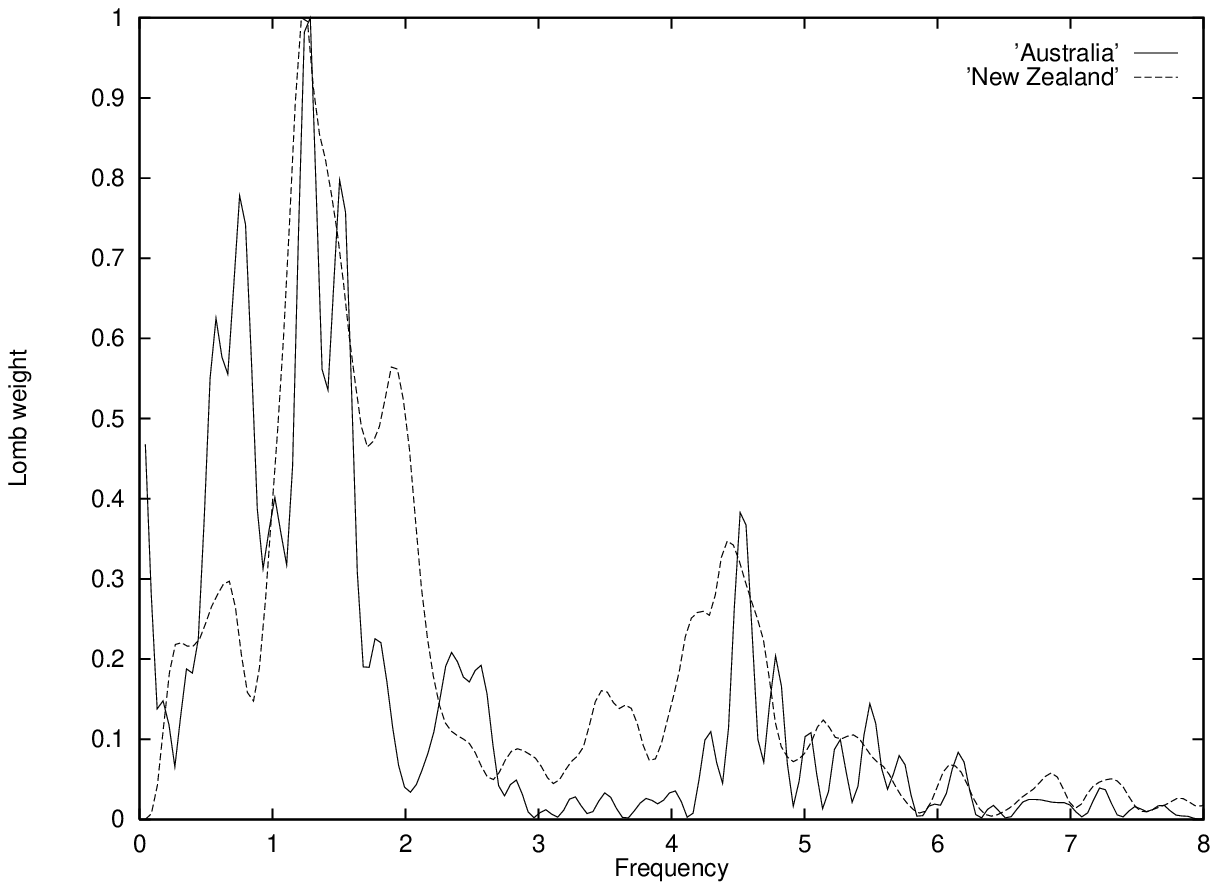,height=7cm,width=8cm} }
\caption{\protect\label{aunz} The Australian  and New Zealand
stock market indices. The line is the best fit with eq.
(\protect\ref{lpdeceq}). See table \protect\ref{antitab2}
for the parameter values of the fits. }

\vspace{5mm}

\parbox[l]{8cm}{
\epsfig{file=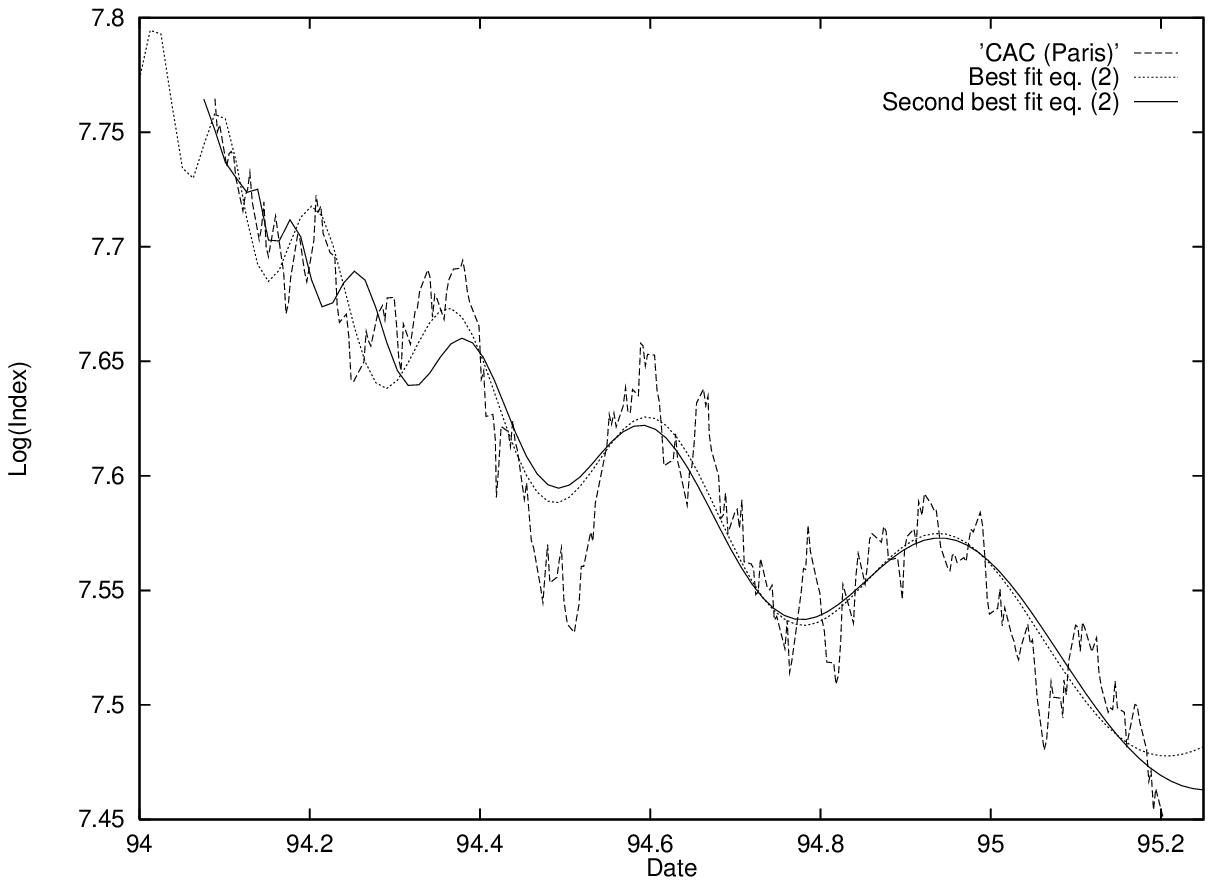,height=7cm,width=8cm}}
\hspace{5mm}
\parbox[r]{8cm}{
\epsfig{file=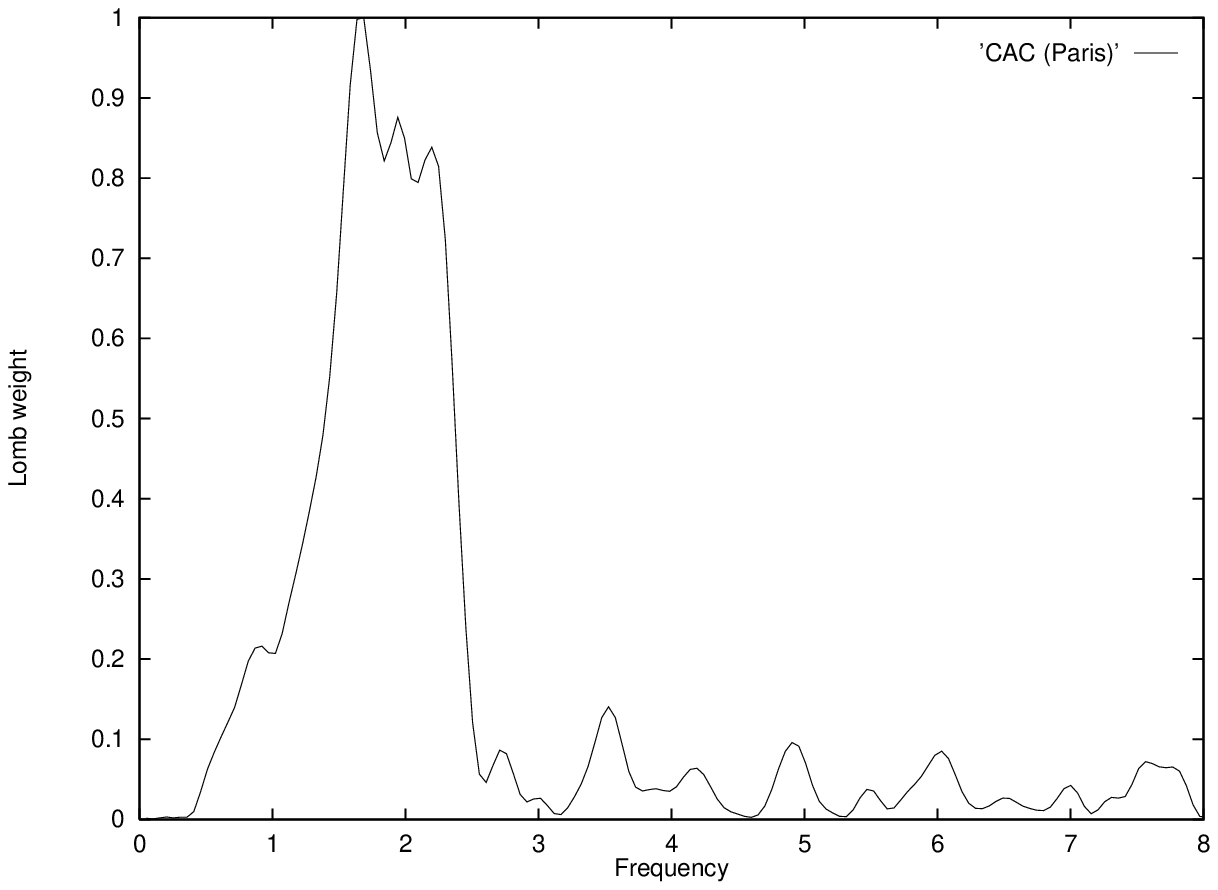,height=7cm,width=8cm} }
\caption{\protect\label{france} The French CAC40. The 2 lines are the best and
the second best fit with eq. (\protect\ref{lpdeceq}). See table
\protect\ref{antitab2} for the parameter values of the fits. Only the second
best fit is used in the Lomb periodogram.}

\end{center}
\end{figure}

\begin{figure}
\begin{center}
\parbox[l]{8cm}{
\epsfig{file=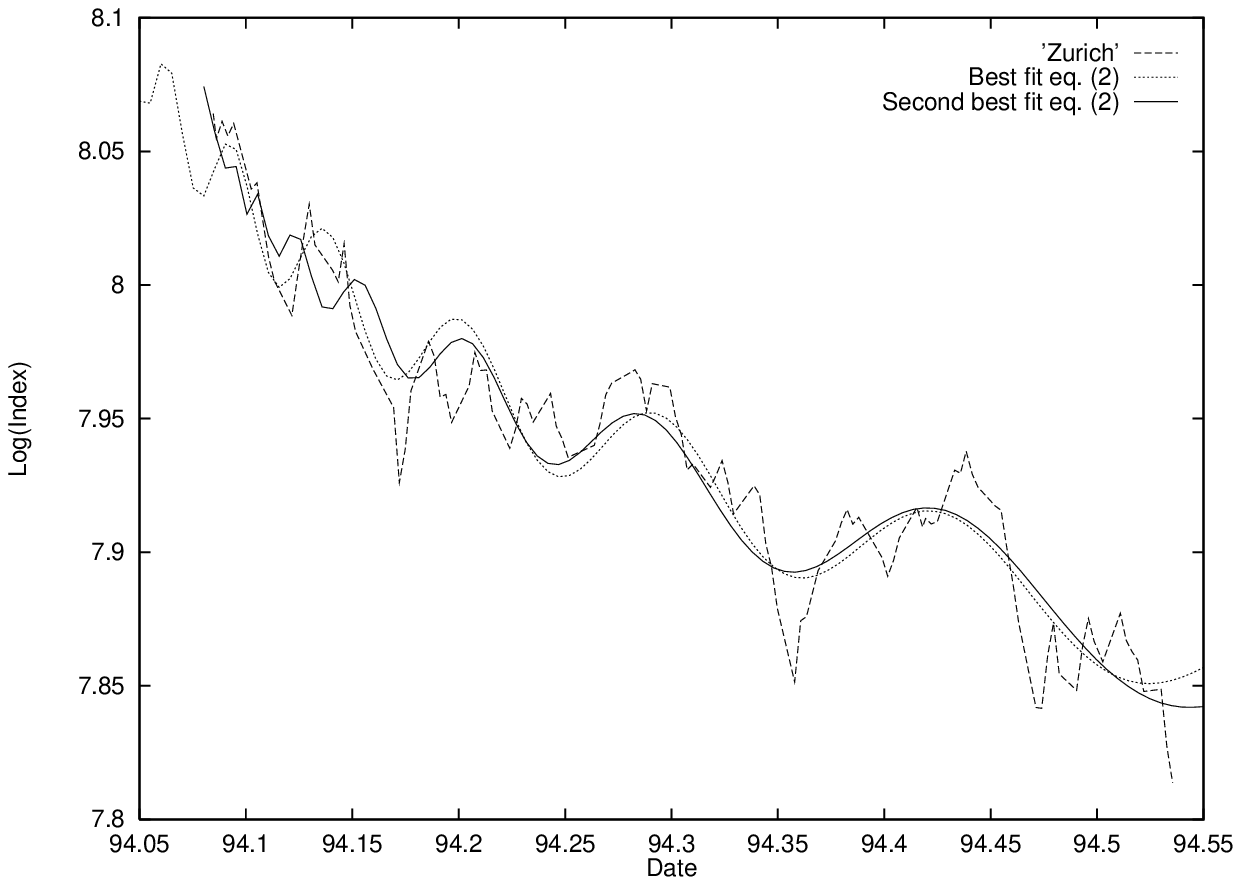,height=7cm,width=8cm}}
\hspace{5mm}
\parbox[r]{8cm}{
\epsfig{file=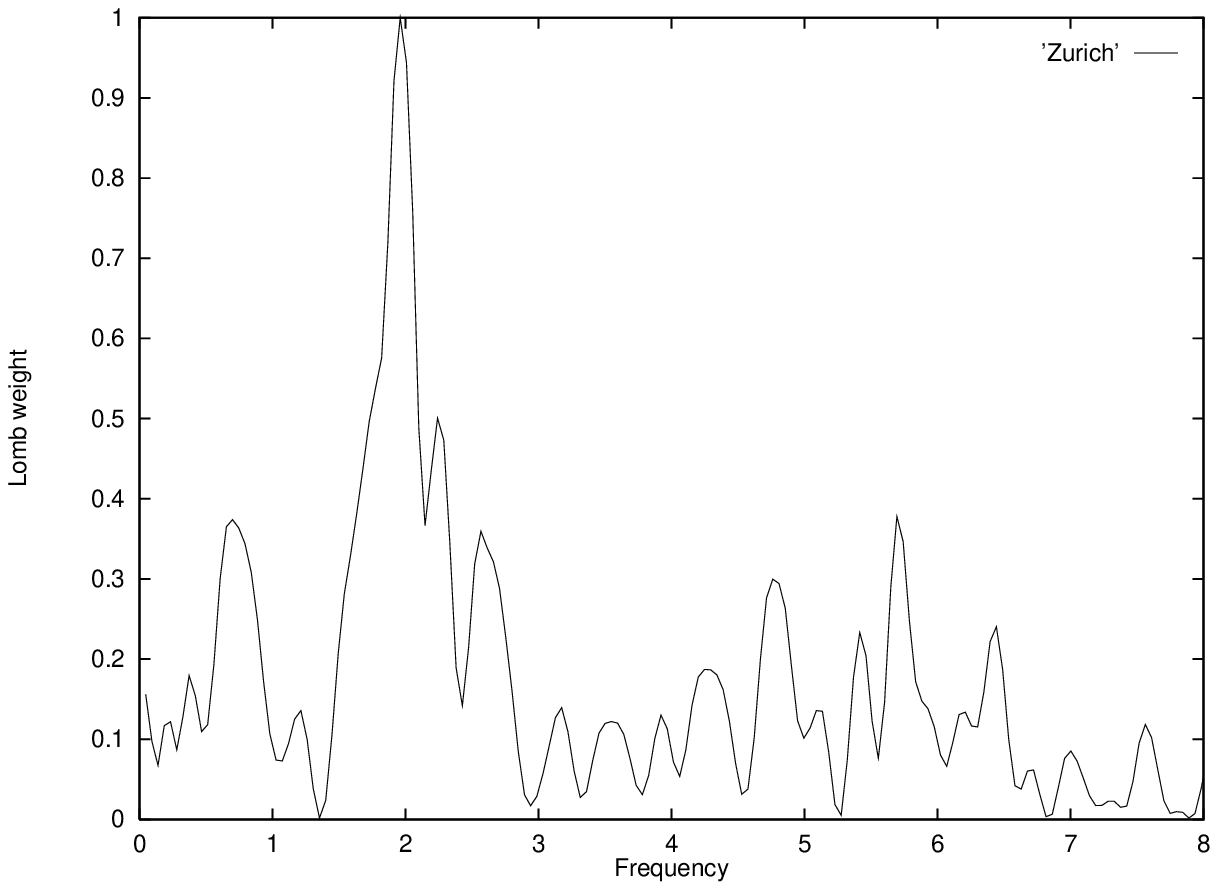,height=7cm,width=8cm} }
\caption{\protect\label{swiss} The Swiss stock market index. The
lines are the two best fit with eq. (\protect\ref{lpdeceq}). See table
\protect\ref{antitab2} for the parameter values of the fits. Only the second
best fit is used in the Lomb periodogram.}

\vspace{5mm}

\parbox[l]{8cm}{
\epsfig{file=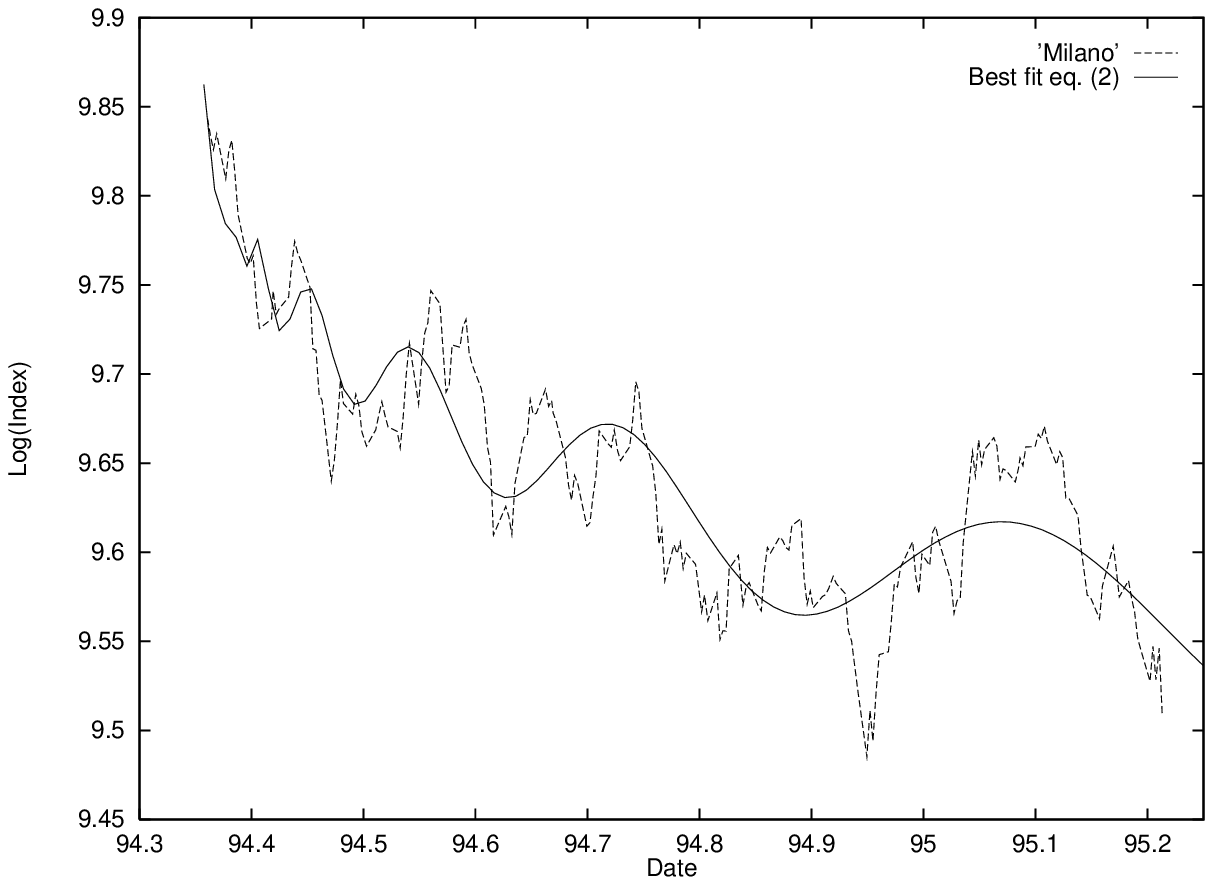,height=7cm,width=8cm}}
\hspace{5mm}
\parbox[r]{8cm}{
\epsfig{file=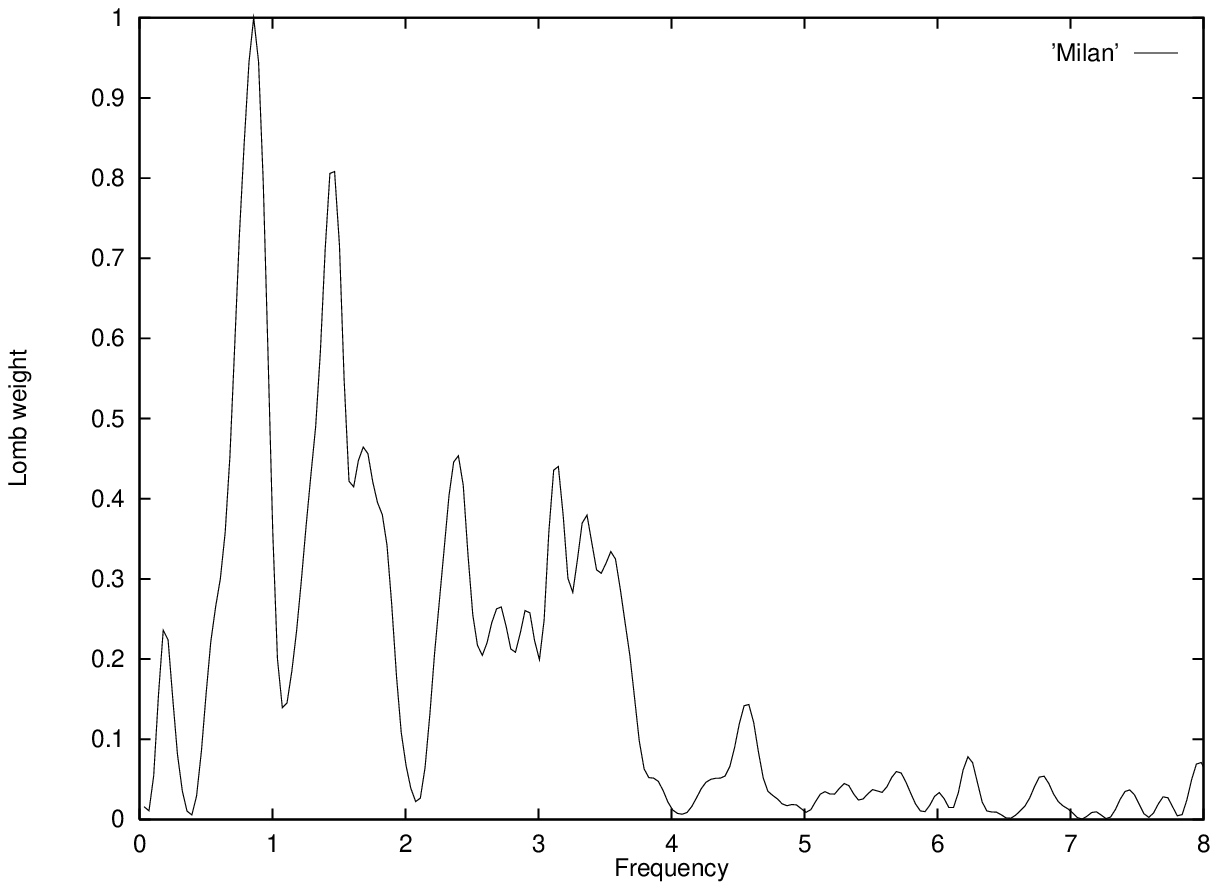,height=7cm,width=8cm}}
\caption{\protect\label{italy} Italian stock market index. The line
is the best fit with eq. (\protect\ref{lpdeceq}). See table
\protect\ref{antitab2} for the parameter values of the fits.}

\end{center}
\end{figure}

\begin{figure}
\begin{center}
\parbox[l]{8cm}{
\epsfig{file=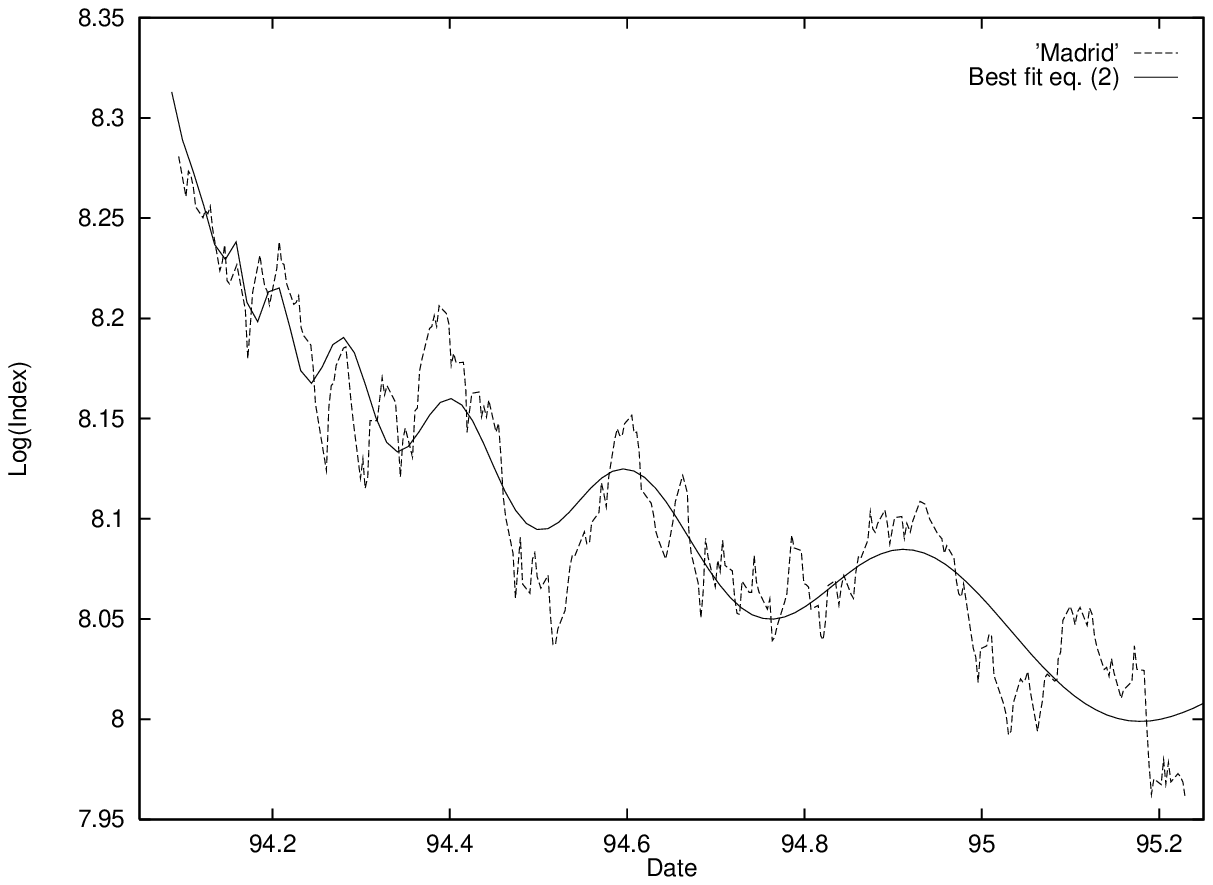,height=7cm,width=8cm}}
\hspace{5mm}
\parbox[r]{8cm}{
\epsfig{file=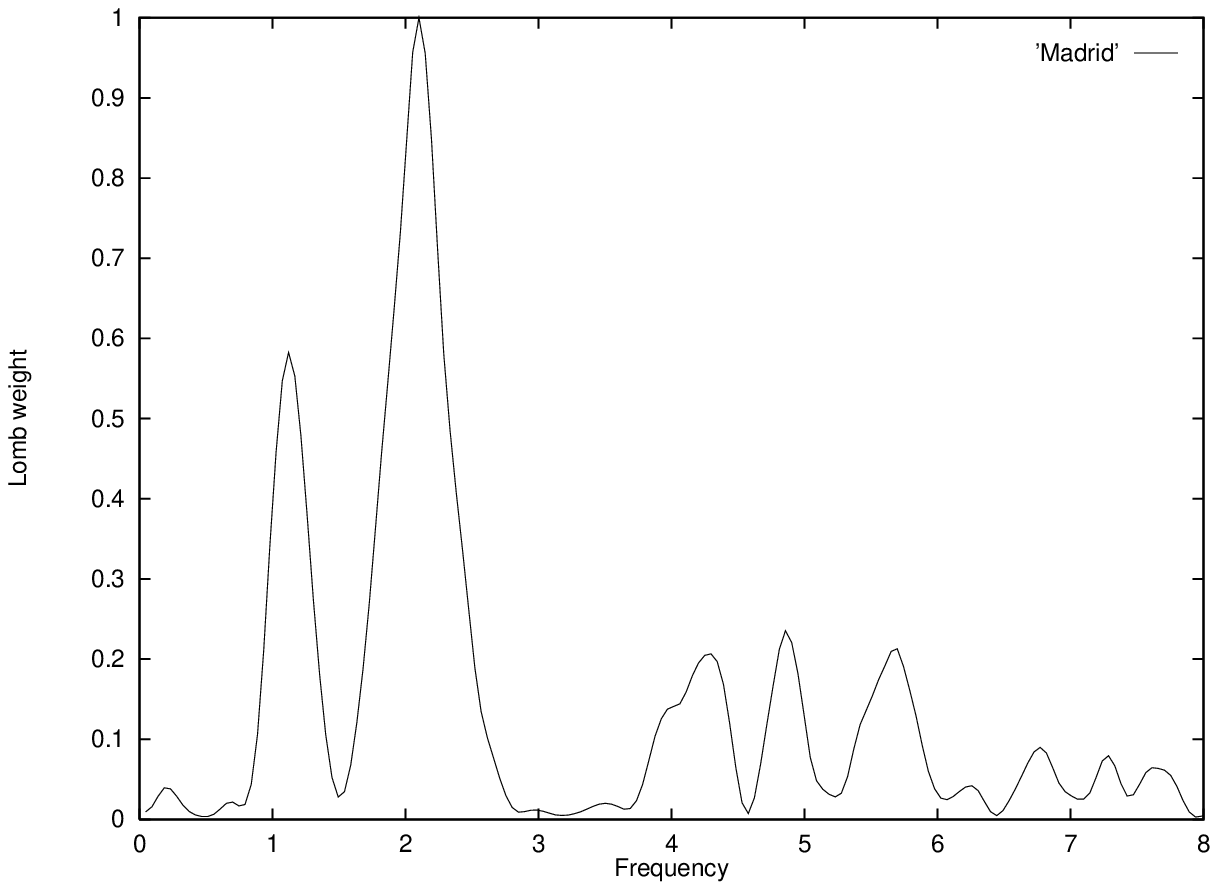,height=7cm,width=8cm}}
\caption{\protect\label{spain} The Spanish (Madrid) stock market index. The
line is the best fit with eq. (\protect\ref{lpdeceq}). See table
\protect\ref{antitab2} for the parameter values of the fits.}
\end{center}
\end{figure}

An example of a decoupling between the West-European stock markets and
Wall Street in the first part of this decade comes from the period
following the crashes/corrections on most
emerging stock markets in early 1994. This rash of crises occurred from
January to June 1994 and concerned the currency markets (Mexico, South Africa,
Turkey, Venezuela) and the stock markets (Chile, Hungary, India, Indonesia,
Malaysia, Philippines, Poland, South Africa, Turkey, Venezuela, Germany,
Hong-Kong, Singapore, U.K.) \cite{Rand}. The period of time is associated to
sharply rising U.S. interest rates. Whereas the S\&P500 dipped less than
$10$\% and recovered within a few months, see figure \ref{sp}, the effect was
much more profound on smaller Western stock markets worldwide. Surprisingly,
the toll on a range of western countries resembled that of a mini-recession
with decreases between $18$\% (London) and $31$\% (Hong Kong) over a period
from $\approx 5$ months (London) to $\approx 13$ months (Madrid), as
summarized in table \ref{antitab1}. For each stock market, the decline in the
logarithm of the index has been fitted with eq. (\ref{lpdeceq}). In figures
\ref{gb} to  \ref{spain}, we see that the decreases in all the stock
markets analyzed can be quantified by eq. (\ref{lpdeceq}) as {\it
log-periodic anti-bubbles} \cite{antibulle}.

Using the second best fit of the CAC40 and the Swiss indices, we see that
the dates of the start of the decline is well-estimated by the value of $t_c$
obtained from the fit. Furthermore, from table \ref{antitab1}, we observe that
the value of the prefered scaling ratio $\lambda = e^{2\pi / \omega}$
is remarkable consistent $\lambda \approx 2.0 \pm 0.3$. This comes as a good
surprise, considering that the stock markets that have been analyzed belong
to three very different geographical regions of the world (Europe, Asia and
Pacific). With respect to the value of the exponent $z$, the fluctuations are
as usual much larger. However, excluding New Zealand and Hong-Kong\footnote{A
possible explanation for the very low value $z\approx 0.03$ may be the
under-representation of trading days in the first part of the data interval
due to holidays. Hence, the last part of the data, where the deceleration is
weaker, is allowed to dominate thus underestimating the trend. A somewhat
less severe under-sampling was also present in the New Zealand index compared
to, {\it e.g.}, the Australian index.}, we obtain $z \approx 0.4 \pm 0.1$,
which again is quite reasonable compared to WMFMs \cite{JLS,JSL}. Furthermore,
the amplitudes $C$ of the log-periodic oscillations are remarkable similar
with $C\approx 0.3-0.4$, except for London ($\approx 0.02$) and Milan
($\approx 0.05$), as shown in table \ref{antitab2}.

\section{Conclusions}

Log-periodic bubbles followed by large crashes/corrections seem to be a
statistical significant feature of Latin-American and Asian stock markets.
Indeed, it seems quite unprobable to attribute the results obtained for the
Latin-American and smaller Asian stock markets to pure noise-fitting because
of the relatively large number of successful cases ($18$) compared to the
number
of unsuccessful cases ($4$) as well as the objective criteria used in
identifying them. Furthermore, removing the extreme value of $\lambda = 8.8$
for one of the Chilean bubbles gives an average of $\left< \lambda\right>
\approx 2.6$ for the remaining $17$ cases, which is very close to the average
value found for the worlds major financial markets
\cite{JS98.2,manisfesto,JLS,JSL}. However, the results obtained for the
Latin-American and smaller Asian markets are as expected less striking
on a one-to-one basis than
those obtained on the major financial markets of the world (WMFMs) that we
analyzed previously with exactly the same methodology
\cite{JS98.2,manisfesto,JLS,JSL}. In this respect, it is quite remarkable that
the bubbles prior to the 3 largest crashes on the Hong-Kong stock market have
the same log-frequency within $\pm 15$\% and quite similar to what has been
found for bubbles on Wall Street and the FOREX.

One important difference lies in the identification of a bubble. For the WMFMs,
the identification of the first and last data point to be used in the fitting
to our formulas was straightforward: The last point was chosen as the highest
value of the index prior to the crash and the first point was the lowest value
prior to the bubble. The results using these criteria have always been
conclusive and a re-run of the fitting algorithm on a different interval was
never necessary. This was not the case for the Latin-American and smaller
Asian stock markets, where the first point had to be changed in approximately
half the cases. This ambiguity is also reflected in the large fluctuations
seen in the parameter values obtained for the meaningful variables $z$ and
$\omega$ (or equivalently $\lambda=e^{2\pi/\omega}$). Weaker signatures
naturally gives larger fluctuations as well as additional sensitivity to
truncation of the fitted interval. The cause for the weaker signatures can
be (at least) three-fold. The signatures can be truly weaker, or they appear
weaker due to the poorer quality of these smaller indices compared to those
the major stock markets. Another hypothesis is the ``finite-size effect''
already mentioned according to which the smallest market size entails larger
fluctuations and possible systematic bias. It seems at present difficult to
distinguish between these different hypotheses.

\begin{figure}
\begin{center}
\epsfig{file=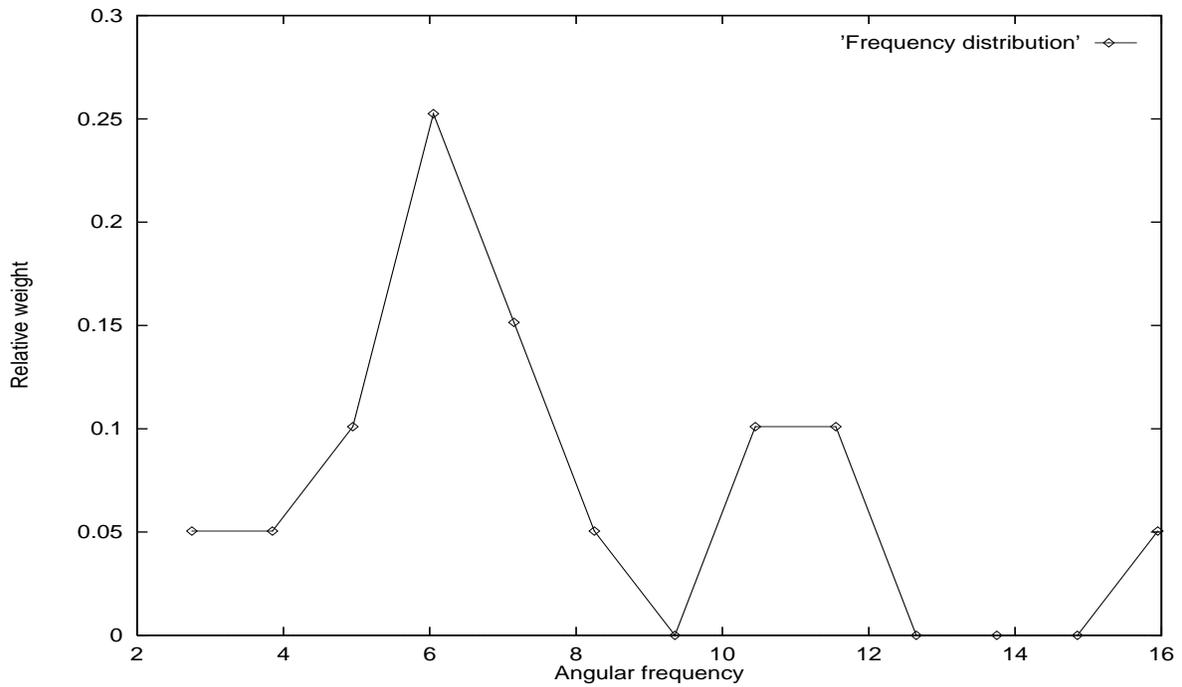,height=9cm,width=16cm}
\caption{\protect\label{histo} Distribution of $\omega$'s for the bubble-fits
given in tables \protect\ref{lattab2} and \protect\ref{asitab2}.}
\end{center}
\end{figure}

With respect to the values obtained for the frequency for the best fits of
the $18$ Latin-American and Asian bubbles that could be quantified using
eq. (\ref{lpeq}), it is rather interesting to see that the fit falls in two
rather distinct clusters one around $\omega \approx 6$ and another around
$\omega \approx 11$ with few values in between as shown in figure
\ref{histo}. It looks like a
frequency-doubling, which correspond to squaring $\lambda$, as allowed by the
theory of critical phenomena \cite{physreports,Saleursor}.

In the second part of this research, we have tried to argue that, in
bullish times, the leading trends on Wall Street will tend to dominate the
smaller Western stock markets. However, it was shown by a quantitative analysis
that these smaller stock markets can collectively decouple their dynamics
from Wall Street. The case we document corresponds to a surprisingly correlated
anti-bubble with log-periodic signatures and power law decay similar to what
has been found on longer times scales for the Nikkei and Gold decays
\cite{antibulle}. In fact, the results obtained for the majority of these
smaller anti-bubbles, {\it i.e.}, excluding the small values of the exponent
$z$ obtained for the New Zealand and Hong Kong stock markets, are quite
compatible with what was obtained for the Gold decay both with respect to the
values for the exponent $z$ and prefered scaling ratio $\lambda$. This
supports the notion that the higher values obtained for $\omega$ is presumably
due to a more rapid dynamics present in smaller market as proposed for the
Gold decay in the early eighties \cite{antibulle}. With respect to the
identification of the data intervals used for the smaller anti-bubbles, we
stress that it did not suffer from the same problems as the Latin-American
and Asian bubbles and could be directly identified unambiguously prior to and
independently from the fitting procedure.

\vskip 0.5cm
We acknowledge stimulating discussions with D. Darcet and encouragements of
D. Stauffer.

\newpage

\mbox{}
\newpage

\mbox{}

\newpage

\begin{table}
\begin{center}
\begin{tabular}{|c|c|c|c|c|c|c|c|c|c|} \hline
Stock market & $t_c$ & $t_{max}$ & $t_{min}$ & $\%$ drop & $z$ & $\omega$ &
$\lambda$ \\ \hline
Argentina I &  $91.80$ & $91.80$ & $91.90$ & $26\%$ & $0.37$ & $4.8$ & $3.7$
\\ \hline
Argentina II&  $92.43$ & $92.42$ & $92.90$ & $59\%$ & $0.22$ & $11.4$ & $1.7$
\\ \hline
Argentina III&  $94.13$ & $94.13$ & $94.30$ & $30\%$ & $0.19$ & $7.2$ & $2.4$
\\ \hline
Argentina IV & $97.89$ & $97.81$ & $97.87$ & $27\%$ & $0.20$ & $10.1$ & $1.9$
\\ \hline
Brazil & $97.58$ & $97.52$ & $97.55$ & $18$\%  & $0.49$ & $5.7$ &
$3.0$ \\ \hline
Chile I   &  $91.77$ & $91.75$ & $91.94$ & $22\%$ & $0.50$ & $7.2$ & $2.4$
\\ \hline
Chile II  &  $94.10$ & $94.09$ & $94.26$ & $20\%$ & $0.30$ & $2.9$ & $8.8$
\\ \hline
Mexico I& $94.10$ & $94.09$ & $94.30$ & $32$\% & $0.12$ & $4.6$ & $3.9$
\\ \hline
Mexico II & $97.93$ & $97.80$ & $97.82$ & $21$\%  & $0.50$ & $6.1$ & $2.8$
\\ \hline
Peru I   & $93.84$ & $93.83$ & $93.88$ & $22$\% & $0.62$ & $11.2$ & $1.8$
\\ \hline
Venezuela & $97.75$ & $97.73$ & $98.07$ & $42$\% & $0.35$ & $3.9$ & $5.0$
\\ \hline
\end{tabular}
\vspace{5mm}
\caption{\label{lattab1}Crash and fit characteristics of the various
speculative
bubbles on the Latin-American market leading to a large draw down
in this decade. $t_c$ is the critical time predicted from the fit of the
market index to eq. (\protect\ref{lpeq}). When multiple fits
exists the fit with the smallest difference between $t_c$ and $t_{max}$ is
chosen. Typically, this will be the best fit, but occasionally it is the
second best fit. The other parameters
$z$, $\omega$ and $\lambda$ of the fit are also shown. The fit is performed
up to the time $t_{max}$,  at which the market index achieved its highest
maximum before the crash. The percentage drop is calculated from the total
loss from $t_{max}$ to $t_{min}$, where the market index achieved it lowest
value as a consequence of the crash. }
\end{center}
\end{table}

\begin{table}
\begin{center}
\begin{tabular}{|c|c|c|c|c|c|c|c|c|c|} \hline
{\small Stock market} & $A$ & $B$ & $C$ & $z$  & $t_c$ & $\omega$ &
$\phi$ \\ \hline
{\small Argentina I} & $36960;23439$ & $-35902;-23319$ & $-18839;-23905$ &
$0.16;0.31$ & $91.81;91.80$ & $4.8;4.5$ & $-0.3;0.0$
\\ \hline
{\small Argentina II} & $30737;36858$ & $-16199;-21792$ & $1034;-834$ & $0.26;0.17$
& $92.42;92.43$ & $11.4;12.1$ & $-0.8;1.6$
\\ \hline
{\small Argentina II} & $34211$ & $-19832$ & $950$ & $0.22$ & $92.43$ & $10.4$ & $0.7$
\\ \hline
{\small Argentina III} & $65798;29095$ & $-53867;-18531$ & $863;1183$ & $0.09;0.35$
& $94.16;94.13$ & $7.2;6.0$ & $-0.9;0.0$
\\ \hline
{\small Argentina IV} & $39906$ & $-21267$ & $-666$ & $0.20$ & $97.89$ & $10.1$ & $1.2$
\\ \hline
Brazil I & $16327$ & $-10717$ & $512$ & $0.33$ & $97.53$ & $5.7$ & $0.1$
\\ \hline
Chile I & $36785;33727$ & $-26853;-24040$ & $-131;-135$ & $0.40;0.45$ &
$91.81;91.79$ & $7.2;6.8$ & $-0.7;-0.7$
\\ \hline
Chile I & $40660$ & $-30726$ & $106$ & $0.38$ & $91.85$ & $5.8$ & $0.4$
\\ \hline
Chile II & $79138$ & $-56312$ & $-203$ & $0.15$ & $94.11$ & $2.9$ & $1.4$
\\ \hline
Mexico I & $3065$ & $-2097$ & $-244$ & $0.65$ & $94.13$ & $4.6$ & $0.6$
\\ \hline
Mexico II & $5637;6764$ & $-2475;-3571$ & $159;-100$ & $0.31;0.20$ &
$97.60;97.62$ & $6.1;12.1$ & $1.1;0.6$
\\ \hline
Peru I & $1770;1151$ & $-1516;-986$ & $-41.5;-62.7$ & $0.32;0.59$ &
$93.91;93.84$ & $11.2;8.6$ & $-1.1;-0.9$
\\ \hline
Peru I & $1435$ & $-1226$ & $47.7$ & $0.50$ & $93.93$ & $6.7$ & $-1.9$
\\ \hline
{\small Venezuela I}& $13460$ & $-8613$ & $801$ & $0.35$ & $97.75$ & $3.9$ & $-0.5$
\\ \hline
\end{tabular}
\vspace{5mm}
\caption{\label{lattab2}Fit parameters of the various
speculative bubbles on the Latin-American financial markets
leading to a crash in this decade. Multiple entries correspond to the two
best fits.}
\end{center}
\end{table}

\begin{table}[b]
\begin{center}
\begin{tabular}{|c|c|c|c|c|c|c|c|c|c|} \hline
Stock market & $A$ & $B$ & $C$ & $z$  & $t_c$ & $\omega$ &
$\phi$ \\ \hline
Argentina & $37072$ & $-31073$ & $-971$ & $0.22$ & $92.44$ & $11.8$ & $0.2$
\\ \hline
Chile & $6599$ & $-1216$ & $166$ & $0.36$ & $95.51$ & $9.7$ & $0.3$
\\ \hline
Venezuela & $10273;9997$ & $-5650;-5539$ & $1002;-1076$ & $0.59;0.63$ &
$97.80;97.81$ & $5.7;5.3$ & $1,8;-1.2$
\\ \hline
Venezuela & $10819;13460$ & $-6064;-8613$ & $884;801$ & $0.58;0.35$ &
$97.75;97.75$ & $6.7;3.9$ & $1.8;-0.5$
\\ \hline
\end{tabular}
\vspace{5mm}
\caption{\label{lattab3}Fit parameters of the anti-bubbles on the
Latin-American financial markets.}
\end{center}
\end{table}

\begin{table}[b]
\begin{center}
\begin{tabular}{|c|c|c|c|c|c|c|c|c|c|} \hline
Stock market & $t_c$ & $t_{max}$ & $t_{min}$ & $\%$ drop & $z$ & $\omega$ &
$\lambda$ \\ \hline
Hong-Kong I & $87.84$ & $87.78$ & $87.85$ & $50\%$ & $0.29$ & $5.6$ & $3.1$
\\ \hline
Hong-Kong II & $94.02$ & $94.01$ & $94.04$ & $17\%$ & $0.12$ & $6.3$ & $2.7$
\\ \hline
Hong-Kong III & $97.74$ & $97.60$ & $97.82$ & $42\%$ & $0.34$ & $7.5$ & $2.3$
\\ \hline
Indonesia I & $94.09$ & $94.01$ & $94.32$ & $26\%$ & $0.44$ & $15.5$ & $1.5$
\\ \hline
Malaysia I  & $94.02$ & $94.01$ & $94.04$ & $22\%$ & $0.24$  & $10.9$ & $1.8$
\\ \hline
Philippines I & $94.02$ & $94.01$ & $94.19$ & $25\%$ & $0.16$ & $8.2$ & $2.2$
\\ \hline
Thailand I & $94.07$ & $94.01$ & $94.05$ & $20\%$ & $0.48$  & $6.1$ & $2.8$
\\ \hline
\end{tabular}
\vspace{5mm}
\caption{\label{asitab1}Crash and fit characteristics of the various
speculative
bubbles on the Asian market leading to a large draw down
in this decade. $t_c$ is the critical time predicted from the fit of the
market index to eq. (\protect\ref{lpeq}). When multiple fits
exists the fit with the smallest difference between $t_c$ and $t_{max}$ is
chosen. Typically, this will be the best fit, but occasionally it is the
second best fit. The other parameters
$z$, $\omega$ and $\lambda$ of the fit are also shown. The fit is performed
up to the time $t_{max}$,  at which the market index achieved its highest
maximum before the crash. The percentage drop is calculated from the total
loss from $t_{max}$ to $t_{min}$, where the market index achieved it lowest
value as a consequence of the crash. }
\end{center}
\end{table}

\begin{table}[b]
\begin{center}
\begin{tabular}{|c|c|c|c|c|c|c|c|c|c|} \hline
Stock market & $A$ & $B$ & $C$ & $z$  & $t_c$ & $\omega$ &
$\phi$ \\ \hline
Hong-Kong I &  $5523;4533$ & $-3247;-2304$ & $171;-174$ & $0.29;0.39$ &
$87.84;87.78$ & $5.6;5.2$ & $-1.6;1.1$
\\ \hline
Hong-Kong II&  $21121$ & $-15113$ & $-429$ & $0.12$ & $94.02$ & $6.3$ & $-0.6$
\\ \hline
Hong-Kong III&  $20077$ & $-8241$ & $-397$ & $0.34$ & $97.74$ & $7.5$ & $0.8$
\\ \hline
Indonesia I & $6.76$ & $-1.11$ & $0.039$ & $0.44$ & $94.09$ & $15.6$ & $-1.3$
\\ \hline
Malaysia I & 7.61$$ & $-1.16$ & $0.038$ & $0.24$  & $94.02$ & $10.9$ & $1.4$
\\ \hline
Philippines I & $9.00$ & $-1.74$ & $-0.078$ & $0.16$ & $94.02$ & $8.2$ & $0.2$
\\ \hline
Thailand I   & $7.81$ & $-1.41$ & $-0.086$ & $0.48$ & $94.07$ & $6.1$ & $-0.2$
\\ \hline
\end{tabular}
\vspace{5mm}
\caption{\label{asitab2}Fit parameters of the various
speculative bubbles on the South-East and East Asian financial
markets leading to a large draw down in this decade. For Hong-Kong, the
crash of 87 has been included for completeness. Multiple entries correspond
tothe two best fits.}
\end{center}
\end{table}

\begin{table}[b]
\begin{center}
\begin{tabular}{|c|c|c|c|c|c|c|c|c|c|} \hline
Stock market & $t_c$ & $t_{max}$ & $t_{min}$ & $\%$ drop & $z$ & $\omega$ &
$\lambda$ \\ \hline
London &  $94.08$ & $94.09$ & $94.48$ & $18\%$ & $0.25$ & $7.6$ & $2.3$
\\ \hline
Hong Kong &  $94.09$ & $94.09$ & $94.53$ & $31\%$ & $0.03$ & $11$ & $1.8$
\\ \hline
Australia&  $94.08$ & $94.09$ & $95.11$ & $22\%$ & $0.46$ & $8.0$ & $2.2$
\\ \hline
New Zealand & $94.08$ & $94.09$ & $94.95$ & $23\%$ & $0.09$ & $7.7$ & $2.3$
\\ \hline
France&  $94.06$ & $94.09$ & $95.20$ & $27\%$ & $0.51$ & $12$ & $1.7$
\\ \hline
Spain&  $94.08$ & $94.09$ & $95.23$ & $27\%$ & $0.28$ & $13$ & $1.6$
\\ \hline
Italy& $94.36$ & $94.36$ & $95.21$ & $28$\% & $0.35$ & $9.2$ & $2.0$
\\ \hline
Switzerland& $94.08$ & $94.08$ & $94.54$ & $22$\%  & $0.45$ & $12$ & $1.7$
\\ \hline
\end{tabular}
\vspace{5mm}
\caption{\label{antitab1}Fit characteristics of the 1994 anti-bubble on the
Western financial markets plus Hong Kong following the emerging markets
collapse in early 1994. $t_c$ is the critical time predicted from the fit of
the market index to eq. (\protect\ref{lpdeceq}). When multiple fits
exists the fit with the smallest difference between $t_c$ and $t_{max}$ is
chosen. Typically, this will be the best fit, but occasionally it is the
second best fit. The other parameters
$z$, $\omega$ and $\lambda$ of the fit are also shown. The fit is performed
from the time $t_{max}$, at which the market index achieved its highest
maximum before the decrease, to the time $t_{min}$, which is the time of the
lowest point of the market before a shift in the trend. The percentage drop is
calculated from the total loss from $t_{max}$ to $t_{min}$ using eq.
(\protect\ref{dropeq}). }
\end{center}
\end{table}

\begin{table}[b]
\begin{center}
\begin{tabular}{|c|c|c|c|c|c|c|c|c|c|} \hline
Stock market & $A$ & $B$ & $C$ & $z$  & $t_c$ & $\omega$ &
$\phi$ \\ \hline
Australia & $7.74$ & $-0.177$ & $-0.035$ & $0.46$ & $94.08$ & $8.0$ & $-0.7$
\\ \hline
France &  $8.36;7.78$ & $-0.810;0.265$ & $0.032;0.036$ & $0.18;0.51$ &
$93.86;94.07$ & $16.5;12.2$ & $1.6;-1.4$
\\ \hline
Hong-Kong & $12.15$ & $-3.14$ & $0.039$ & $0.03$ & $94.09$ & $10.8$ & $0.72$
\\ \hline
Italy & $9.88$ & $-0.340$ & $-0.050$ & $0.35$ & $94.36$ & $9.2$ & $0.2$
\\ \hline
London & $8.29;8.21$ & $-0.35;-0.29$ & $-0.016;-0.019$ & $0.25;0.35$ &
$94.08;94.08$ & $7.6;7.1$ & $-0.6;-0.1$
\\ \hline
New Zealand & $8.15$ & $-0.551$ & $-0.028$ & $0.09$ & $94.08$ & $7.8$ & $-0.3$
\\ \hline
Spain & $8.41$ & $-0.370$ & $-0.030$ & $0.28$ & $94.08$ & $13.2$ & $1$
\\ \hline
Switzerland & $8.79;8.09$ & $-0.986;-0.313$ & $0.024;0.036$ & $0.12;0.45$ &
$93.99;94.08$ & $17.1;12.2$ & $-1.7;-0.3$
\\ \hline
\end{tabular}
\vspace{5mm}
\caption{\label{antitab2}Fit parameters of the various anti-bubbles on the
western markets and Hong-Kong}
\end{center}
\end{table}


\newpage

\begin{thebibliography}{}

\bibitem[Ball and Holt, 1998]{BallHolt} Ball S.B. and Holt C.A.,
{\it Classroom games: Speculation and bubble in an asset market},
J. of Economic Perspectives 12, 207-218.

\bibitem[Barro {\it et al.}, 1989]{krach87} Barro R.J., Fama E.F., Fischel D.R.,  Meltzer A.H.,
Roll R. and Telser L.G., {\it Black monday and the future of financial markets}, edited by
R.W. Kamphuis, Jr., R.C. Kormendi and J.W.H. Watson (Mid American Institute for
Public Policy Research, Inc. and Dow Jones-Irwin, Inc..

\bibitem[Blanchard, 1979]{Blanchard} Blanchard O.J.,
{\it Speculative bubbles, crashes and rational expectations},
Economics Letters. {\bf 3}, 387-389.

\bibitem[Busshaus and Rieger, 1999]{StaufferlargeN1} Busshaus C. and Rieger H., 
{\it A prognosis oriented microscopic stock market model},
Physica A {\bf 267}, 443.


\bibitem[Cardy, 1998]{finitesize} Cardy J. L.,
{\it ed., Finite-size scaling},
North-Holland,  New York, NY, USA.


\bibitem[Chan {\it et al.}, 1998]{Chanetal} Chan K., McQueen G. and Thorley S.,
{\it Are there rational speculative bubbles in Asian stock markets?},
Pacific-Basin Finance Journal {\bf 6}, 125-151.

\bibitem[Drozdz {\it et al.}, 1999]{Drozdz} Drozdz S., Ruf F., Speth J., Wojcik M.,
{\it Imprints of log-periodic self-similarity in the stock market},
in press in Eur. Phys. J. B, http://xxx.lanl.gov/abs/cond-mat/9901025

\bibitem[Egenter {\it et al.}. 1999]{StaufferlargeN2} Egenter E.,  Lux T. and Stauffer D., 
{\it Finite size effects in Monte Carlo simulations of two stock market models}, 
Physica A {\bf 268}, 250.

\bibitem[Feigenbaum and Freund, 1996]{FF96} Feigenbaum J.A. and Freund P.G.O., 
{\it Discrete scale invariance in stock markets before crashes},
Int. J. Mod. Phys. C {\bf 10}, 3737-3745.

\bibitem[Feigenbaum and Freund, 1998]{97crash} Feigenbaum J.A. and Freund P.G.O.,
{\it Discrete Scale Invariance and the ``Second Black Monday''}.
Mod. Phys. Letters {\bf 12}, 57-60.

\bibitem[Flannery {\it et al}, 1992]{Lomb} Press W.H., Flannery B.P., Teukolsky S.A. and Vetterling W.T.,
{\it Numerical Recipes},
Cambridge University Press, Cambridge UK.

\bibitem[Galbraith, 1997]{Galbraith} Galbraith J.K.,
{\it The great crash, 1929},
Boston : Houghton Mifflin Co., 1997.

\bibitem[Gluzman and Yukalov, 1998]{Gluzman} Gluzman S. and Yukalov V. I.,
{\it Renormalization Group Analysis of October Market Crashes},
Mod. Phys. Lett. B {\bf 12}, 75-84.

\bibitem[Hellthaler, 1995]{StaufferlargeN3}  Hellthaler T.,
{\it The influence of investor number on a microscopic market model},
Int. J. Mod. Phys.C {\bf 6}, 845.

\bibitem[Intriligator, 1998]{Intriligator} Intriligator M.D.,
{\it Russia\,: Lessons of the economic collapse},
New York Times, Aug. 8, 1998. Presented to the World Bank, September 4.

\bibitem[Johansen, 1997]{Thesis} Johansen A.,
{\it Discrete scale invariance and other cooperative phenomena in
spatially extended systems with threshold dynamics}, Ph.D. Thesis, Niels Bohr
Inst. (Dec. 1997). Available on www.nbi.dk/\~~johansen/pub.html

\bibitem[Johansen and Sornette, 1998]{predic} Prediction communicated on sept. 17, 1997
to the French office for the  protection of
proprietary softwares and inventions under number registration 94781 (see also
footnote 13 of D. Stauffer and D. Sornette,
Log-periodic oscillations for biased diffusion on random lattice,
Physica A 252, 271-277 (1998).


\bibitem[Johansen and Sornette, 1999a]{JS98.2} Johansen A. and  Sornette D.,
{\it Critical Crashes}, Risk Magazine {\bf 12}, 91-94.

\bibitem[Johansen and Sornette, 1999b]{manisfesto} Johansen A. and  Sornette D.,
{\it Modeling the stock market prior to large crashes},
Eur. Phys. J. B {\bf 9}, 167-174.

\bibitem[Johansen {\it et al.}, 1999c]{JSL} Johansen A., Sornette D. and Ledoit O.,
{\it Predicting Financial Crashes using discrete scale invariance},
in press in J. of Risk, preprint 
(http://econwpa.wustl.edu/eprints/fin/papers/9903/9903006.abs
and http://xxx.lanl.gov/abs/cond-mat/9903321)

\bibitem[Johansen and Sornette, 1999d]{antibulle} Johansen A. and Sornette D.,
{\it Financial ``anti-bubbles'': log-periodicity in Gold and Nikkei collapses},
Int. J. Mod. Phys. B  in press, preprint
http://xxx.lanl.gov/abs/cond-mat/9901268.

\bibitem[Johansen and Sornette, 1999e]{predicNikkei} see http://alf.nbi.dk/~~johansen/pub.html for
comparison between the early Jan.99 prediction and the recent Nikkei
quotes; see also
D. Stauffer, Monte-Carlo-simulation mikroskopischer B\"{o}rsenmodelle,
Physikalische Blatter 55 N5, 49-51 (May 1999) where our prediction is
explicitely quoted.

\bibitem[Johansen {\it et al.}, 2000]{JLS} Johansen A., Ledoit O. and Sornette D.,
{\it Crashes as critical points},
Int. J. Theo. \& Appl. Finance Vol. 3 No. 1, http://xxx.lanl.gov/abs/cond-mat/9810071.

\bibitem[Keynes, 1964]{Keynes} Keynes J.M.,
{\it The general theory of employment, interest and money},
New York, Harcourt, Brace \& World, Inc., 1964 (originally 1936).

\bibitem[Kohl, 1997]{StaufferlargeN4} Kohl R., 
{\it The influence of the number of different stocks on the
Levy-Levy-Solomon model}, Int. J. Mod. Phys. C {\bf 8}, 1309;

\bibitem[Lowell {\it et al.}, 1998]{Rand} Lowell J., Neu C.R. and Tong D.,
{\it Financial crises and contagion in emerging market countries}, RAND.

\bibitem[Montroll and Badger, 1974]{Montroll} Montroll E.W. and Badger W.W.,
{\it Introduction to quantitative aspects of social phenomena},
Gordon and Breach Science Publishers, New York.

\bibitem[Saleur and Sornette, 1996]{Saleursor} Saleur H. and Sornette D.,
{\it Complex exponents and log-periodic corrections in frustrated systems},
J.Phys.I France {\bf 6}, 327-355.


\bibitem[Sornette, 1998]{physreports} Sornette D.,
{\it Discrete scale invariance and complex dimensions},
Physics Reports {\bf 297}, 239-270.


\bibitem[Sornette {\it et al.}, 1996]{SJB96}Sornette D., Johansen A. and Bouchaud J.P.,
{\it Stock market crashes, Precursors and Replicas},
J. Phys. I France {\bf 6}, 167-175.

\bibitem[Sornette and Johansen, 1997]{SJ97} Sornette D. and Johansen A.,
{\it Large financial crashes}, 
Physica A {\bf 245}, 411-422.

\bibitem[Sornette and Johansen, 1998]{SJ98} Sornette D. and Johansen A.,
{\it A Hierarchical Model of Financial Crashes},
Physica A {\bf 261}, 581-598.

\bibitem[Stauffer and Sornette, 1994]{percostausor} Stauffer D. and Sornette D.,
{\it Self-Organized Percolation Model for Stock Market Fluctuations},
submitted Physica A (http://xxx.lanl.gov/abs/cond-mat/9906434)

\bibitem[Vandewalle {\it et al.}, 1998a]{Van1} Vandewalle N., Boveroux Ph., Minguet A. and Ausloos M.,
{\it The krach of October 1987 seen as a phase transition: amplitude and
universality.} Physica A {\bf 255}, 201-210.

\bibitem[Vandewalle {\it et al.}, 1998b]{Van2} Vandewalle N., Ausloos M., Boveroux Ph. and Minguet A.,
{\it How the financial crash of October 1997 could have been predicted},
Eur. Phys. J. B {\bf 4}, 139-141.


\end{thebibliography}
\end{document}